\pdfoutput=1
\documentclass[a4paper,11pt]{article}
\usepackage[mathlines]{lineno}
\usepackage{jcappub}

\usepackage[table,svgnames,dvipsnames]{xcolor}

\usepackage[normalem]{ulem}
\usepackage{aas_macros}
\usepackage{subfigure}
\usepackage{graphicx}
\usepackage{graphics}
\usepackage{dcolumn}
\usepackage{bm}
\usepackage{cases}
\usepackage{booktabs}
\usepackage{comment}
\usepackage{multirow}
\usepackage{makecell}
\usepackage{siunitx}
\usepackage{tabularx}
\usepackage{xspace}
\usepackage{soul} 
\graphicspath{{figs/}}

\usepackage{orcidlink} 
\usepackage{hanging} 

\def\be{\begin{equation}}
\def\ee{\end{equation}}

\def\ba#1\ea{\begin{align*}#1\end{align*}}

\renewcommand{\emph}[1]{\textit{#1}}
\definecolor{RoyalBlue}{rgb}{0.25,.41,.88}
\definecolor{WildStrawberry}{HTML}{EE2967}
\definecolor{RedWine}{rgb}{0.743,0,0}
\definecolor{bittersweet}{rgb}{1.0, 0.44, 0.37}
\definecolor{burntorange}{rgb}{0.8, 0.33, 0.0}
\definecolor{midnightgreen}{rgb}{0.0, 0.29, 0.33}
\definecolor{otherblue}{rgb}{0.20, 0.73, 0.92}

\usepackage[nameinlink,noabbrev]{cleveref}
\crefname{equation}{Eq.}{Eqs.}
\crefname{section}{Section}{Sections}
\crefname{figure}{Figure}{Figures}
\crefname{table}{Table}{Tables}
\crefname{appendix}{Appendix}{Appendices}
\Crefname{figure}{Figure}{Figures}
\Crefname{equation}{Equation}{Equations}
\Crefname{section}{Section}{Sections}
\Crefname{table}{Table}{Tables}

\newcommand{\mksym}[1]{\ifmmode {\rm #1}\else #1\fi}
\newcommand{\dataplus}{\allowbreak+}

\newcommand{\leftparbox}[2]{\parbox{#1}{\begin{flushleft} #2 \end{flushleft}}}
\newcommand{\oneonesig}[4][5cm]{
\begin{equation}
\left.
#2 \quad\mbox{\text{\leftparbox{#1}{(#3)#4}}}
  \right.
\end{equation}
}
\newcommand{\onetwosig}[4][5cm]{
\begin{equation}
\left.
  #2 \quad\mbox{\text{\leftparbox{#1}{(95\,\%,~#3)#4}}}
  \right.
\end{equation}
}
\newcommand{\twoonesig}[4][\pbwidth]{
\begin{equation}
\left.
 \begin{aligned}
#2 \\ #3
 \end{aligned}
\ \right\} \ \ \mbox{\text{\leftparbox{#1}{#4}}}
\end{equation}
}

\newcommand{\threeonesig}[5][\pbwidth]{
\begin{equation}
\left.
 \begin{aligned}
#2 \\ #3 \\ #4
 \end{aligned}
\ \right\} \ \ \mbox{\text{\leftparbox{#1}{#5}}}
\end{equation}
}

\newcommand{\Om}{\Omega_\mathrm{m}}

\newcommand{\Ob}{\Omega_\mathrm{b}}
\newcommand{\Ok}{\Omega_\mathrm{K}}
\newcommand{\Or}{\Omega_\mathrm{R}}
\newcommand{\OL}{\Omega_\Lambda}
\newcommand{\ob}{\omega_\mathrm{b}}
\newcommand{\ocdm}{\omega_\mathrm{cdm}}
\newcommand{\om}{\omega_\mathrm{m}}

\newcommand{\Neff}{N_{\mathrm{eff}}}

\newcommand{\lcdm}{$\Lambda$CDM} 
\newcommand{\wcdm}{$w$CDM} 
\newcommand{\wowacdm}{$w_0w_a$CDM} 
\newcommand{\lya}{Ly$\alpha$\xspace}

\newcommand{\DVrd}{D_\mathrm{V}/r_\mathrm{d}}
\newcommand{\DMrd}{D_\mathrm{M}/r_\mathrm{d}}
\newcommand{\DHrd}{D_\mathrm{H}/r_\mathrm{d}}
\newcommand{\DM}{D_\mathrm{M}}
\newcommand{\DV}{D_\mathrm{V}}
\newcommand{\Hrd}{H_0r_\mathrm{d}}
\newcommand{\rd}{r_\mathrm{d}}
\newcommand{\zd}{z_\mathrm{d}}
\newcommand{\cs}{c_\mathrm{s}}
\newcommand{\FAP}{F_\mathrm{AP}}
\newcommand{\sumnu}{\sum m_\nu}

\newcommand{\Planck}{\emph{Planck}}
\newcommand{\zpiv}{z_\mathrm{p}}
\newcommand{\wpiv}{w_\mathrm{p}}

\newcommand{\Mpc}{\, \text{Mpc}}
\newcommand{\hinvMpccubed}{\, h^{-3} \, \text{Mpc}^{3}}
\newcommand{\hMpcinvcubed}{\, h^{3} \, \text{Mpc}^{-3}}

\newcommand{\hmpcinv}{\,h\,{\rm Mpc^{-1}}}
\newcommand{\kmsMpc}{\,{\rm km\,s^{-1}\,Mpc^{-1}}}
\newcommand{\eV}{{\,\rm eV}}

\newcommand{\planckact}{\Planck+ACT}

\arxivnumber{2404.03002} 
\title{DESI 2024 VI:  Cosmological Constraints from the Measurements of Baryon Acoustic Oscillations}

\author{{DESI Collaboration}:}
\emailAdd{spokespersons@desi.lbl.gov}
\affiliation{Affiliations are in Appendix \ref{sec:affiliations}}

\author[1]{{A.~G.~Adame},}
\author[2]{{J.~Aguilar},}
\author[3]{{S.~Ahlen}\orcidlink{0000-0001-6098-7247},}
\author[4]{{S.~Alam}\orcidlink{0000-0002-3757-6359},}
\author[5,6]{{D.~M.~Alexander}\orcidlink{0000-0002-5896-6313},}
\author[2]{{M.~Alvarez},}
\author[7]{{O.~Alves},}
\author[2]{{A.~Anand}\orcidlink{0000-0003-2923-1585},}
\author[8,7]{{U.~Andrade}\orcidlink{0000-0002-4118-8236},}
\author[9]{{E.~Armengaud}\orcidlink{0000-0001-7600-5148},}
% ...

\date{\today}

\abstract{
We present cosmological results from the measurement of baryon acoustic oscillations (BAO) in galaxy, quasar and Lyman-$\alpha$ forest tracers from the first year of observations from the Dark Energy Spectroscopic Instrument (DESI), to be released in the DESI Data Release 1. DESI BAO provide robust measurements of the transverse comoving distance and Hubble rate, or their combination, relative to the sound horizon, in seven redshift bins from over 6 million extragalactic objects in the redshift range $0.1 <z<4.2$. To mitigate confirmation bias, a blind analysis was implemented to measure the BAO scales. DESI BAO data alone are consistent with the standard flat \lcdm\ cosmological model with a matter density $\Om=0.295\pm0.015$.  Paired with a baryon density prior from Big Bang Nucleosynthesis and the robustly measured acoustic angular scale from the cosmic microwave background (CMB), DESI requires $H_0=(68.52\pm0.62)\,\kmsMpc$. In conjunction with CMB anisotropies from \Planck\ and CMB lensing data from \Planck\ and ACT, we find $\Om=0.307\pm 0.005$ and $H_0=(67.97\pm0.38)\,\kmsMpc$. Extending the baseline model with a constant dark energy equation of state parameter $w$, DESI BAO alone require $w=-0.99^{+0.15}_{-0.13}$. In models with a time-varying dark energy equation of state parametrised by $w_0$ and $w_a$, combinations of DESI with CMB or with type Ia supernovae (SN~Ia) individually prefer $w_0>-1$ and $w_a<0$. This preference is 2.6$\sigma$ for the DESI+CMB combination, and persists or grows when SN~Ia are added in, giving results discrepant with the \lcdm\ model at the $2.5\sigma$, $3.5\sigma$ or $3.9\sigma$ levels for the addition of the Pantheon+, Union3, or DES-SN5YR supernova datasets respectively. For the flat \lcdm\ model with the sum of neutrino mass $\sumnu$ free, combining the DESI and CMB data yields an upper limit $\sumnu < 0.072$ $(0.113)\eV$ at 95\% confidence for a $\sumnu>0$ $(\sumnu>0.059) \eV$ prior. These neutrino-mass constraints are substantially relaxed if the background dynamics are allowed to deviate from flat \lcdm. 
}

\begin{document}
\maketitle
\flushbottom

\section{Introduction}
\label{sec:intro}

The discovery of the accelerated expansion of the universe \cite{SupernovaSearchTeam:1998fmf,SupernovaCosmologyProject:1998vns} 
demonstrated that the dynamics of the universe are presently dominated by dark energy, a component with negative pressure \cite[for reviews, see, e.g.,][]{Frieman:2008sn,Weinberg:2013agg}. Over the past quarter-century, a wide variety of cosmological measurements have lent further support for what has become the standard cosmological model: a spatially flat universe with an energy budget today composed of about 5\% baryonic matter, 25\% cold dark matter (CDM), 70\% dark energy in the form of Einstein's cosmological constant ($\Lambda$) and smaller contributions provided by massive neutrinos and radiation.

Increasingly accurate data from type Ia supernovae (SN~Ia) have confirmed and significantly strengthened the original ground-breaking results for the accelerated expansion of the universe (e.g.,\ \cite{SupernovaSearchTeam:2004lze,ESSENCE:2007acn,SNLS:2011lii,SupernovaCosmologyProject:2011ycw,Betoule:2014}). Meanwhile, mapping out the anisotropies in the cosmic microwave background (CMB) radiation, starting with the ground-breaking COBE results \cite{Bennett:1996}, and continuing with increasingly precise measurements \cite{Miller:1999,Hanany:2000qf,Boomerang:2001tye} that led to the full-sky maps by WMAP \cite{Hinshaw:2013} and Planck \cite{Planck-2018-cosmology} experiments, as well as mapping of the CMB anisotropy on smaller angular scales 
\cite{Hou:2012xq,Aiola:2020}, revolutionised the field of cosmology by providing percent-level constraints on key cosmological parameters.
Recent cosmological constraints from probes of the large-scale structure \cite{2021A&A...646A.140H,2022PhRvD.105b3520A,2023PhRvD.108l3520M,2023OJAp....6E..36D}, CMB \cite{Madhavacheril:ACT-DR6,2023PhRvD.108l2005P}, and distance measurements from SN~Ia \cite{Brout:2022,Rubin:2023,DES:2024tys} have largely confirmed and sharpened constraints on the  cosmological model while establishing an increasingly sophisticated methodology framework focused on stringent control of systematic errors.

Current constraints on key cosmological parameters are largely consistent with the $\Lambda$CDM cosmological model with cold dark matter and dark energy described by the cosmological constant $\Lambda$. However, dark energy dynamics has not been stringently tested.
In addition, within the $\Lambda$CDM model,  tensions have appeared at various degrees of statistical significance (the two most popular being referred to as the ``Hubble tension" and the ``$\sigma_8$ tension"). Such tensions, if not due to unaccounted systematics, may indicate new physics beyond $\Lambda$CDM.

The use of galaxies as tracers of large-scale structure has traditionally played an essential role in cosmology and provides important complementary information to that from SN~Ia and CMB. Galaxy clustering was pioneered half a century ago and further developed and applied to larger and better galaxy catalogs in the intervening years \cite{1973ApJ...185..413P,1974ApJS...28...19P,Groth:1977gj,Davis:1982gc,1988ApJ...332...44D,1993MNRAS.265..145B,2001Natur.410..169P,2001MNRAS.327.1297P, 2003AJ....126.2081A,2004ApJ...606..702T,2006PhRvD..74l3507T}. It has established itself as a key cosmological measurement that provides a direct link between observations and properties of dark matter and dark energy.  

The clustering of matter encodes a preferred scale, the sound horizon at the baryon drag epoch of the early universe \cite{1998ApJ...496..605E}.  This feature, which is imprinted on the matter distribution of the early universe by physics around recombination and earlier, is stretched with the expansion of the universe, appearing at a comoving galaxy separation of $\rd \sim 150 \Mpc$. Hence $\rd$ is a standard ruler, whose length is dictated by early-time physics. In particular, the length of the standard ruler may be calibrated with high accuracy, e.g., by CMB observations \cite{Planck-2018-cosmology}.  Since galaxies trace the matter content of the universe, the BAO feature is transferred into galaxy clustering, where it manifests as a single localised peak in the galaxy correlation function and an oscillatory signature, or ``wiggles", in the galaxy power spectrum.  Furthermore, the signature is also visible in other tracers of mass such as fluctuations in the Lyman-$\alpha$ forest --- spectral features that indicate the radial distribution of neutral hydrogen clouds between the observer and distant quasars. Because the BAO feature has a distinct signature and resides in the linear-clustering regime, measurements of this scale are relatively free from systematic errors associated with nonlinear physics. BAO provides a key cosmological probe sensitive to the cosmic expansion history, with well-controlled systematics. Using BAO we can test for dark energy dynamics and spatial curvature, and in combination with other probes constrain the Hubble constant, the sum of neutrino masses, and the number of light species.  Several reviews of the BAO as a cosmological probe are available \cite[e.g.,][]{2005NewAR..49..360E, 2010deot.book..246B, 2013PhR...530...87W, 2021PhRvD.103h3533A}.

The accuracy with which the BAO feature may be measured from a galaxy survey is principally limited by sample variance and Poisson noise in the galaxy clustering measurements, necessitating galaxy surveys with effective volume of at least $1 \, h^{-3} \, \text{Gpc}^3$ \cite{2006MNRAS.365..255B, 2007ApJ...665...14S, 2008MNRAS.383..755A}.\footnote{An exception is the \lya BAO measurement, limited by the number of spectra and their signal-to-noise, with only a minor contribution from cosmic variance \cite{2007PhRvD..76f3009M,McQuinn2011,2014JCAP...05..027F}.} The BAO feature was first detected in 2005  by the Sloan Digital Sky Survey (SDSS) \cite{2005ApJ...633..560E} and the Anglo-Australian Telescope Two-degree Field Galaxy Redshift Survey \cite{20052dFBAO}. Subsequent measurements, leading to distance determinations accurate to within a few percent, were performed using the SDSS-III Luminous Red Galaxy Sample \cite{ 2007MNRAS.381.1053P, 2010MNRAS.401.2148P}, the WiggleZ Dark Energy Survey \cite{2011MNRAS.415.2892B, 2011MNRAS.418.1707B, 2014MNRAS.441.3524K} and the 6-degree Field Galaxy Survey \cite{2011MNRAS.416.3017B, 2018MNRAS.481.2371C}. Further extensions of the SDSS yielded more accurate percent-level BAO measurements using the Baryon Oscillation Spectroscopic Survey \cite[BOSS,][]{2012MNRAS.427.3435A, 2014MNRAS.441...24A, 2017MNRAS.470.2617A} at $z < 0.7$ and its extension \cite[eBOSS,][]{2018MNRAS.473.4773A, 2021MNRAS.500..736B, 2021MNRAS.500.1201H} at $z > 0.7$.  BAOs have also been detected within the Lyman-$\alpha$ forest mapped by quasar surveys, allowing distance and expansion measurements at $z > 2$ \cite{2013A&A...552A..96B, 2014JCAP...05..027F, 2015A&A...574A..59D, 2017A&A...603A..12B, 2020ApJ...901..153D}. Prior to DESI these different BAO measurements, considered together, had mapped out the cosmological distance-redshift relation with $1\%$-$2\%$ accuracy at a series of redshifts in the range $z < 2.5$ \cite{2015PhRvD..92l3516A, 2021PhRvD.103h3533A}.

The Dark Energy Spectroscopic Instrument (DESI) is carrying out a Stage IV survey \cite{DESI2016a.Science,DESI2022.KP1.Instr} that was designed to significantly improve cosmological constraints through measurements of the clustering of galaxies, quasars, and the Lyman-$\alpha$ forest. DESI is conducting a five-year survey over $14,200$ square degrees in the redshift range  $0.1<z<4.2$ with a spectroscopic sample size that will be ten times that of the previous SDSS surveys. DESI covers six different classes of tracers, including low redshift galaxies of the bright galaxy survey (BGS), luminous red galaxies (LRG), emission line galaxies (ELG), quasars as direct tracers, and Lyman-$\alpha$ (\lya) forest quasars to trace the distribution of neutral hydrogen. Additionally, a sample of stellar targets is being observed to a high density in an overlapping Milky Way Survey \cite{MWS.TS.Cooper.2023} to explore the stellar evolution and kinematics of the Milky Way. For cosmology, DESI is designed to impose stringent constraints on both the expansion history and the growth rate of large scale structure. 
Promising detections of the BAO feature at the few percent level \cite{BAO.EDR.Moon.2023} from the DESI early data release \cite{DESI2023b.KP1.EDR} have confirmed that DESI is on target to meet the top-level science requirements on BAO measurements. Specifically, DESI will tightly constrain matter density in the universe, the equation of state of dark energy, spatial curvature, the amplitude of primordial fluctuations, and neutrino mass. It will also sharply test modifications to the general theory of relativity proposed to explain the accelerated expansion of the universe \cite{Koyama:2015vza,Joyce:2016vqv,Ishak:2018his,2021JCAP...11..050A}.

In this paper, we report the constraints derived from the measurements of baryon acoustic oscillations with the first year of data from DESI as part of a larger series of papers. The results of this paper build upon the two-point measurements and validation, as detailed in \cite{DESI2024.II.KP3}; the BAO measurement from galaxies and quasars as summarised in \cite{DESI2024.III.KP4}; and the Ly-$\alpha$ forest BAO measurements and validation described in \cite{DESI2024.IV.KP6}. In a follow-up paper, the clustering analysis of the first year of DESI data is also performed over a wider range of scales than just the BAO feature including the effects of redshift space distortions (RSD) \cite{DESI2024.V.KP5}. The Data Release 1 (DR1) of the Dark Energy Spectroscopic Instrument \cite{DESI2024.I.DR1}, comprising spectroscopic data obtained in the first year of observations, will be made public at a later stage. 

The paper is structured as follows: \cref{sec:data} summarises our data and methodology, including a brief overview of the BAO data analysis and the integration of external datasets. \cref{sec:desi_cosmo} presents the results of the BAO analysis and discusses internal consistencies within the DESI data and external consistencies with SDSS data. \cref{sec:lcdm} presents our constraints on the \lcdm\ model, including a comparison of our results with external data. \cref{sec:DE} discusses DESI constraints on extended dark energy models both independently and in combination with CMB and SN~Ia data. In \cref{sec:Hubble}, we explore the constraints on the Hubble parameter for a variety of cosmological models and dataset combinations. \cref{sec:neutrinos} summarises our constraints on the neutrino sector. Lastly, we summarise our findings in \cref{sec:conclusions}.

\section{Data and methodology}
\label{sec:data}

In this section, we introduce the cosmological quantities relevant for this analysis, review the basics of the BAO as a cosmological probe, then describe  the DESI data we analyze, the external datasets we combine with DESI data, and the analysis methods we adopt. 

\subsection{Cosmological background} 
\label{sec:cosmo}

In a homogeneous and isotropic cosmology (i.e.,\ with the Friedmann-Lema\^{i}tre-Robertson-Walker (FLRW) metric), the transverse comoving distance is
\be 
\DM(z)=\frac{c}{H_0\sqrt{\Ok}}\, \sinh\left[\sqrt{\Ok}\int_0^z \frac{dz^\prime}{H(z^\prime)/H_0}\right]\ . 
\ee
Here $z$ is redshift, $H_0$ is the Hubble constant with $H(z)$ the Hubble parameter, $\Ok$ is the curvature density parameter, and $c$ is the speed of light. In $\Lambda$CDM, $\Ok = 1- \Om - \OL-\Or$, where $\Om$, $\Or$, and $\OL$ are the energy densities relative to critical in matter, radiation, and the cosmological constant, respectively, all evaluated at the present time. Since $\sinh$ is a complete function, the expression above holds for positive, negative, or zero $\Ok$. 
The Hubble parameter in $\Lambda$CDM is 
\begin{equation}
    H(z) = H_0\sqrt{\Om(1+z)^3 +\Or(1+z)^4 +\Ok(1+z)^2 +\OL }.
\end{equation}
The Hubble parameter also contains the contribution from neutrinos given by their energy density relative to critical $\Omega_\nu$. At sufficiently high redshifts all neutrinos behave as relativistic species, but at the low redshifts that DESI tracers probe, their contribution is dominated by that of massive neutrino species which contribute to $H(z)$ like matter.

BAO measurements depend on the sound horizon at the drag epoch $\rd$. This is the distance that sound can travel between the Big Bang and the drag epoch which indicates the time when the baryons decoupled from photons. At redshifts higher than the drag-epoch redshift $\zd$, the photons were coupled to electrons via Compton scattering, and they in turn had Coulomb interactions with baryons. The redshift when the baryons were released from the ``Compton drag" of the photons, $\zd$, can be obtained with a recombination calculation (a fitting formula is given in \cite{1998ApJ...496..605E}). Note that the drag epoch occurs slightly later (at $\zd\simeq 1060$ in the standard model) than photon decoupling ($z_\ast\simeq 1090$) simply because there are so few baryons relative to the number of photons. The expression for $\rd$ is given by 
\begin{equation}
    \rd=\int_{\zd}^\infty\frac{\cs (z)}{H(z)}dz\,,
    \label{eq:rd}
\end{equation}
where $\cs$ is the speed of sound which, prior to recombination, is given by
\begin{equation}
\cs(z) = \frac{c}
{\sqrt{3\displaystyle\left (1+\frac{3\rho_\mathrm{B}(z)}{4\rho_\gamma(z)}\right )}}\,
\end{equation}
where $\rho_\mathrm{B}$ and $\rho_\gamma$ are the baryon and radiation energy densities, respectively. The speed of sound evaluates to approximately $\cs\simeq c/\sqrt{3}$ well before recombination, but then decreases, and finally drops sharply at recombination. For standard early-time physics assumptions, the drag-epoch sound horizon can be approximated with \cite{Brieden2h}
\begin{equation}
    \rd=147.05\left(\frac{\om}{0.1432}\right)^{-0.23}\left(\frac{\Neff}{3.04}\right)^{-0.1}\left(\frac{\ob}{0.02236}\right)^{-0.13}\;{\rm Mpc},
    \label{eq:rdinLCDM}
\end{equation}
where $\om\equiv \Om h^2$ and $\ob\equiv \Ob h^2$ are the matter and baryon physical energy densities today, respectively, and $\Neff$ is the effective number of extra relativistic degrees of freedom.

The sound horizon at the drag epoch leaves an imprint in the distribution of matter that serves as a cosmological standardised ruler \cite{2003ApJ...594..665B, 2003ApJ...598..720S, 2003PhRvD..68h3504L, 2007PhRvD..76f3009M}.
To illustrate the concept, consider an ensemble of galaxy pairs at a given redshift $z$:  
if the separation vectors of the pairs are oriented perpendicular to the observer's line-of-sight, a preferred angular separation of galaxies $\Delta \theta$ may be observed, measuring the comoving distance $D_\mathrm{M}(z) \equiv \rd/\Delta \theta$ to this redshift. 
With the separation vector oriented parallel to the line-of-sight, a preferred redshift separation $\Delta z$ may be observed, measuring a comoving distance interval that for small intervals gives the Hubble parameter at that redshift, represented in this paper by the equivalent distance variable $D_\mathrm{H}(z)\equiv c/H(z) = r_\mathrm{d}/\Delta z$. BAO measurements hence constrain the quantities $D_\mathrm{M}(z)/r_\mathrm{d}$ and $D_\mathrm{H}(z)/r_\mathrm{d}$. This interpretation holds true for standard assumptions and models not too dissimilar from $\Lambda$CDM, given the statistical power of the data \cite{2015JCAP...01..034T}.
For those BAO measurements in certain redshift bins with low signal-to-noise ratio, we instead quote constraints on the angle-averaged quantity, $\DV(z) / \rd$, where $\DV(z)$ is the angle-average distance that quantifies the average of the distances measured along, and perpendicular to, the line of sight to the observer \cite{2005ApJ...633..560E}:

\begin{equation}
\DV(z) \equiv \left (z\DM(z)^2 D_H(z)\right )^{1/3}.
\label{eq:DV}
\end{equation}

BAO measurements are performed at a series of redshifts, allowing constraints to be obtained on the cosmological parameters governing the distance-redshift relation including curvature and dark energy, and the Hubble constant if external information on the absolute BAO scale is provided.

\subsection{DESI BAO data from Data Release 1}
\label{sec:DESIY1}

DESI spectroscopic targets are selected from photometric catalogs of the 9th public data release of the DESI Legacy Imaging Surveys \citep{schlegel22a},\footnote{\url{https://www.legacysurvey.org/dr9/}} drawn from three optical surveys in the $grz$ optical bands: DECaLS using the DECam camera \cite{decam} (which includes data from the Dark Energy Survey (DES) \cite{DES_overview}) south of declination $32.375^\circ$, and north of this limit the Beijing-Arizona Sky Survey (BASS) \cite{bass_overview}, and the Mosaic z-band Legacy Survey (MzLS) \cite{dey16}. Four different classes of extragalactic targets are defined: the bright galaxy sample (BGS, \cite{BGS.TS.Hahn.2023}), luminous red galaxies (LRG, \cite{LRG.TS.Zhou.2023}), emission line galaxies (ELG \cite{ELG.TS.Raichoor.2023}), and quasars (QSO \cite{QSO.TS.Chaussidon.2023}).

Spectroscopic observations of these targets are carried out with the DESI instrument~\cite{DESI2022.KP1.Instr} mounted on the Nicholas U. Mayall Telescope at Kitt Peak National Observatory in Arizona. Each observation field is covered by a ``tile", consisting in a set of targets located within that sky area~\cite{TS.Pipeline.Myers.2023} and assigned to each of the 5000 fibers in the focal plane of the telescope. Each fiber is placed at the celestial coordinates of its assigned target by a robotic positioner \cite{FocalPlane.Silber.2023,FBA.Raichoor.2024} andcarries the target's light from the focal plane to one of the ten DESI spectrographs. DESI observing time is dynamically separated into a ``bright" time program (when BGS are observed) and ``dark" time observations (when LRG, ELG, and QSO are targeted) depending on observing conditions. Redshift distributions, exposure times, calibration and observation procedures were optimised during a period of survey validation \cite{DESI2023a.KP1.SV} that included a visual inspection campaign \cite{VIGalaxies.Lan.2023,VIQSO.Alexander.2023}.
The DR1 spectroscopic dataset is built from main survey operations starting from May 14, 2021 through June 14, 2022, using an observing strategy meant to prioritise depth \cite{SurveyOps.Schlafly.2023}, resulting in $2,744$ ``dark" tiles and $2,275$ ``bright" tiles. The covered tile surface area is of order $7,500 \; \deg^{2}$, just over half of the expected final coverage of $14,200 \; \deg^{2}$. However, the completeness within this area will significantly increase, as one can infer given an expected final number of 9,929 dark and 5,676 bright observed tiles \cite{SurveyOps.Schlafly.2023}. The combined effective volume is expected to increase by more than a factor 3 \cite{DESI2024.III.KP4}. The observed data are processed by the DESI spectroscopic pipeline \cite{Spectro.Pipeline.Guy.2023} on a daily basis for immediate quality checks. The redshift catalogs used for this analysis and released with DESI DR1 are obtained from a spectroscopic reduction run with a fixed pipeline version internally denoted as ``iron".

Large-scale structure catalogs of galaxy and quasar positions suitable for the clustering analysis are built from the redshift and parent target catalogs and their two-point function measurements are obtained with all DR1 specific details presented in \cite{DESI2024.II.KP3}; a technical overview of the general process is presented in \cite{desilss}. The selection function is defined and correction weights are designed to compensate for systematic variations due to the effects of imaging anisotropies on the input target samples~\cite{KP3s2-Rosado,ChaussidonY1fnl}, redshift measurement efficiency~\cite{KP3s4-Yu,KP3s3-Krolewski}, and incompleteness in target assignment~\cite{KP3s7-Lasker,KP3s6-Bianchi,KP3s5-Pinon}. Simulations of the DR1 dataset are presented in \cite{ZhaoCY1sim}. 

Studies of the \lya forest are based on a quasar catalog with alternative redshift estimates to minimise possible biases caused by the same \lya absorption~\cite{KP6s4-Bault}.
The method to extract \lya fluctuations from the quasar spectra is described in \cite{2023MNRAS.tmp.3626R}, including the masking of pixels contaminated by Broad Absorption Line features~\cite{2023arXiv230903434F} and Damped Lyman-$\alpha$ systems~\cite{KP6s7-Zou}. Below, we provide details on each of the target samples studied and the properties of the particular samples that are used for this analysis. 

\paragraph{The Bright Galaxy Sample (BGS, $0.1 < z < 0.4$):}

The nominal target selection (BGS Bright) relies on $r$-magnitude cuts tuned to achieve uniform density across the photometric samples. Gaia catalogs \cite{gaia} are additionally used to remove point-like sources. The resulting sample has 854 targets per square degree, which are assigned fibers with high priority during bright time observations.

The nominal BGS sample is flux-limited and high density, with strong evolution in redshift. Over $5.5$ million reliable BGS redshifts were measured in DR1. However, in order to produce a more homogeneous sample, a cut was engineered to produce a sample of roughly constant comoving number density of $5 \times 10^{-4} \;\hMpcinvcubed$ using $k$-corrected $r$-band absolute magnitudes from \cite{fastspecfit,fastspecfit_code} and a redshift-dependent correction for evolution (matching the sample used for \cite{DESI2023b.KP1.EDR}). The BGS number density is similar to that of LRGs at $z=0.4$ and high enough to make shot-noise a minor contribution to the BAO statistical uncertainty (see \cite{DESI2024.II.KP3} for more details). The final BGS clustering sample used for the BAO measurement comprises $300,017$ redshifts in $0.1 < z < 0.4$, with an assignment completeness of $61.6\%$, which is expected to increase to $\sim 80\%$ in the finalised survey. 

\paragraph{The Luminous Red Galaxy Sample (LRG, $0.4 < z < 0.6$ and $0.6 < z < 0.8$):} The LRG target selection \cite{LRG.TS.Zhou.2023} uses photometry from the Wide-field Infrared Survey Explorer (WISE) \cite{wise} to select red objects with a nearly constant comoving number density of $5 \times 10^{-4} \;\hMpcinvcubed$ over $0.4 < z < 0.8$. An additional cut on the $z$-magnitude measured within the aperture of a DESI fiber further selects targets with high S/N spectra. The obtained $624\;\deg^{-2}$ target sample is assigned fibers with priority higher than ELG targets but lower than QSO targets.

The DESI DR1 LRG clustering sample used for BAO measurements consists of $2,138,600$ redshifts in the redshift interval $0.4 < z < 1.1$. The DR1 assignment completeness is $69.2\%$, which is expected to increase to $90\%$ in the completed survey. The lower redshift bound was chosen to separate the sample from BGS, as most low-redshift LRG targets are also BGS targets, while the upper bound was designed to ensure a minimum density of $10^{-4} \;\hMpcinvcubed$. This sample is further split for the clustering analysis into 3 disjoint redshift ranges, $0.4 < z < 0.6$, $0.6 < z < 0.8$, and $0.8 < z < 1.1$ (for brevity, we sometimes refer to these as LRG1, LRG2 and LRG3, respectively), however we do not use any BAO measurement from the highest-redshift LRG bin alone. It is instead combined with the overlapping lowest-redshift ELG bin, as described in the combined LRG and ELG sample section below.

\paragraph{The Emission Line Galaxy Sample (ELG, $1.1 < z < 1.6$):} ELG targets are selected with colour cuts in the $(g-r)$ vs $(r-z)$ plane to prioritise objects with [OII] emission in the desired redshift range \cite{ELG.TS.Raichoor.2023} for secure redshift measurements. Low-redshift objects are filtered out with a $g$-magnitude cut and another cut on the $g$-magnitude within the aperture of a DESI fiber further favors targets with high S/N spectra. While a random $10\%$ fraction of the ELG sample receives the same fiber assignment priority as LRGs to facilitate small-scale clustering measurements in the redshift range where the samples overlap, the remaining $90\%$ of ELGs are assigned fibers at a lower priority than LRGs and QSOs.

The DR1 ELG sample defined for clustering analysis in \cite{DESI2024.II.KP3} comprises $2,432,022$ reliable redshifts in the interval $0.8 < z < 1.6$. The lower redshift bound aligns with that of the high redshift LRG sample, while the upper bound rejects objects whose [OII] doublet falls outside of the spectrograph wavelength coverage. In DR1, the fiber completeness is only $35\%$, which should increase to over $60\%$ in the final dataset. The sample is split into two disjoint redshift ranges for clustering analysis, $0.8 < z < 1.1$ and $1.1 < z < 1.6$ (similarly referred to as ELG1 and ELG2 for brevity). We do not use any standalone BAO measurement from the low-redshift ELG bin; it is instead combined with the overlapping high-redshift LRG bin.

\paragraph{The combined LRG and ELG Sample (LRG+ELG, $0.8 < z < 1.1$):} The high-redshift LRG and the low-redshift ELG samples overlap in the range $0.8 < z < 1.1$. These two samples are concatenated with inverse-variance weights to obtain a combined LRG+ELG catalog, following the methodology presented in \cite{KP4s5-Valcin}. The increased combined density is expected to improve the reconstruction efficiency, while the inverse-variance weights are designed to maximise the measurement of the BAO precision. As described in \cite{DESI2024.III.KP4}, the obtained combined LRG+ELG BAO measurement is $\sim 10\%$ more precise, while being fully consistent with the LRG-alone measurement in this redshift range (and with that of ELG alone, although this has a larger uncertainty). We therefore use the combined LRG+ELG $0.8 < z < 1.1$ BAO measurement for the cosmological inference.

\paragraph{The Quasar Sample (QSO, $0.8 < z < 2.1$):} The QSO target selection relies on identifying a flux excess in the near-infrared, which is assessed by comparing near-infrared WISE $W1$, $W2$ magnitudes to Legacy Imaging Surveys optical $grz$ magnitudes \cite{QSO.TS.Chaussidon.2023}. Sources with stellar morphology (PSF type) are selected, and a $r$-magnitude cut is applied to eliminate bright stars and faint targets. Quasars are then selected from the $r$ magnitude and 10 colors using $grzW1W2$ with a Random Forest (RF) algorithm, designed to produce a target sample of density $310 \deg^{-2}$. The resulting QSO target sample is assigned fibers with maximum priority. 

The DR1 QSO sample used for BAO measurements in $0.8 < z < 2.1$ consists of $856,652$ redshifts. Quasars at $z>2.1$ are not used as part of this sample. Because of their maximum priority, the assignment completeness of QSO in DR1 is already high, $87.6\%$.

\paragraph{The Lyman-$\alpha$ Forest Sample (\lya, $1.77 < z < 4.16$):} 
The highest-redshift BAO measurement from DESI DR1 is described in \cite{DESI2024.IV.KP6}, and is obtained from a combined analysis of correlations of three different datasets. 
The first dataset consists of the positions (celestial coordinates and redshifts) of $709,565$ quasars in the redshift range $1.77 < z < 4.16$. 
In addition, the \lya forest in the spectra of high-redshift quasars at $z>2.1$ constitute an alternative tracer of matter density fluctuations. 
We use two \lya forest datasets, consisting of fluctuations in two different rest-frame wavelength regions of the background quasar spectra: $428,403$ in region A ($1040 < \lambda_r < 1205$ \AA) and $137,432$ in region B ($920 < \lambda_r < 1020$ \AA).\footnote{Region B is treated separately from region A as it is also affected by higher order Lyman lines.}
In order to increase the signal-to-noise of the spectra, quasars identified as having $z > 2.1$ after a single epoch of observation are prioritised for up to four observations.

\cref{tab:Y1data} summarises the properties of the different samples used in this analysis, as well as the BAO measurements obtained from them using the methods described below. Overall, this analysis of the DESI Data Release 1 relies on over $6$ million unique redshifts, more than twice the 
number of redshifts considered in the final SDSS cosmological analysis~\cite{2021PhRvD.103h3533A}. The aggregate precision on the BAO isotropic scale is $0.49\%$ to be compared to $0.60\%$ for the final SDSS measurements.

\subsection{BAO analysis}
\label{sec:BAOanalysis}

We now briefly describe how DESI BAO analyses of DR1 data implement the standard ruler measurement.

BAO analyses usually need to assume a fiducial cosmology although the resulting distance and expansion measurements remain independent of these assumptions in a leading-order approximation, unless the assumed fiducial model differs significantly from the model to be tested, or the truth \cite{2020JCAP...01..038H,Bernal:2020vbb,Pan:2023zgb}. First, a fiducial cosmological model is adopted to convert tracer angular positions and redshifts to comoving coordinates, and the two-point clustering pattern is measured as a function of comoving separation assuming this fiducial model. The clustering pattern perpendicular and parallel to the line-of-sight changes in a predictable way between fiducial and trial cosmologies according to the Alcock-Paczynski (AP) effect \cite{1979Natur.281..358A, 2008PhRvD..77l3540P}, allowing a wide range of cosmological models to be tested \cite{2020MNRAS.494.2076C}.

Second, the theoretical calibration of the BAO scale is encoded by a primordial matter power spectrum generated assuming a fiducial cosmological model \cite{1998ApJ...496..605E, 2000ApJ...538..473L}. This matter power spectrum is used to produce a template tracer clustering model, indexed by the sound horizon at baryon drag, which is scaled by two free scaling parameters---along and perpendicular to the line-of-sight---during data fitting.  In order to measure the location of the BAO feature alone, the model also includes ``nuisance" parameters for the broad-band shape of the tracer power spectrum, which are marginalised over in the analysis \cite{2012MNRAS.427.3435A}.  Residual theoretical systematics arising from nonlinear matter growth, tracer bias, and redshift-space distortions are reduced by this procedure, and are generally considered to be present at a level of $\sim 0.1\%$ \cite{2004PhRvD..70j3523E, 2007ApJ...664..660E, 2008PhRvD..77b3533C, 2011ApJ...734...94M}. The modeling and the procedure for performing BAO galaxy and quasar fits have been updated for DESI from the broadband polynomial marginalization procedure used in previous surveys, according to the theoretical framework laid out in \cite{KP4s2-Chen}.

Finally, the amplitude of the BAO feature imprinted in the early universe is diluted by the bulk-flow motion of matter from its true locations in the density field, driven by the growth of structure, which changes tracer separations by a few $h^{-1} \, \text{Mpc}$. This effect may be partially reversed using density-field reconstruction to estimate these displacements based on the gravitational field inferred from the observed tracer distribution \cite{Eisenstein2007b,Padmanabhan2012}. Applying these reversed displacements to tracers sharpens the BAO signature\footnote{Note that the \lya\ analysis, discussed further below in this subsection, does not make use of the density-field reconstruction.} and mitigates the small ($<0.3\%$) nonlinear shift of the BAO-scale in late-time clustering, allowing for more accurate cosmological inferences \cite{2007ApJ...664..675E, 2012MNRAS.427.2132P}. The reconstruction algorithm used has been updated relative to that used for eBOSS \cite{Alam-eBOSS:2021} and the specific choice of algorithm and settings were validated as described in \cite{KP4s4-Paillas}. 

For the galaxy and quasar BAO data points considered in this paper we make use of the post-reconstruction clustering measurements in configuration space presented in \cite{DESI2024.III.KP4}. Anisotropic BAO measurements are obtained for LRG and ELG samples, but we restrict to isotropic fits for BGS and QSO samples due to the lower signal-to-noise ratio of the two-point function measurements. The uncertainties in these clustering measurements are described by covariance matrices whose construction and validation are outlined in \cite{KP4s7-Rashkovetskyi,KP4s8-Alves,KP4s6-Forero-Sanchez}. Various contributions to the total systematic error budget arising from theoretical modeling uncertainties (which are at most $0.1\%$ and $0.2\%$ of the isotropic and anisotropic BAO parameters, respectively \cite{KP4s2-Chen}), uncertainties due to the galaxy-halo connection (of $\lesssim 0.2\%$ \cite{KP4s10-Mena-Fernandez,KP4s11-Garcia-Quintero}) and observational systematic effects (which are negligible \cite{KP3s2-Rosado, KP3s4-Yu, DESI2024.III.KP4}) have been estimated in a series of supporting papers. These are added together in quadrature as summarised in \cite{DESI2024.III.KP4} to compute the total error budget for the BAO measurements in the BGS, LRG, ELG, and QSO samples. Together, these steps represent the state-of-the-art in the measurement of BAO from galaxy tracers and are described in detail in the companion DESI paper \cite{DESI2024.III.KP4}.

In addition to measuring the clustering of galaxies and quasars, we also make use of the \lya forest. The \lya BAO measurement is presented in \cite{DESI2024.IV.KP6}, and it combines a measurement of the auto-correlation of fluctuations in the \lya forest and of its cross-correlation with the density of quasars at $z>1.77$. The method to measure and model the correlations is described in \cite{2023JCAP...11..045G}, and it is based on previous \lya BAO measurements from eBOSS \cite{2020ApJ...901..153D}. In a supporting study \cite{KP6s6-Cuceu} we validate the analysis using synthetic datasets, simulated using the methodology described in \cite{2024arXiv240100303H}. An important improvement in the methodology is that we now take into account the small correlation between the measurements of the auto- and the cross-correlation. Finally, the (minor) impact of correlated noise in the \lya auto-correlation is studied and characterised in \cite{KP6s5-Guy}. Besides the validation tests using synthetic datasets, in \cite{DESI2024.IV.KP6} we also present several data splits and a long list of consistency tests that show that our \lya BAO measurement are robust to changes in the measurement methodology or in the modeling. For instance, we show that adding an ad-hoc smooth component to the model with up to 48 extra free parameters does not vary the BAO parameters by more than 0.1\%.

The BAO measurements obtained from the various samples used are shown in \cref{tab:Y1data}. As already mentioned, for the BGS and QSO tracers, we only measure the angle-averaged $\DVrd$ quantity, due to the lower signal-to-noise achieved. For all other tracers, we quote results in $\DMrd$ and $\DHrd$, and provide the value of the correlation between them, $r$.

\begin{table*}
    \centering
    \small
    \resizebox{\columnwidth}{!}{
    \begin{tabular}{|l|c|c|c|c|c|c|c|} 
    \hline
    \multirow{2}{*}{tracer} & \multirow{2}{*}{redshift} & \multirow{2}{*}{$N_{\rm tracer}$} & \multirow{2}{*}{$z_{\rm eff}$} & \multirow{2}{*}{$\DMrd$}  & \multirow{2}{*}{$\DHrd$} & \multirow{2}{*}{$r$ or $\DVrd$} & $V_{\rm eff}$ \\[-0.1cm]  
           &          &                  &               &          &         &                &  (Gpc$^3$)    \\  \hline

    BGS  & $0.1-0.4$ & 300,017 & $0.295$ &---&--- & $7.93 \pm 0.15$ & $1.7$ \\
    LRG1  & $0.4-0.6$ & 506,905 & $0.510$ & $13.62 \pm 0.25$ & $20.98 \pm 0.61$ & $-0.445$ & $2.6$\\
    LRG2  & $0.6-0.8$ & 771,875 & $0.706$ & $16.85 \pm 0.32$ & $20.08 \pm 0.60$ & $-0.420$ & $4.0$\\
    LRG3+ELG1  & $0.8-1.1$ & 1,876,164 & $0.930$ & $21.71 \pm 0.28$ & $17.88 \pm 0.35$  & $-0.389$ & $6.5$\\
    ELG2  & $1.1-1.6$ & 1,415,687 & $1.317$ & $27.79 \pm 0.69$ & $13.82 \pm 0.42$ & $-0.444$ & $2.7$\\
    QSO  & $0.8-2.1$ & 856,652 & $1.491$ &---&--- & $26.07 \pm 0.67$ & $1.5$\\
    Lya QSO  & $1.77-4.16$ & 709,565 & $2.330$ & $39.71 \pm 0.94$ & $8.52 \pm 0.17$ & $-0.477$ & ---\\
    \hline
    \end{tabular}
    }
    \caption{\label{tab:Y1data}Statistics for the DESI samples used for the DESI DR1 BAO measurements used in this paper. For each tracer and redshift range we quote the number of objects ($N_{\rm tracer}$), the effective redshift ($z_{\rm eff}$) and effective volume ($V_{\rm eff}$). Note that for each sample we measure either both $\DM/\rd$ and $D_\mathrm{H}/\rd$, which are correlated with a coefficient $r$, or $\DV/\rd$. Redshift bins are non-overlapping, except for the shot-noise-dominated measurements that use QSO (both as tracers and for \lya{} forest). 
    }
\end{table*}

\subsubsection{Blinding framework}
\label{sec:blinding}

An imperative of science with modern cosmology experiments, and especially for a survey with the statistical precision that DESI is able to achieve, is to mitigate against the possibility of the results being affected by observer confirmation bias. This can manifest itself through the attribution of unexpected results, or results which do not match the prior prejudices of the observer, to systematic effects which must be corrected for in the data analysis. Thus the observer may---consciously or subconsciously---continue to search for and correct real or imagined systematic effects in the data only until they agree with some pre-conceived desired result, and no further.

To mitigate against the possibility of such biases unknowingly entering into our analyses, we apply a system of \emph{blinding} our data, to conceal the true cosmological results during the period where systematic errors were being investigated and analysis pipelines finalised. This blinding was applied in two different ways:
\begin{itemize}
    \item For the discrete galaxy and quasar samples, blinding was applied \emph{at the catalog level} \cite{Brieden:2020}, according to the procedure described and validated in \cite{KP3s9-Andrade}. The redshifts of the tracers in the catalogs were shifted in such a way to change the position of the BAO feature as well as the redshift-space distortion (RSD) signal, and the weights applied to the tracers were adjusted in order to blind the measurement of primordial non-Gaussianities from the large-scale clustering. True redshifts and weights were not made available for the analysis before unblinding. 
    \item For the \lya sample, catalog-level blinding was challenging due to sky lines and Galactic absorption features located at known observed wavelengths in the spectra. Instead, the apparent position of the BAO peak was shifted by directly applying an additive component to the measured correlation functions, as described in \cite{DESI2024.IV.KP6}.
\end{itemize}
In both cases, all analyses used the blinded data during initial testing and systematic errors were identified and corrected. Only after the analysis pipeline had been finalised, the systematic error budget determined, and a series of strictly prescribed validation tests passed (see \cite{DESI2024.III.KP4,DESI2024.IV.KP6} for a detailed description) was the data blinding removed and the unblind results revealed. For discrete tracers, it was decided prior to unblinding that the final choice of covariance matrices would be made after unblinding, and that the LRG and ELG samples would be concatenated in $0.8 < z < 1.1$ for a combined BAO measurement in this redshift range.

We mention two unplanned updates in the BAO analysis pipeline for discrete tracers made after unblinding and decided independently of the actual BAO measurements: a completeness correction was fixed in the clustering catalogs, and the implementation of the BAO theoretical model was corrected to fully match the theoretical framework laid out in~\cite{KP4s2-Chen}. Corrections in the theoretical model changed BAO measurements by less than $0.3 \sigma$ (differences in most redshift bins under $0.1 \sigma$), and updates to the catalogs by at most $0.6 \sigma$ (variations in most redshift bins being under $0.2 \sigma$). We also report two minor post-unblinding changes in the \lya BAO analysis: a bug fix that led to a $< 0.1\sigma$ change in the BAO measurement, and an update in a bias parameter that had no effect on the results.

Two additional points are worth noting. For the discrete galaxy and quasar samples, the analyses related to the full shape of the broadband clustering signal (described in \cite{DESI2024.V.KP5}) and those pertaining to the BAO results presented in \cite{DESI2024.III.KP4} and included in this paper had separate pipelines and robustness criteria to be satisfied before unblinding, and were not unblinded at the same time. 
Additionally, it was decided that cosmological model inference from the BAO and full shape measurements would remain outside of the scope of the blinding, and would be performed once BAO and full shape unblinded results are obtained.\footnote{Internal checks of the inference pipeline were performed with synthetic data prior to unblinding.}

\subsubsection{Summary of DESI DR1 BAO likelihoods}
\label{sec:likelihoods}

The results used in this paper are derived from the BAO measurements and fits described above applied to each of the five tracer samples across 7 distinct redshift bins covering a total redshift range from 0.1 to 4.2. 

The BGS and LRG samples chosen cover disjoint redshift ranges. As a result of the lack of overlap, the correlation between the BAO measurements in these samples
is small, and we assume it to be negligible and do not include any covariance between them in our cosmological analysis. In the $0.8<z<1.1$ redshift bin, where the LRG and ELG tracers overlap, they are analysed together using a single multi-tracer approach, described in \cite{KP4s5-Valcin,DESI2024.III.KP4}. The QSO sample overlaps with this redshift bin, and with the ELGs in the range $1.1<z<1.6$. In a general case where two tracer samples overlap in volume, the correlation coefficient between two estimates of the power spectrum obtained from these tracers may be estimated as:
\begin{equation}
    \label{eq:sample_correlation}
    C = \frac{\mathrm{Cov}(\hat{P}_1, \hat{P}_2)}{\sqrt{\mathrm{Var}(\hat{P}_1)\mathrm{Var}(\hat{P}_2)}}
    =\frac{\int dV\;\displaystyle\frac{X_1X_2}{X_{12}}}{(\int dV X_1)^{1/2}(\int dV X_2)^{1/2}}\,,
\end{equation}
where $X_i\equiv \left(\frac{n_iP_i}{1+n_iP_i}\right)^{2}$ for $i=1,2$, $X_{12}\equiv\left(\frac{n_1n_2P_1P_2}{n_{12}\sqrt{P_1P_2} + n_1n_2P_1P_2}\right)^{2}$, $n_1$ and $n_2$ are the mean tracer number densities of the two samples and $n_{12}$ is the common sample density, and $P_1$ and $P_2$ are the corresponding typical amplitudes of the power spectra of the two tracers at the BAO scale $k\simeq0.1\;\hmpcinv$. For different tracer types, such as the overlapping ELG and QSO samples of relevance in the $1.1<z<1.6$ redshift bin, there are no common tracers so $n_{12}=0$.\footnote{In this case and in the limit of a thin redshift slice, \cref{eq:sample_correlation} simplifies to $C=\frac{V_{12} P_1P_2}{\sqrt{V_1V_2}(P_1+1/\bar{n}_1)(P_2+1/\bar{n}_2)}$, as used in \cite{Alam-eBOSS:2021}.} This correlation between power spectrum estimates can be roughly regarded as an approximation to the correlation between the BAO scale measurements from the two overlapping tracer samples, although we regard this only as a guide and caution against using this estimate for any precise combination of results from different tracers.

For the ELG and QSO tracers in $1.1<z<1.6$, using representative values in \cref{eq:sample_correlation} gives a very small correlation, $C<0.1$. The correlation between QSOs and LRGs and ELGs in the lower redshift $0.8<z<1.1$ bin is even smaller. We therefore ignore this correlation and treat the QSO BAO measurement as effectively independent of those from LRG and ELG tracers. We also ignore the correlation between the BAO measurement from the QSO sample (at $z_{\rm eff}=1.49$) and the \lya BAO measurement (at $z_{\rm eff}=2.33$). Even though quasars in the range $1.77 < z < 2.1$ contribute to both measurements, in the \lya BAO measurement they are only used in cross-correlation with \lya absorption at $z>2$. Therefore, these measurements would only be correlated by cosmic variance, but both measurements are dominated by shot-noise. In this redshift range, the number density of the quasars is less than $3\times10^{-5}\hMpcinvcubed$ and given the $P(k)$ at the scales most relevant for BAO is $<7500\hinvMpccubed$, $nP \ll 1$ and any correlation between the measurements will be small. 

As described in \Cref{sec:cosmo}, the cosmological quantities measured by the BAO analysis in each redshift bin are either the two (correlated) distance ratios ($\DMrd$, $\DHrd$) or just a single one, $\DVrd$, depending on the signal-to-noise ratio. The posteriors in these quantities from the BAO fits are sampled using MCMC, as described in \cite{DESI2024.III.KP4}; we find that the recovered posteriors are in all cases well approximated by Gaussian distributions. Refs. \cite{DESI2024.III.KP4,DESI2024.IV.KP6} and the supporting papers cited therein describe the methods used to estimate the systematic uncertainties and their magnitudes. These systematic uncertainties are added in quadrature to the statistical uncertainties determined from the posterior sampling with the final combination presented as a block diagonal Gaussian covariance matrix, with off-diagonal correlations only between the ($\DMrd$, $\DHrd$) elements within the same redshift bin. We use this covariance matrix or its individual components together with the vectors of the mean ($\DMrd$, $\DHrd$) values for the cosmological inference presented hereafter. 

\subsection{External Datasets}
\label{sec:external_data}

In this paper, we compare cosmological results from our data to those from a number of other recent experiments and provide joint constraints from different combinations of datasets. We briefly describe these data and likelihoods below.

\subsubsection{Big Bang Nucleosynthesis and $\Ob h^2$}
\label{sec:BBN}

In a flat \lcdm\ background model, the BAO distance ladder results determine $\Om$ and $\Hrd$. From \cref{eq:rdinLCDM}, the sound horizon depends on the physical matter and baryon densities, $\om=\Om h^2$ and $\ob=\Ob h^2$. Therefore, assuming standard neutrino content of the universe ($\Neff=3.044$ \cite{Froustey:2020,Bennett:2021}), we require prior knowledge of $\Ob h^2$ to break the degeneracy between $H_0$ and $\rd$.

Big Bang Nucleosynthesis (BBN) theory predicts the abundances of light elements including D and $^4$He in the early universe, and these abundances depend on the baryon to photon ratio $\eta_{10}$. Thus the observational determination of the primordial deuterium abundance D/H \cite{Cooke:2018} and the helium fraction $Y_{\rm P}$ \cite{Aver:2015,Aver:2022} can be used to deduce $\Ob h^2$.

The resulting constraints on $\Ob h^2$ depend on the details of the theoretical predictions and their treatment of underlying nuclear interaction cross-sections, in particular for the deuterium burning reactions. For a review of the details we refer interested readers to \cite{PDG:2022}. A recent analysis \cite{Schoeneberg:2024} makes use of the new \texttt{PRyMordial} code \cite{Burns:2024} to recompute the predictions while correctly marginalizing over uncertainties in the reaction rates, and reports the conservative constraints
\be
\label{eq:BBN_LCDM}
\Ob h^2= 0.02218 \pm 0.00055
\ee
in the standard \lcdm{} model, and 
\be
\label{eq:BBN_Neff}
\Ob h^2=  0.02196 \pm 0.00063
\ee
when allowing for additional light relics (\lcdm+$N_\mathrm{eff}$ model), where in the latter case there is also a covariance between $\Ob h^2$ and $N_\mathrm{eff}$ which we account for.\footnote{This covariance is available from \url{https://github.com/schoeneberg/2024_bbn_results}.} We adopt these values as BBN priors in the following sections. Note that these BBN priors are more conservative than some previously proposed in the literature (e.g., \cite{Mossa:2020}); this is a direct result of the marginalization over reaction rate uncertainties performed in \cite{Schoeneberg:2024}.

\subsubsection{Cosmic Microwave Background}
\label{sec:CMB}

The power spectra of anisotropies in the cosmic microwave background (CMB) have been exquisitely measured by \Planck{} \cite{Planck-2018-overview,Planck-2018-cosmology}. A fundamental quantity that is determined from the oscillations in these power spectra is the measured sky angular scale of acoustic fluctuations, i.e.\ $\theta_\ast=r_\ast/D_\mathrm{M}(z_\ast)$, where $r_\ast$ is the comoving sound horizon at recombination and $D_\mathrm{M}(z_\ast)$ is the transverse comoving distance to the redshift of recombination. This angular scale is a very sharp feature and can be measured precisely independently of possible systematic effects affecting the high-$\ell$ part of the CMB power spectrum (in particular the $A_\mathrm{L}$ parameter).

Within the \lcdm{} model \Planck\ reports \cite{Planck-2018-cosmology}
\be
\label{eq:theta_star}
100\theta^{\Planck}_\ast=1.04109 \pm 0.00030,
\ee
a $0.03\%$ precision measurement through the combination of temperature and polarization data alone. The central value of $\theta^{\Planck}_\ast$ is almost entirely independent of the specific cosmological model assumed, although the uncertainties can increase significantly in some model extensions affecting early-time physics. This is a consequence of the simple geometrical interpretation: $\theta_\ast$ is the BAO scale imprinted in the CMB anisotropies at recombination. DESI, on the other hand, measures the same BAO features imprinted in the galaxy distributions at lower redshifts. It is therefore natural to combine DESI BAO results with the constraint on $\theta_\ast$ from \Planck, which can be regarded as a minimal and highly robust purely geometric version of the full CMB information. To achieve this, we implement a Gaussian external prior on the quantity $100\theta_\ast$ with mean $1.04110$ and variance $0.00053^2$. The mean here matches the reported value in \cite{Planck-2018-cosmology} when including CMB lensing, and is very close to that in \cref{eq:theta_star}. We have conservatively increased the width of the prior by $\sim75\%$ relative to the baseline result in order to accommodate the increased uncertainties in models beyond standard flat \lcdm, especially in models that allow for additional light relics (\lcdm+$\Neff$), based on the results for that model from~\cite{Planck-2018-cosmology}.

Of course, the CMB power spectra also contain vastly more information than just $\theta_\ast$, and we also explore the consequences for cosmological models from the combination of this full CMB information with DESI BAO. We use as our baseline the temperature (TT) and polarisation (EE) auto-spectra, plus their cross-spectra (TE), as incorporated in the \texttt{simall}, \texttt{Commander} (for multipoles $\ell<30$) and \texttt{plik} (for $\ell\geq30$) likelihoods from the official PR3 release \cite{Planck-2018-likelihoods}. Subsequently, a new \Planck{} data release PR4 has been made available, which involves a consistent reprocessing of the data from both the LFI and HFI instruments on \Planck{} using the new common pipeline \texttt{NPIPE}, leading to slightly more data, lower noise, and better consistency between frequency channels \cite{Planck-2020-NPIPE}. Various teams have released additional high-$\ell$ likelihood packages using PR4 \cite{Efstathiou:2021,Rosenberg:2022,Tristram:2023}, which are updated versions of likelihoods included in earlier \Planck{} data releases, and use alternative methods to derive information from the high-$\ell$ power. However, in most scenarios we consider in this paper, results from these updated likelihoods are very similar to those obtained from the official PR3 \texttt{plik} likelihood. Therefore, we choose to keep \texttt{plik} as our baseline, and note any variations in the results due to differences between this and the newer PR4 likelihoods\footnote{\texttt{CamSpec} is bundled as an additional likelihood in the \texttt{cobaya} sampling code \cite{Torrado:2021,Torrado:2019}, at \url{https://cobaya.readthedocs.io/en/latest/likelihood_planck.html}; \texttt{HiLLiPoP} is available from \url{https://github.com/planck-npipe/hillipop}.} in \cref{sec:appendix_PR4}.

In addition to the primary temperature and polarization anisotropy power spectra, modern CMB experiments are also able to measure the power spectrum of the gravitational lensing potential, $C_L^{\phi\phi}$, from the connected 4-point function \cite{Planck-2013-lensing,Planck-2015-lensing}. The latest and most precise CMB lensing data comes from the combination of \texttt{NPIPE} PR4 \Planck{} CMB lensing reconstruction \cite{Carron:2022} and the Data Release 6 of the Atacama Cosmology Telescope (ACT) \cite{Madhavacheril:ACT-DR6,Qu:2023,MacCrann:2023}.\footnote{The likelihood is available from \url{https://github.com/ACTCollaboration/act_dr6_lenslike}; we use the \texttt{actplanck\_baseline} option.} We adopt this combined CMB lensing likelihood from both experiments as our baseline. For the sake of brevity, in the following text and figures, we will denote results obtained using temperature and polarisation information from \Planck, and CMB lensing information from the \planckact\ combination, simply as ``CMB". Where occasionally necessary, we will explicitly label results that do not use CMB lensing as ``CMB (no lensing)".

\subsubsection{Type Ia supernovae}
\label{sec:SNIa}

Type Ia supernovae (SN~Ia) serve as standardizable candles which offer an alternative way to measure the expansion history of the universe. Historically, SN~Ia led to the discovery of the accelerating expansion \cite{SupernovaSearchTeam:1998fmf,SupernovaCosmologyProject:1998vns}, following earlier, more complex arguments for $\Lambda$-dominated models based on observations of large-scale structure \cite{1990Natur.348..705E,White:1993wm}. Within the \lcdm{} model SN~Ia have lower statistical power than modern BAO measurements, but provide important information on dark energy when analyzing the less restricted models considered in  \cref{sec:DE}. 

In this paper, we make use of three different SN~Ia datasets. The Pantheon+ compilation \cite{Scolnic:2021amr} consists of 1550 spectroscopically-confirmed SN~Ia in the redshift range $0.001<z<2.26$. We use the public likelihood from \cite{Brout:2022} incorporating the full statistical and systematic covariance, imposing a bound $z>0.01$ in order to mitigate the impact of peculiar velocities in the Hubble diagram \cite{Peterson:2021hel}. More recently, the Union3 compilation of 2087 SN~Ia, many (1363 SN~Ia) in common with Pantheon+, was presented in \cite{Rubin:2023},\footnote{Data provided by the Union3 team,  private communication.} and includes a different treatment of systematic errors and uncertainties based on Bayesian Hierarchical Modelling. Finally, the Dark Energy Survey, as part of their Year 5 data release, recently published results based on a new, homogeneously selected sample of 1635 photometrically-classified SN~Ia with redshifts $0.1<z<1.3$, which is complemented by 194 low-redshift SN~Ia (which are in common with the Pantheon+ sample) spanning $0.025<z<0.1$ \cite{DES:2024tys}.\footnote{Data available at \url{https://github.com/des-science/DES-SN5YR}.}
The contribution of SN~Ia in general, and the differences between the different datasets\footnote{We note that these SN~Ia data sets are not independent of each other and we therefore do not combine their results: Pantheon+ and Union3 share $\sim1360$ supernovae but differ in their analysis methodology and marginalization over astrophysical and systematic parameters, while DES contributes a new high-$z$ data set of $\sim1500$ photometrically-classified SN~Ia but still uses about 194 historical SN~Ia at $z<0.1$ in common with the other two.} are discussed in \cref{sec:DE}. In the rest of the paper, we will denote the Pantheon+ dataset as PantheonPlus, and the DES-SN5YR dataset as DESY5, in order to provide a consistent, concise set of labels suitable for tables and figure legends, and to avoid ambiguities with the `+' symbol used to denote the combinations of datasets.

\begin{table}[t] 
    \centering
    \begin{tabular}{|llll|}
    \hline
    parametrisation & parameter & default & prior\\  
    \hline 
    \textbf{background-only} & $\Om$ &---& $\mathcal{U}[0.01, 0.99]$\\
    no $\rd$ calibration & $\rd h \; (\Mpc)$ &---& $\mathcal{U}[10, 1000]$  \\
    with $\rd$ calibration & $H_{0} \; (\kmsMpc)$ &---& $\mathcal{U}[20, 100]$  \\
     & $\ob$ &---& $\mathcal{U}[0.005, 0.1]$  \\
    \hline 
    \textbf{CMB} & $\ocdm$ &---& $\mathcal{U}[0.001, 0.99]$ \\
    & $\ob$ &---& $\mathcal{U}[0.005, 0.1]$ \\
    & $100 \theta_{\mathrm{MC}}$ &---& $\mathcal{U}[0.5, 10]$ \\
    & $\ln(10^{10} A_{s})$ &---& $\mathcal{U}[1.61, 3.91]$ \\
    & $n_{s}$ &---& $\mathcal{U}[0.8, 1.2]$ \\
    & $\tau$ &---& $\mathcal{U}[0.01, 0.8]$ \\
    \hline 
    \textbf{extended} & $\Ok$ & $0$ & $\mathcal{U}[-0.3, 0.3]$ \\
    & $w_0$ or $w$ & $-1$ & $\mathcal{U}[-3, 1]$ \\
    & $w_{a}$ & $0$ & $\mathcal{U}[-3, 2]$ \\
    & $\sumnu \; (\eV)$ & $0.06$ & $\mathcal{U}[0, 5]$ \\
    & $\Neff$ & $3.044$ & $\mathcal{U}[0.05, 10]$  \\
    \hline
    \end{tabular}
    \caption{
    Parameters and priors used in the analysis. All of the priors are flat in the ranges given. We consider two parametrisations,``background-only" when using BAO and SN data only, and ``CMB" where data from \Planck\ and ACT are involved. In both cases, the same priors are used for parameter extensions. A single massive neutrino of mass $\sumnu = 0.06 \; \eV$ is assumed, except in the $\Lambda$CDM+$m_\nu$ model, for which we consider three degenerate massive neutrino species ($N_{\nu}=3$ in \texttt{CAMB}). Note that $\Om$ includes the contribution of any massive neutrinos. In addition to the flat priors on $w_0$ and $w_a$ listed in the table, we also impose the requirement $w_0+w_a<0$ in order to enforce a period of high-redshift matter domination.
    }
    \label{tab:priors}
\end{table}

\subsection{Cosmological inference}
\label{sec:inference}

We included in the cosmological inference code \texttt{cobaya} \cite{Torrado:2019,Torrado:2021} the PantheonPlus, Union3 and DESY5 SN~Ia likelihoods, as well as our new DESI BAO likelihoods.
The CMB likelihoods used are based on public packages that are either included in the public \texttt{cobaya} version or available directly from the respective teams.

When running \texttt{cobaya}, we rely on the Boltzmann code \texttt{CAMB}~\cite{LewisCAMB:2000, HowlettCAMB:2012} for theoretical cosmology calculations. When using the combined \Planck+ACT lensing likelihood, we use higher precision settings as recommended by ACT.

All Bayesian inference is performed using the Metropolis-Hastings MCMC sampler \cite{LewisMCMC:2002, LewisMCMC:2013} in \texttt{cobaya}. \Cref{tab:priors} summarises the cosmological parameters that are sampled over in different runs, and the priors that are placed on them. In the base \lcdm\ model, for data combinations that only probe the background evolution (i.e., BAO and SN~Ia), we either sample in the parameters $\Om$ and $\rd h$ or---when using external data such as from BBN in order to help calibrate the $\rd$ value and break the $\rd h$ degeneracy in BAO data---in $\Om$, $H_0$ and the physical baryon density $\ob$. When also including CMB likelihoods, we sample instead in the standard six-parameter basis $\left(\omega_{\rm cdm},\ob, 100\theta_{\rm MC}, \ln{(10^{10}A_s)}, n_s, \tau\right)$, where $\theta_{\rm MC}$ is an approximation to the acoustic angular scale $\theta_\ast$, $A_s$ is the amplitude of the primordial scalar power spectrum and $n_s$ is its spectral index, and $\tau$ is the reionization optical depth. We also consider extensions of the minimal \lcdm\ model in which other parameters are allowed to vary: the spatial curvature $\Ok$, the sum of neutrino masses $\sumnu$, the single constant dark energy equation of state parameter $w$ or the two parameters $(w_0,w_a)$ when allowing a time-varying equation of state $w(a)$ \cite{Chevallier:2001, Linder2003}, and the effective number of relativistic species $\Neff$. Priors on these extended parameters are shown in the bottom section of \cref{tab:priors}; in any instance where one of these parameters is not varied, it is held fixed at the default value listed.

When including CMB data, we take advantage of the hierarchy between (fast) nuisance parameters and (slow) cosmological parameters by taking more steps along the latter, setting the oversampling parameter \texttt{oversample\_power} to $0.4$. We also use the so-called dragging method that enables taking larger steps in the slow parameter subspace with the help of intermediate transitions of fast parameters~\cite{NealDragging:2005}. For each dataset and model combination, we run four chains in parallel, starting from proposal covariance matrices built from preliminary runs. Chains are stopped when the Gelman-Rubin~\cite{GelmanRubin} criterion ${\rm R}-1<0.01$ is satisfied, where ${\rm R}$ is the largest eigenvalue of the ratio of the inter- to intra-chain covariance matrices. We further require the effective sample size of the chains (defined as the maximum of the weighted chain length divided by the autocorrelation length for all parameters) to be $\gtrsim 10^3$ to achieve percent precision on the moments of the marginal posteriors. In the case of symmetric 1D marginalised posteriors we report the mean and standard deviation of the samples, while the $68 \%$ minimal credible interval is quoted in other cases, except when otherwise stated.
We use \texttt{getdist}\footnote{\url{https://github.com/cmbant/getdist}}~\cite{Lewis:2019xzd} to derive the constraints presented in this paper.
Where appropriate, we also compute the best-fit with the \texttt{iminuit}~\cite{iminuit,James:1975dr} algorithm, as implemented in \texttt{cobaya}, starting from the four maximum a posteriori (MAP) points of the corresponding four MCMC chains. In our analysis, we use $\Delta \chi_\mathrm{MAP}^2 \equiv -2 \Delta \log{\mathcal{L}}$, defined to represent the difference (times $-2$) of log-posteriors at the maximum posterior points.


\section{DESI Distance Measurements}
\label{sec:desi_cosmo} 

\subsection{Distance-Redshift Results}
\label{sec:expansion_history}

Our summary \cref{tab:Y1data} provides the BAO distance scales measured from each of the DESI tracers and redshift bins. As noted above, for the BGS and QSO samples, we report a single result for the angle-averaged quantity $\DVrd$ \cite{DESI2024.III.KP4}. For all other redshift bins, we quote marginalised constraints on both $\DMrd$ and $\DHrd$ individually; note however that these two quantities are correlated with each other, with a correlation factor that depends on redshift, also provided in \cref{tab:Y1data} and fully accounted for in the BAO likelihoods.

\begin{figure}[t]
    \centering
    \includegraphics[width = \columnwidth]{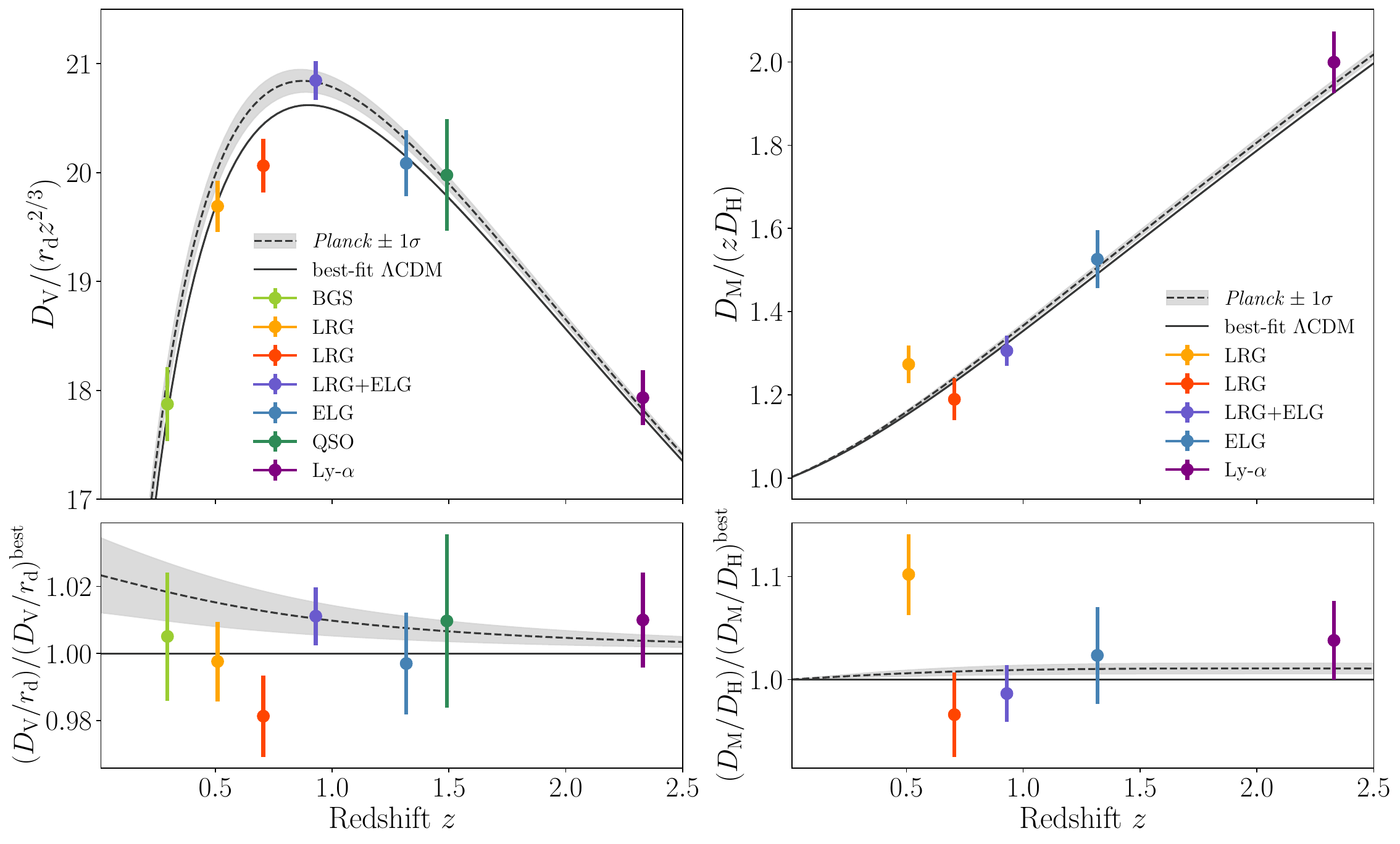}
    \caption{Top row: DESI measurements of the BAO distance scales at different redshifts, parametrised as (\emph{left}) the ratio of the angle-averaged distance $\DV\equiv(z\DM^2D_\mathrm{H})^{1/3}$ to the sound horizon at the baryon drag epoch, $r_\mathrm{d}$, and (\emph{right}) the ratio of transverse and line-of-sight comoving distances $\FAP\equiv\DM/D_\mathrm{H}$, from all tracers and redshift bins as labeled. For visual clarity and to compress the dynamic range of the plot, an arbitrary scaling of $z^{-2/3}$ has been applied on the left, and $z^{-1}$ on the right. The solid and dashed grey lines show model predictions from, respectively, the flat \lcdm\ model that best fits this data, with $\Om=0.294$ and $\Hrd=1.0194\times10^4\;\mathrm{km\,s}^{-1}$, and from a \lcdm\ model with parameters matching the \Planck\ best-fit cosmology. The BGS and QSO data points appear only in the left panel and not the right one because the signal-to-noise ratio of the data is not yet sufficient to measure both parameters for these tracers. Bottom row: The same data points and models as in the top row, but now shown as the ratio relative to the predictions for the best-fit flat \lcdm\ model. 
    }
    \label{fig:D_vs_z}
\end{figure}

\Cref{fig:D_vs_z} shows a summary of these DESI BAO results in the form of a Hubble diagram. In order to conveniently display the results from all redshift bins together, we convert  {results between the  $\left(\DMrd, \DHrd\right)$ and $\left(\DVrd, \DM/D_\mathrm{H}\right)$ bases} for all tracers,  {by mapping the change in parameter basis at each posterior sample and recalculating the marginalised constraints}.\footnote{ {In practice, our BAO analysis of measurements from discrete tracers uses} the $\left(\DVrd, \DM/D_\mathrm{H}\right)$ basis,  {but also reports $\left(\DMrd, \DHrd\right)$ as derived parameters, as the former are} much more independent of each other (though not totally so). The \lya{} BAO measurements still retain a significant correlation in this basis.} The top row then shows $\DVrd$ (scaled by an arbitrary factor of $z^{-2/3}$ for visualisation purposes) in the left panel, and $\DM/D_\mathrm{H}$ (similarly arbitrarily scaled by $z^{-1}$) in the right panel. The solid and dashed grey lines in each panel indicate the corresponding model predictions for the \lcdm{} model that best fit the DESI data (\cref{sec:flat-lcdm}), and the \Planck{} best-fit \lcdm{} model, respectively. The lower panel shows the same data again but now as the ratio of the $\DVrd$ and $\FAP\equiv\DM/D_\mathrm{H}$ values to those for the best-fit \lcdm{} model to DESI data. The solid and dashed grey lines in these panels therefore represent the same two models as in the top row.

\subsection{Internal consistency of DESI results}
\label{sec:internal_consistency}

\Cref{fig:D_vs_z} shows visually that the flat \lcdm\ model provides a good fit to the DESI BAO results: quantitatively, the $\chi^2$ value for this fit is $12.66$ for 10 degrees of freedom (dof), as we have 12 data points and 2 free parameters, namely $\Om$ and $\Hrd$ (\cref{tab:priors}). These two parameters have a direct relationship to the BAO data points shown in \cref{fig:D_vs_z}, since in the flat \lcdm\ model $\Om$ fully determines $\FAP(z)$ and fixes the shape of $\DVrd$ as a function of redshift, while $\Hrd$ sets a redshift-independent constant normalization term for $\DVrd$. The single most anomalous result is the measurement of $\FAP(z=0.51)$, which is mildly offset (at the $\sim2\sigma$ significance level) from the \lcdm\ expectation. This appears to be a simple statistical fluctuation and we do not regard it as significant. In \cref{sec:appendix_DESI+SDSS} we explicitly confirm that this fluctuation does not alter any of the conclusions drawn about any cosmological models in \cref{sec:lcdm,sec:DE,sec:Hubble,sec:neutrinos} below. 
Moreover, since our submission, other papers have appeared discussing further the statistical significance of the DESI data points and their impact on the fit, see e.g. \cite{Efstathiou:2024,Colgain:2024,Wang:2024arXiv240413833W,Wang:2024arXiv240502168W,Liu:2024}.

\begin{figure}
    \centering
    \includegraphics[width = 0.48\columnwidth]{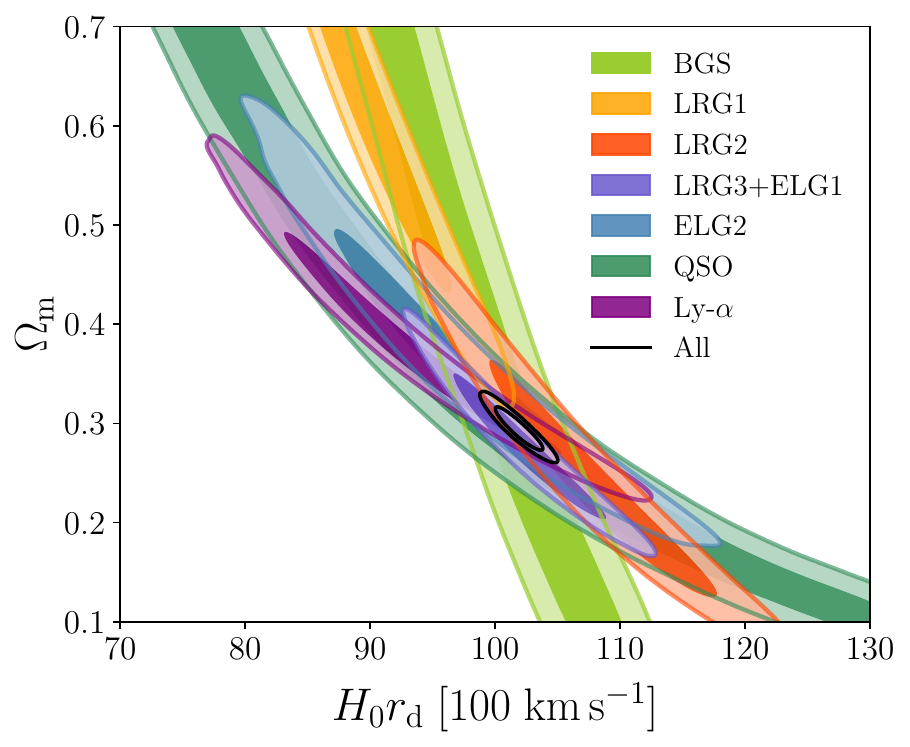} 
    \includegraphics[width=.47\columnwidth]{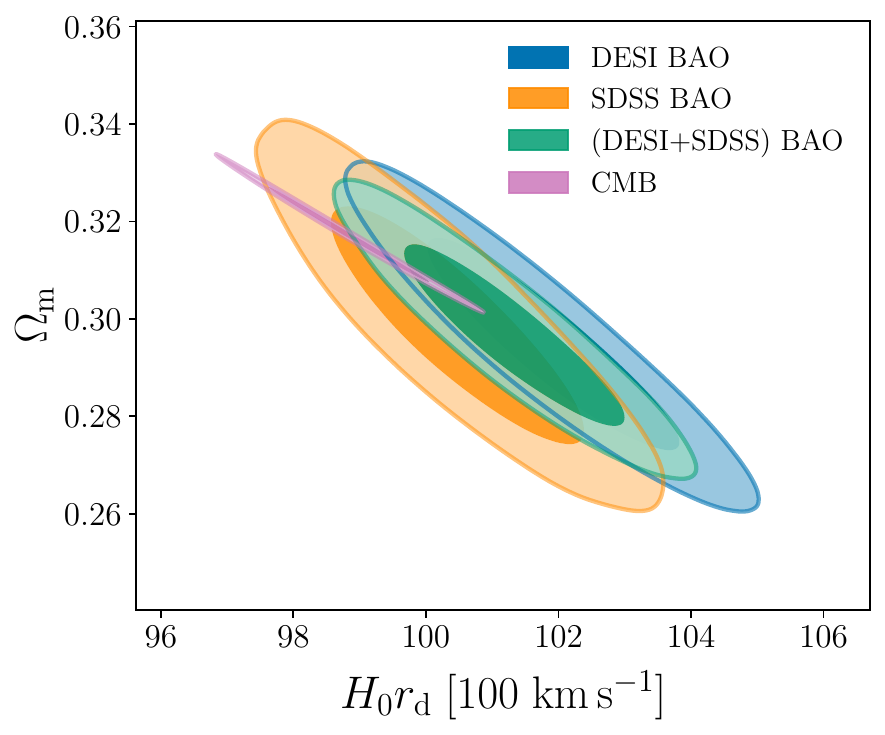}
    \caption{\emph{Left panel:} 68\% and 95\% credible-interval contours for parameters $\Om$ and $\rd h$ obtained for a flat \lcdm\ model from fits to BAO measurements from each DESI tracer type individually, as labeled. Results from all tracers are consistent with each other and the change in the degeneracy directions arises from the different effective redshifts of the samples. \emph{Right panel:} the corresponding results in flat \lcdm\ for fits to BAO results from all DESI redshift bins (blue), the final SDSS results from \cite{Alam-eBOSS:2021} (orange), and the combination of these two as described in the text (green). The corresponding result from the CMB (including CMB lensing) is shown in pink.}
    \label{fig:Om_H0_rd}
\end{figure}

It is also worth confirming that the parameters inferred from each individual redshift bin are consistent with each other. This is shown in \cref{fig:Om_H0_rd}, where in the left panel we show the 68\% and 95\% credible intervals on $\Om$ and $\Hrd$ obtained from fitting a flat \lcdm\ model to each DESI tracer type individually, using the priors described in \cref{tab:priors}.
The slightly anomalous $\FAP(z=0.51)$ value results in a mild shift in the contour for the $0.4<z<0.6$ LRG1 bin, but we do not regard this as significant.
The individual contours have characteristic degeneracy directions that depend on redshift \cite{Cuceu:2019}, and show a strong degree of overlap, confirming the internal consistency of the DESI data. A consequence of the change in the degeneracy directions with redshift is that the combination of all tracers provides a tight final constraint, shown in the right panel of \cref{fig:Om_H0_rd}. 

\subsection{Comparison to BAO results from SDSS}
\label{sec:DESI_SDSS_consistency}

The region of the sky and the range of redshifts observed by DESI partially overlap with those observed by the previous generation of BOSS \cite{Dawson:2013} and eBOSS \cite{Dawson:2016} survey programs of SDSS \cite{York:2000}, whose final BAO results were presented in \cite{Alam-eBOSS:2021}. It is therefore pertinent to compare our BAO results to those from SDSS. A visual comparison of the distance scale measurements obtained from DESI and SDSS is shown in Figure 15 of \cite{DESI2024.III.KP4}.\footnote{\cite{DESI2024.III.KP4} also shows a comparison to BAO measurements from 6dFGS \cite{Beutler:2011} and WiggleZ \cite{Kazin:2014}, although these have lower precision.} As a rough guide, $\sim70\%$ of the DESI DR1 footprint was covered by BOSS; conversely, $\sim65\%$ of the BOSS footprint has been covered by DESI DR1. Therefore the input catalog data used in the DESI and SDSS BAO analyses, while different due to the details of the instrument performance and observing strategy, nevertheless have a fraction of shared objects, introducing a correlation that depends on the redshift and tracer type. 

This correlation is greatest at low redshifts, where the DESI BGS and LRG samples overlap the SDSS Main Galaxy Sample and LRGs from BOSS (LOWZ and CMASS) and eBOSS. There is also a substantial overlap in volume between the QSO samples of DESI and eBOSS in the redshift range $0.8<z<2.1$. Despite this, however, fewer than 15\% of the DESI quasars are also present in the eBOSS catalog. As the measurements of QSO power spectra are shot-noise dominated in both DESI and eBOSS, this results in a negligible overall correlation between the results.

For redshifts below $z=0.6$, the DESI DR1 data currently covers a smaller effective volume than SDSS (compare the $V_{\rm eff}$ values in \cref{tab:Y1data} to those reported in ~\cite{Alam-eBOSS:2021}), but this will change in future data releases. The redshift binning used differs between DESI and SDSS, which makes a direct comparison of BAO results complicated at $z<0.4$, but the $0.4<z<0.6$ and $0.6<z<0.8$ redshift bins closely match the redshift ranges used by BOSS and eBOSS.\footnote{eBOSS actually used the redshift range $0.6<z<1.0$ to define the target sample, but the number density of LRGs in the catalog fell off very sharply beyond $z>0.85$ so this approximation holds.} The SDSS results in these bins are \twoonesig[3cm]
{\DMrd(z=0.51)=13.36\pm0.21,}
{\DHrd(z=0.51) = 22.33\pm0.58,}
{SDSS~LRG $0.4<z<0.6$,}
and
\twoonesig[3cm]
{\DMrd(z=0.70)=17.86\pm0.33,}
{\DHrd(z=0.70) = 19.33\pm0.53,}
{SDSS~LRG $0.6<z<0.85$,}
respectively, which can be compared to the DESI results in \cref{tab:Y1data}. While the results at effective redshift $z=0.51$ are in good agreement, a larger difference can be seen in the $0.6<z<0.8$ redshift bin, particularly in comparison to the DESI result $\DMrd(z=0.71)=16.85\pm0.32$.

To gauge the significance of this disagreement, we can estimate the degree of correlation between power spectrum measurements from DESI and those of SDSS. To do this, we adopt \cref{eq:sample_correlation}, using approximate values for the total volume overlap, the number density of galaxies in common between the two surveys, and assuming a representative amplitude of the power spectrum $P_0=10^4\;(h^{-1}{\rm Mpc})^3$ for both samples. This gives indicative values of $C=0.35$ for $0.4<z<0.6$ and $C=0.21$ for $0.6<z<0.8$. Assuming that this degree of correlation also applies to BAO results derived from the power spectra, and accounting for the predicted changes in $\DMrd$ and $\DHrd$ between $z=0.70$ and $z=0.71$ in the DESI fiducial cosmology, the discrepancy between the DESI and SDSS results in the $0.6<z<0.8$ redshift range is at roughly the $\sim3\sigma$ level.

The cause of this difference is not clear. As discussed in \cite{DESI2024.III.KP4}, numerous improvements in the data processing and BAO fitting methods have been introduced for DESI, and reprocessing the raw BOSS/eBOSS catalog data using the DESI BAO pipeline gives small shifts compared to the results published in \cite{Alam-eBOSS:2021}. However, this accounts for a shift of at most a small fraction of the published uncertainties in $\DMrd$ and $\DHrd$. We note that in this redshift bin the SDSS catalog was a composite formed from the BOSS CMASS galaxy sample extending to $z\simeq0.75$ and the deeper eBOSS LRG sample extending to $z=1$ over a much smaller sky footprint; in contrast the DESI LRG sample is much more uniform. The cause of the difference may simply an unlucky sample variance fluctuation---if so, this will soon become clear with later DESI data releases, which will have a much larger $V_{\rm eff}$ in this redshift range.

While the degree of correlation between BAO results from the galaxy and quasar samples for DESI and SDSS has only been approximately estimated as described above, in the \lya\ forest BAO analysis, the degree of correlation has been more thoroughly quantified in \cite{DESI2024.IV.KP6} and shown to agree very well. \cite{DESI2024.IV.KP6} also provides results for a combined ``DESI+SDSS" \lya\ BAO measurement:
\be
\label{eq:DESI+SDSSLya1}
\DMrd(z=2.33)=38.80\pm0.75,
\ee
and
\be
\label{eq:DESI+SDSSLya2}
\DHrd(z=2.33) = 8.72\pm0.14,
\ee
with an anticorrelation of $\rho=-0.48$ between $\DMrd$ and $\DHrd$.

The $\DMrd(z=0.71)$ measurement from LRGs yields the greatest difference between SDSS and DESI at any redshift. Nevertheless, using the BAO measurements from all redshifts together, the BAO distance ladders obtained from DESI and SDSS result in consistent inference of cosmological parameters. The right panel of \cref{fig:Om_H0_rd} shows the posterior credible intervals for the parameters $\Om$ and $\rd h$ in a flat \lcdm\ cosmology from fitting to the DESI (blue) and SDSS (orange) BAO data. The DESI values are described in \cref{sec:lcdm} and \cref{tab:parameter_table1}. The results are clearly in very good agreement with each other, with no significant difference in $\Om$ and a shift of just $\sim1\sigma$ in $\rd h$.

In this paper our primary focus will be on the cosmological consequences of the DESI DR1 BAO results alone, as a homogeneously analysed dataset across redshift and tracer type. However, it is also possible to construct a dataset of BAO distance measurements from the combination of DESI and SDSS results in order to maximise measurement precision across the entire redshift range $0.1<z<4.2$. Bearing in mind that the degree of correlation between BAO results from discrete galaxy and quasar tracers in the two surveys has not been precisely quantified---to avoid double-counting information---this combined sample should be selected by choosing the result from the survey covering the larger effective volume $V_{\rm eff}$ at a given redshift. Thus, the composite BAO dataset can be constructed as follows:
\begin{itemize}
\item at $z<0.6$ where SDSS currently has a larger $V_{\rm eff}$, we use the SDSS results at $z_{\rm eff}=0.15, 0.38$ and $0.51$ in place of the DESI BGS and lowest-redshift LRG points;
\item at $z>0.6$ where DESI has $V_{\rm eff}$ larger than that of SDSS, we use the DESI results from LRGs over $0.6<z<0.8$, the LRG+ELG combination over $0.8<z<1.1$, and ELGs and QSOs at higher redshifts; and
\item for the \lya\ BAO we use the combined DESI+SDSS result from \cref{eq:DESI+SDSSLya1,eq:DESI+SDSSLya2} above.
\end{itemize}
We use the label ``DESI+SDSS" to refer to this composite BAO dataset, while reiterating that this is not the same as simply combining the likelihoods from each survey individually due to the overlap in volumes. The posterior in parameters $(\Om, \rd h)$ inferred from fitting a flat \lcdm\ model to this dataset is shown in green in the right panel of \cref{fig:Om_H0_rd}. In \cref{sec:appendix_DESI+SDSS} we present constraints using the DESI+SDSS data combination in various cosmological models, specifically comparing them to constraints using DESI data alone.

\section{Cosmological constraints in the \lcdm\ model}
\label{sec:lcdm}

\begin{table}
\centering
\resizebox{\columnwidth}{!}{%
    \small
    \begin{tabular}{lcccccc}
    \toprule
    \midrule
    \multirow{2}{*}{model/dataset} & \multirow{2}{*}{$\Om$} & $H_0$ & \multirow{2}{*}{$10^3\Ok$} & \multirow{2}{*}{$w$ or $w_0$} & \multirow{2}{*}{$w_a$}\\
     & & [$\kmsMpc$] & & & \\
    \midrule
    {\bf Flat} $\boldsymbol{\Lambda}${\bf CDM} &&&&\\
    DESI & $0.295\pm 0.015$ &---&---&---&---\\
    DESI+BBN & $0.295\pm 0.015$ & $68.53\pm 0.80$ &---&---&---\\
    DESI+BBN+$\theta_\ast$ & $0.2948\pm 0.0074$ & $68.52\pm 0.62$ &---&---&---\\
    DESI+CMB & $0.3069\pm 0.0050$ & $67.97\pm 0.38$ &---&---& --\\
    \midrule
    $\boldsymbol{\Lambda}${\bf CDM+}$\boldsymbol{\Ok}$ &&&&\\
    DESI & $0.284\pm 0.020$ &---& $65^{+68}_{-78}$ &---&---\\
    DESI+BBN+$\theta_\ast$ & $0.296\pm 0.014$ & $68.52\pm 0.69$ & $0.3^{+4.8}_{-5.4}$ &---&---\\
    DESI+CMB & $0.3049\pm 0.0051$ & $68.51\pm 0.52$ & $2.4\pm1.6$ &---&---\\
    \midrule
    $\boldsymbol{w}${\bf CDM} &&&&\\
    DESI & $0.293\pm 0.015$ &---&---& $-0.99^{+0.15}_{-0.13}$ &---\\
    DESI+BBN+$\theta_\ast$ & $0.295\pm 0.014$ & $68.6^{+1.8}_{-2.1}$ &---& $-1.002^{+0.091}_{-0.080}$ &---\\
    DESI+CMB & $0.281\pm 0.013$ & $71.3^{+1.5}_{-1.8}$ &---& $-1.122^{+0.062}_{-0.054}$ &---\\
    DESI+CMB+Panth. & $0.3095\pm 0.0069$ & $67.74\pm 0.71$ &---& $-0.997\pm 0.025$ &---\\
    DESI+CMB+Union3 & $0.3095\pm 0.0083$ & $67.76\pm 0.90$ &---& $-0.997\pm 0.032$ &---\\
    DESI+CMB+DESY5 & $0.3169\pm 0.0065$ & $66.92\pm 0.64$ &---& $-0.967\pm 0.024$ &---\\
    \midrule
    $\boldsymbol{w_0w_a}${\bf CDM} &&&&\\
    DESI & $0.344^{+0.047}_{-0.026}$ &---&---& $-0.55^{+0.39}_{-0.21}$ & $< -1.32$ \\
    DESI+BBN+$\theta_\ast$ & $0.338^{+0.039}_{-0.029}$ & $65.0^{+2.3}_{-3.6}$ &---& $-0.53^{+0.42}_{-0.22}$ & $< -1.08$ \\
    DESI+CMB & $0.344^{+0.032}_{-0.027}$ & $64.7^{+2.2}_{-3.3}$ &---& $-0.45^{+0.34}_{-0.21}$ & $-1.79^{+0.48}_{-1.0}$ \\
    DESI+CMB+Panth. & $0.3085\pm 0.0068$ & $68.03\pm 0.72$ &---& $-0.827\pm 0.063$ & $-0.75^{+0.29}_{-0.25}$ \\
    DESI+CMB+Union3 & $0.3230\pm 0.0095$ & $66.53\pm 0.94$ &---& $-0.65\pm 0.10$ & $-1.27^{+0.40}_{-0.34}$ \\
    DESI+CMB+DESY5 & $0.3160\pm 0.0065$ & $67.24\pm 0.66$ &---& $-0.727\pm 0.067$ & $-1.05^{+0.31}_{-0.27}$ \\
    \midrule
    $\boldsymbol{w_0w_a}${\bf CDM+}$\boldsymbol{\Ok}$ \\
    DESI & $0.313\pm 0.049$ &---& $ 87^{+100}_{-85}$ & $-0.70^{+0.49}_{-0.25}$ & $< -1.21$\\
    DESI+BBN+$\theta_\ast$ & $0.346^{+0.042}_{-0.024}$ & $65.8^{+2.6}_{-3.5}$ & $ 5.9^{+9.1}_{-6.9}$ & $-0.52^{+0.38}_{-0.19}$ & $< -1.44$\\
    DESI+CMB & $0.347^{+0.031}_{-0.025}$ & $64.3^{+2.0}_{-3.2}$ & $ -0.9\pm2$ & $-0.41^{+0.33}_{-0.18}$ & $< -1.61$ \\
    DESI+CMB+Panth. & $0.3084\pm 0.0067$ & $68.06\pm 0.74$ & $0.3\pm1.8$ & $-0.831\pm 0.066$ & $-0.73^{+0.32}_{-0.28}$ \\
    DESI+CMB+Union3 & $0.3233^{+0.0089}_{-0.010}$ & $66.45\pm 0.98$ & $-0.4\pm1.9$ & $-0.64\pm 0.11$ & $-1.30^{+0.45}_{-0.39}$ \\
    DESI+CMB+DESY5 & $0.3163\pm 0.0065$ & $67.19\pm 0.69$ & $-0.2\pm1.9$ & $-0.725\pm 0.071$ & $-1.06^{+0.35}_{-0.31}$ \\
    \midrule
    \bottomrule
    \end{tabular}
}
\caption{
    Cosmological parameter results from DESI DR1 BAO data in combination with external datasets and priors, in the baseline flat \lcdm\ model and extensions including spatial curvature and two parametrisations of the dark energy equation of state, as listed. Results are quoted for the marginalised means and 68\% credible intervals in each case, including for upper limits. Note that DESI data alone measures $\rd h$ and not $H_0$, but for reasons of space this result is omitted from the table and provided in the text instead. In this and other tables, the shorthand notation ``CMB" is used to denote the addition of temperature and polarisation data from \Planck\ and CMB lensing data from the combination of \Planck\ and ACT.
    \vspace{0.1em}
    \label{tab:parameter_table1}
}
\end{table}

\subsection{Flat background}
\label{sec:flat-lcdm}

In this section, we present cosmological constraints for the flat \lcdm\ cosmological model in which $\Ok=0$, $\OL=1-\Om$. As discussed in \cref{sec:internal_consistency,sec:DESI_SDSS_consistency} above, in this model BAO distance scales alone constrain only two free parameters: the matter density parameter $\Om$ and the combination $\Hrd$ of the Hubble constant $H_0$ and the sound horizon $\rd$. The 68\% credible-interval results for these parameters are
\twoonesig[1.5cm]
{\Om &= 0.295\pm 0.015,}
{\rd h &= (101.8\pm 1.3) \Mpc,}
{DESI~BAO, \label{eq:DESI_LCDM_results}} 
where to simplify units and notation we quote results for $\rd h\equiv\Hrd/(100\;\kmsMpc)$ instead of $\Hrd$. The posterior constraints are shown in the right panel of \cref{fig:Om_H0_rd}, and it is clear that they are in very good agreement with the previously reported SDSS values of $\Om=0.299\pm 0.016$ and $\rd h = (100.4\pm 1.3) \Mpc$ \cite{Alam-eBOSS:2021}. 

It is apparent from \cref{fig:Om_H0_rd} that DESI BAO prefer a somewhat larger value of $\rd h$ compared to the value $\rd h=(98.82\pm 0.82) \Mpc$ obtained from the combination of CMB temperature, polarisation and lensing (or $\rd h=(98.9\pm 1.0) \Mpc$ from the CMB when excluding CMB lensing).
It is therefore of interest to
quantify the level of consistency between these two data sets. We do so by calculating the relative $\chi^2$ between the two datasets as 
\begin{equation}
    \label{eq:2D_chi2}
   \chi^2 = (\mathbf{p}_A - \mathbf{p}_B)^T(\mathrm{Cov}_A+\mathrm{Cov}_B)^{-1}(\mathbf{p}_A - \mathbf{p}_B), 
\end{equation}
and converting it to the probability-to-exceed between two datasets (e.g.\ \cite{Abbott:2017smn,DES:2020hen}). Here $\mathbf{p}_A$ and $\mathbf{p}_B$ refer to ($\Om, \rd h$) parameter vectors obtained from fits to the two datasets, and $\mathrm{Cov}_A$ and $\mathrm{Cov}_B$ to the corresponding $2\times 2$ covariances. Applying this statistic to DESI DR1 BAO and CMB, we find that the differences in this parameter space amount to a mild $1.9\sigma$-level discrepancy when including CMB lensing, or $1.6\sigma$ without. We thus consider that DESI BAO data are consistent with the CMB in this parameter space.

Since BAO distance measurements alone are sensitive to the combination $\Hrd$ \cite{Sutherland:2012ys}, an external calibration of the sound horizon $\rd$ is required in order to break the $H_0$--$\rd$ degeneracy and obtain a constraint on the Hubble constant $H_0$. This method of calibrating the BAO distance scales using the sound horizon at early times is known as the ``inverse distance ladder" approach \cite{PhysRevD.92.123516, cuestainverseladder}. Directly calibrating the BAO standard ruler using the value $\rd=147.09\pm 0.26\Mpc$ obtained from using all CMB and CMB lensing information \cite{Planck-2018-cosmology} gives
\begin{equation} 
\label{eq:H0_DESI-rd}
H_0= (69.29\pm 0.87)\kmsMpc
\qquad
(\mbox{DESI BAO\,+\,$\rd$ from CMB}). 
\end{equation}
An alternative approach that is more independent of CMB information is to use a prior on $\Ob h^2$ obtained from BBN (\cref{sec:BBN}), which is sufficient to determine $\rd$ and break the degeneracy \cite{Addison:2013, 2015PhRvD..92l3516A}. Within flat \lcdm, DESI DR1 BAO + BBN give 
\begin{equation}
\label{eq:H0_DESI-bbn}
H_0= (68.53\pm 0.80)\kmsMpc
\qquad
(\mbox{DESI~BAO\,+\,BBN}).    
\end{equation}
This constraint is 20\% more precise than the equivalent one obtained from SDSS BAO data, $H_0=(67.35\pm 0.97)\kmsMpc$ \cite{Alam-eBOSS:2021}, which is a result of the improved precision of the DESI BAO measurements at higher redshifts, leading to narrower posterior constraints perpendicular to the main degeneracy direction in the $\Om$--$\rd h$ plane (\cref{fig:Om_H0_rd}). The central value is also shifted by about $\sim1\sigma$ to higher $H_0$, a direct consequence of the similar shift in $\Hrd$ shown in \cref{fig:Om_H0_rd}. However, the two results are fully consistent with each other.

We can also use a conservative model-independent prior 
on the acoustic angular scale $\theta_\ast$ seen from the CMB (\cref{sec:CMB}), without including additional CMB information. While on its own knowledge of $\theta_\ast$ is insufficient to break the $\rd$--$h$ degeneracy (because the baryon density $\Ob h^2$ remains unknown, so $r_d$ cannot be fixed), it is extremely robust and, in combination with BBN information, further tightens the $H_0$ result to
\begin{equation}
\label{eq:H0_DESI-bbn-thetastar}
H_0= (68.52\pm 0.62)\;\kmsMpc
\qquad
(\mbox{DESI BAO\,+\,BBN\,+\,$\theta_\ast$}),    
\end{equation}
a 0.9\% measurement of the Hubble constant. As discussed further in \cref{sec:Hubble} below, this more precise constraint can also be extended to be robust to many early-universe extensions of the base \lcdm\ model, in particular to assumptions about the effective number of relativistic species $\Neff$. 

It is interesting to note that the direct $\rd$ calibration from the CMB produces the weakest constraint on $H_0$ despite apparently using the most external information and being the most model-dependent. This is because BBN and $\theta_\ast$ information affect $\Om$ and are applied consistently in the posterior sampling, and thus leverage the degeneracy between $\Om$ and $\rd h$ from BAO data, while a prior directly applied on $\rd$ cannot do so.

\begin{figure}
    \centering 
    \includegraphics[width=0.48\columnwidth]{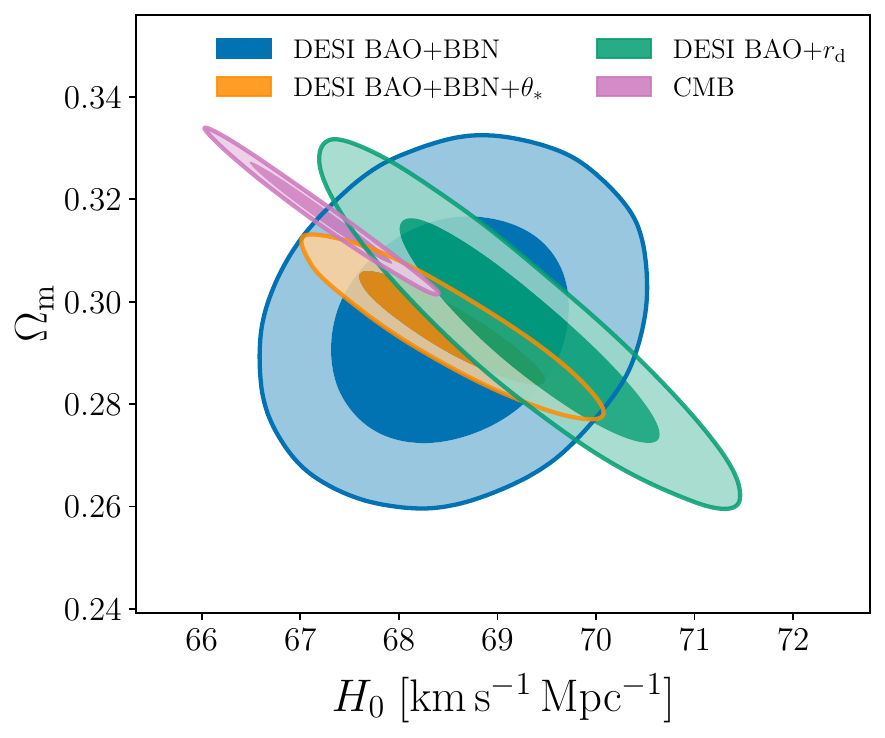}
    \includegraphics[width=0.48\columnwidth]{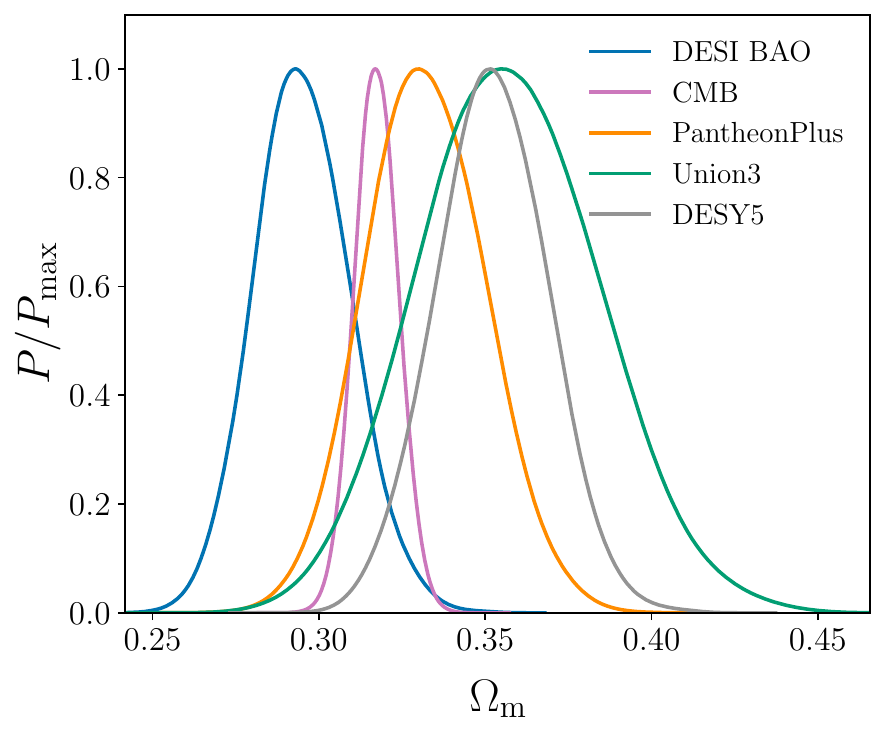}
    \caption{\emph{Left panel}: marginalised posterior constraints on matter density $\Om$ and the Hubble constant $H_0$, obtained from combining DESI BAO data with external data used to calibrate the sound horizon $\rd$, in a flat \lcdm\ cosmological model. The combinations shown use a prior on $\ob$ determined from BBN (blue), the combination of a BBN $\ob$ prior and measurement of the acoustic angular scale $\theta_\ast$ (orange), and $\rd$ directly calibrated from CMB results from \Planck\ (green). The pink contour shows the corresponding constraints from the combination of CMB and CMB lensing. \emph{Right panel}: The marginalised 1D posteriors on $\Om$ in flat \lcdm, from DESI BAO, CMB and the three SN datasets, as labelled.}
    \label{fig:Om-H0}
\end{figure}

The left panel of \cref{fig:Om-H0} summarises the constraints in the $\Om$--$H_0$ plane obtained from the combination of DESI BAO data with each of the external priors discussed so far, and compares them to the combined CMB result from \Planck\ and ACT. All combinations including DESI data favour somewhat higher values of $H_0$ and lower values of $\Om$ than the mean values for the CMB posterior. However, the results are not in serious tension: using the metric outlined in \cref{eq:2D_chi2} above, the biggest discrepancy is at the $\sim2.1\sigma$ level, for DESI+BBN+$\theta_\ast$ compared to CMB. Given this level of agreement, there is no issue with combining DESI and CMB data to obtain joint constraints; doing so we find
\twoonesig[2cm]
{\Om &= 0.3069\pm 0.0050,}
{H_0 &= (67.97\pm 0.38) \kmsMpc}
{DESI~BAO+ CMB. \label{eq:DESI+CMB_LCDM_results}} 
These results are summarised in \cref{tab:parameter_table1}, which also shows parameter constraints obtained in other extended models.

A final instructive comparison within the context of the flat \lcdm\ is between the constraints on the matter density $\Om$ offered by DESI and SN~Ia. These are shown in the right panel of \cref{fig:Om-H0}. PantheonPlus reports $\Om=0.331\pm 0.018$, Union3 gives $\Om=0.359^{+0.025}_{-0.028}$, and DESY5 gives $\Om=0.353\pm 0.017$. There is therefore a moderate variation in both central values and quoted uncertainties across different SN~Ia compilations, but all of them prefer higher values of $\Om$ than DESI and the CMB. The statistical significance of the differences compared to DESI, calculated as described above, stands at $1.6\sigma$ for PantheonPlus, $2.0\sigma$ for Union3, and $2.6\sigma$ for DESY5. While they do not meet a $3\sigma$ threshold for significant tension, these numbers indicate a degree of disagreement between these datasets and DESI results when interpreted in the flat \lcdm\ model. Should these mismatches persist and become more significant when more data is acquired, they will require further investigation.

\subsection{\lcdm\ model with free spatial curvature}
\label{sec:curved_lcdm}

\begin{figure} [t!] 
    \centering
    \includegraphics[width = 0.47\columnwidth]{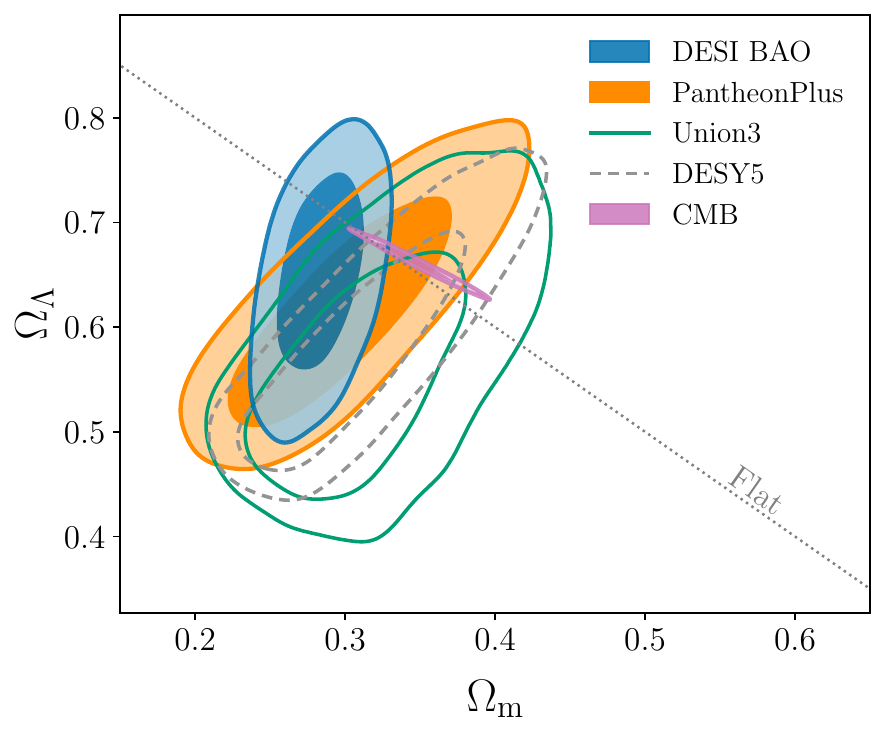} 
    \includegraphics[width = 0.48\columnwidth]{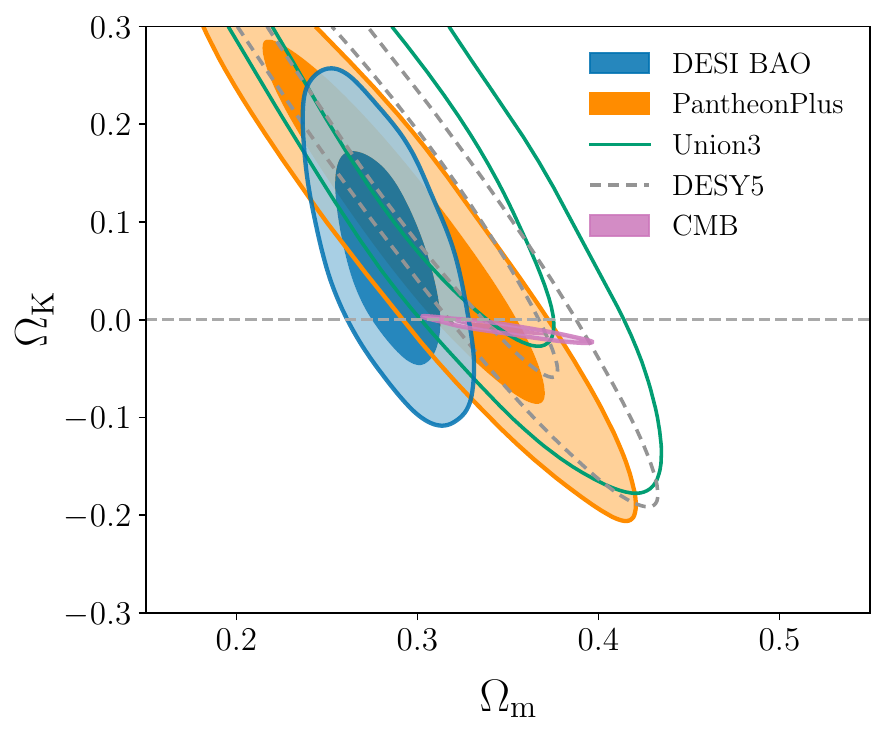} 
    \caption{68\% and 95\% marginalised posterior constraints on $\Om$--$\OL$ plane (left) and $\Om$--$\Ok$ (right) in the one-parameter extension of the \lcdm\ model with free curvature, \lcdm$+\Ok$. In the left panel the supernova contours are truncated at the lower-left by the $\mathcal{U}[-0.3, 0.3]$ prior on $\Ok$.}
    \label{fig:Omegam_OmegaL}
\end{figure}

Relaxing the condition of spatial flatness in the $\Lambda$CDM model, we allow the curvature parameter $\Ok$ to vary. In an FLRW background, this is equivalent to allowing the dark energy density $\OL=1-\Om-\Ok$ to vary independently from the matter density $\Om$, while still keeping the dark-energy equation of state fixed at $w=-1$. Because DESI provides relative measures of the BAO scale at multiple redshifts, it can determine the expansion rate as a function of redshift and thus measure $\OL$ independent of any calibration of the sound horizon from external data \cite{2020PhRvL.124v1301N}. In this model DESI alone thus measures the parameters $\left(\Om, \OL,\rd h\right)$, and finds the following 68\% credible-interval constraints
\threeonesig[3.5cm]{\Om &= 0.284\pm 0.020,}{\OL &= 0.651^{+0.068}_{-0.057},}{\rd h &= (100.9\pm 1.6)\,{\rm Mpc},}{DESI~BAO. \label{eq:DESI_kLCDM}}
Expressed in terms of curvature, the DESI result is $\Ok= 0.065^{+0.068}_{-0.078}$. 
\Cref{fig:Omegam_OmegaL} shows the corresponding 2D credible intervals in both $(\Om,\OL)$ and $(\Om, \Ok)$ plane.

The addition of model-independent information on the acoustic angular scale $\theta_\star$ effectively extends the redshift range of the BAO distance ladder, reaching all the way to last scattering at $z\sim1090$. This dramatically improves the curvature constraints, giving $\Ok=0.0108^{+0.015}_{-0.0056}$. Adding a BBN prior this is further tightened to 
\oneonesig[4.5cm]{\Ok=0.0003^{+0.0048}_{-0.0054}}{DESI~BAO\,+\,BBN\,+\,$\theta_\ast$}{. \label{eq:DESI-bbn-thetastar_OmegaK}} 

The corresponding curvature constraint from the CMB is 
\oneonesig[1.5cm]{\Ok= -0.0102\pm 0.0054}{CMB}{. \label{eq:CMB_OmegaK}} 
This value is $\sim2\sigma$ away from $\Ok=0$, and the difference is even larger (over $3\sigma$) when not including CMB lensing. This is a well-known issue in the \Planck\ PR3 likelihood (see discussion in \cite{Planck-2018-cosmology}, as well as, e.g., \cite{DiValentino:2020,Handley:2021}) caused by the combination of the geometric degeneracy limiting CMB measurements of curvature, and the observed slight excess of smoothing of high-$\ell$ peaks in the temperature power spectrum. Later \Planck\ PR4 likelihoods somewhat alleviate, but do not completely remove, this CMB preference for $\Ok<0$ \cite{Efstathiou:2021,Rosenberg:2022,Tristram:2023}. Thus BAO results are crucial in providing a curvature constraint that is independent of the CMB.
The combined result from DESI BAO and all CMB information is 
\oneonesig[3.5cm]{\Ok= 0.0024\pm 0.0016}{DESI~BAO+CMB}{. \label{eq:DESI_CMB_OmegaK}} 

\section{Dark energy}
\label{sec:DE}

Over the past quarter century, an impressive variety of cosmological observations have confirmed and vastly strengthened the evidence that the expansion of the universe is accelerating. The standard cosmological model has a flat background geometry, but a sub-critical total (cold and baryonic) matter density.  While the measurements thus far are in good agreement with the simplest flat \lcdm\ scenario---a vacuum energy described by the cosmological constant dominating the energy density and responsible for the late-time acceleration---much effort is dedicated to measuring the expansion and growth history of the universe and looking for any departures from this model. In this regard, constraining dark energy models beyond the simplest \lcdm\ model is a principal goal of DESI. 


\subsection{Flat $w$CDM model}
\label{sec:wcdm}

\begin{figure}
    \centering
    \includegraphics[width = 0.68\columnwidth]{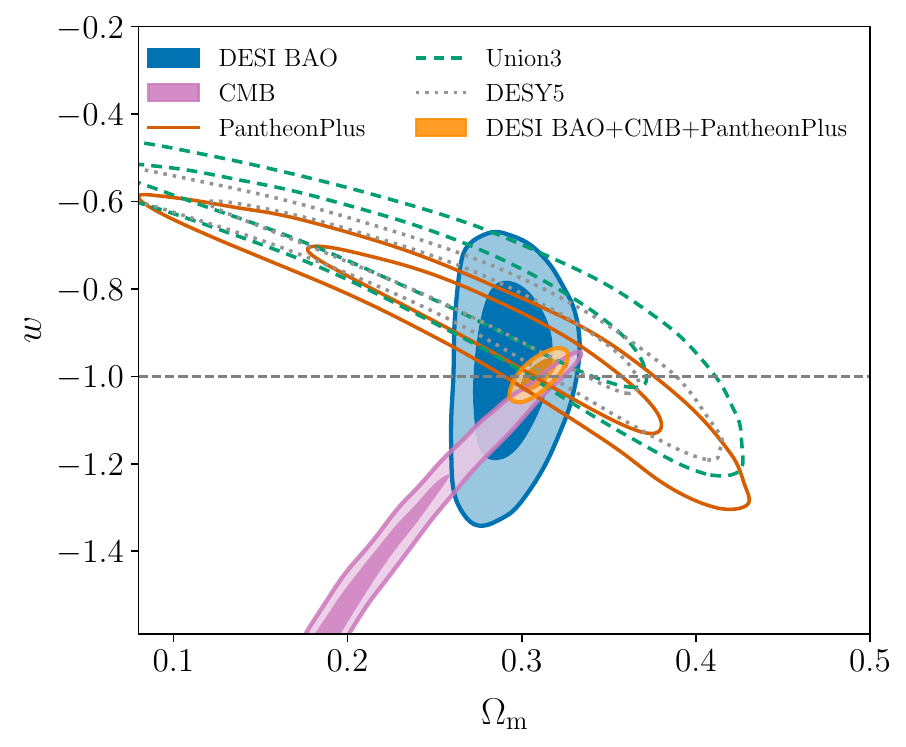}
    \caption{Constraints on $\Om$ and $w$ in the flat \wcdm\ model. The constraints from DESI BAO alone are shown in blue, those from the CMB in pink, and different SN~Ia compilations in solid and dashed green. The orange contour shows the combined constraint from DESI, CMB and PantheonPlus SN~Ia. All contours show 68\% and 95\% credible intervals. Note the remarkable complementarity of cosmological probes in this plane.}
    \label{fig:omegam_w}
\end{figure}

Although a cosmological constant fits existing data well, the tiny observed value of $\Lambda$ relative to typical scales in particle physics poses great theoretical challenges \cite{Martin2012,Burgess2013,Padilla2015}. 
Acceleration physics beyond $\Lambda$ necessarily has dynamics – time dependence (and spatial perturbations, though these have diminishing effect the closer the time dependence is to $\Lambda$, i.e.\  constant).

At the background cosmology level entering cosmic distances, the acceleration physics can be treated as an effective dark energy density and pressure. Thus we have a dark energy equation of state parameter, or pressure to energy density ratio, $w(a)$, and a current dark energy density value, $\Omega_{\rm de}$, to describe the dark energy component. In the early 2000s, with the time variation $w(a)$ inaccessible to observations, analyses often fixed $w =\,$const ($w$CDM), but this is insensitive to crucial dynamics that might be indicated by data. 
$w$CDM can however still be useful as an alert if the recovered constraint on constant $w$ has statistically significant deviation from $w=-1$. Note the converse is not true: measuring $w=-1$ assuming $w$ is constant does not indicate $\Lambda$ is correct (known as the ``mirage of $\Lambda$"  \cite{Linder2007}). 

\Cref{fig:omegam_w} shows the constraints on $\Om$ and $w$ from a variety of different data and combinations. In this plane the DESI contours are close to vertically aligned, providing a tight constraint on $\Om$ that is largely independent of $w$, in contrast to SN~Ia and CMB probes, which show distinctive degeneracy directions corresponding to the transverse comoving distances that each of these probes constrains \cite{2003PhRvD..67h3505F}. We find
\twoonesig[1.5cm]
{\Om &= 0.293\pm 0.015,}
{w   &= -0.99^{+0.15}_{-0.13},}
{DESI~BAO, \label{eq:DESI_wCDM}}
from DESI alone, while combining DESI BAO with BBN and $\theta_\ast$ significantly tightens the constraint on $w$ to $w=-1.002^{+0.091}_{-0.080}$. Adding CMB data shifts the contours slightly along the CMB degeneracy direction, giving 
\twoonesig[3.5cm]
{\Om &= 0.281\pm 0.013,}
{w   &= -1.122^{+0.062}_{-0.054},}
{DESI~BAO+CMB. \label{eq:DESI_CMB_wCDM}}
Finally, the tightest constraints are obtained from the combination of these data with SN~Ia. For example for the PantheonPlus SN~Ia dataset:
\twoonesig[3.5cm]
{\Om &= 0.3095\pm 0.0069,}
{w   &= -0.997\pm 0.025,}
{DESI+CMB\dataplus PantheonPlus. \label{eq:DESI_SN_CMB_wCDM}}
Similar constraints are obtained when substituting PantheonPlus SN~Ia for DESY5 or Union3 (though with slightly larger uncertainties in the latter case). These results are summarised in \cref{tab:parameter_table1}. In summary, DESI data, both alone and in combination with other cosmological probes, do not show any evidence for a constant equation of state parameter different from $-1$ when a flat $w$CDM model is assumed. 

\subsection{Flat $w_0w_a$CDM model}
\label{sec:w0wa}

\begin{figure}
    \centering
    \includegraphics[width = 0.48\columnwidth]{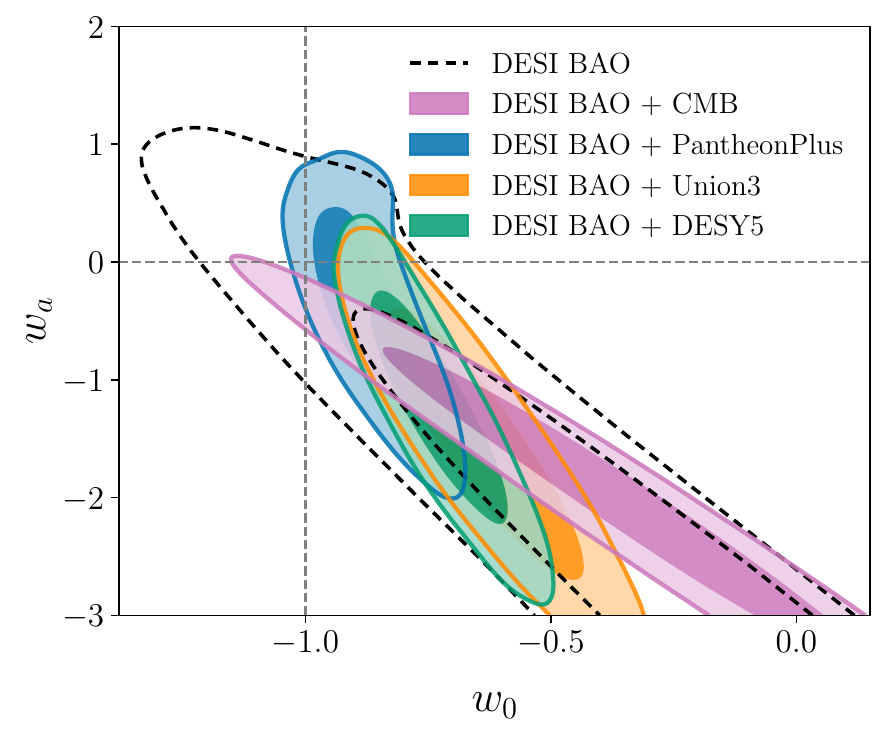} 
    \includegraphics[width = 0.47\columnwidth]{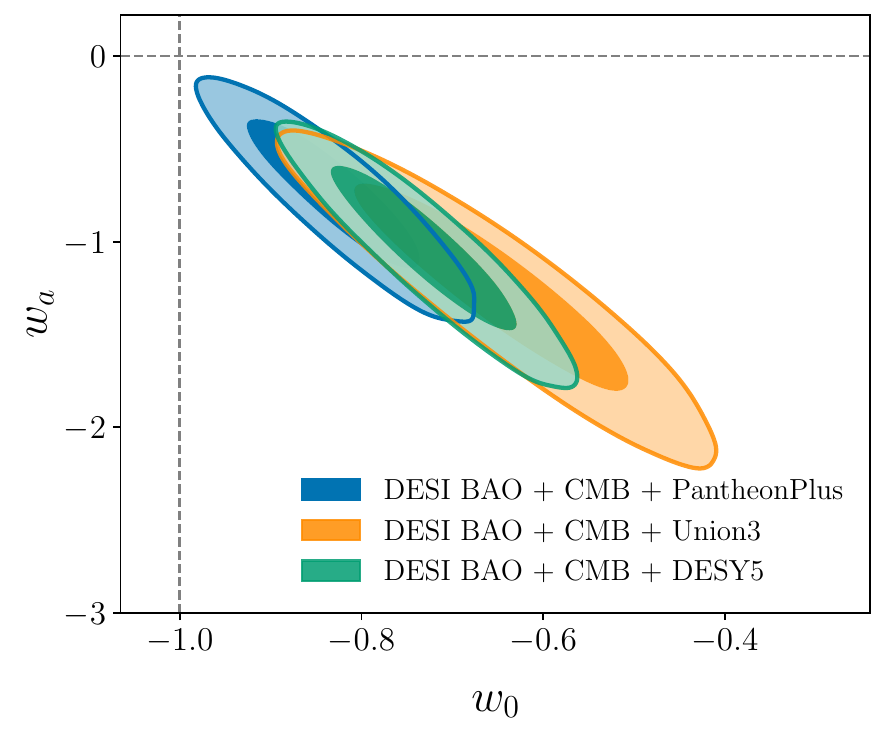}
    \caption{\emph{Left panel}: 68\% and 95\% marginalised posterior constraints in the $w_0$--$w_a$ plane for the flat $w_0w_a$CDM model, from DESI BAO alone (black dashed), DESI + CMB (pink), and DESI + SN~Ia, for the PantheonPlus \cite{Brout:2022}, Union3 \cite{Rubin:2023} and DESY5 \cite{DES:2024tys} SNIa datasets in blue, orange and green respectively. Each of these combinations favours $w_0>-1$, $w_a<0$, with several of them exhibiting mild discrepancies with \lcdm\ at the $\gtrsim2\sigma$ level. However, the full constraining power is not realised without combining all three probes. \emph{Right panel}: the 68\% and 95\% marginalised posterior constraints from DESI BAO combined with CMB and each of the PantheonPlus, Union3 and DESY5 SN~Ia datasets. The significance of the tension with \lcdm\ ($w_0=-1$, $w_a=0$) estimated from the $\Delta \chi_\mathrm{MAP}^{2}$ values is $2.5\sigma$, $3.5\sigma$ and $3.9\sigma$ for these three cases respectively.}
    \label{fig:w0wa_1}
\end{figure}

Taking into account the physical dynamics of dark energy, the parametrisation $w(a)=w_0+w_a\,(1-a)$ was derived and has been demonstrated to match the background evolution of distances arising from exact dark energy equations of motion to an accuracy of $\mathcal{O}(0.1\%)$ for viable cosmologies over a wide range of physics -- scalar fields, modified gravity, and phase transitions \cite{Linder2003,dePutter2008}.\footnote{ {The converse is however not true and not all values in the $\left(w_{0}, w_{a}\right)$ parameter space provide an  equation of state that can be mapped up to high accuracy to a viable physical models.}} In this section, we present constraints on this model, referred to as \wowacdm, which reduces to \lcdm\ for $w_0=-1$, $w_a=0$. Constraining the $w_0$--$w_a$ parameter space and its corresponding behavior as well as distinguishing it from $\Lambda$ is a key science goal of DESI. 

 We adopt wide flat priors on $w_0$ and $w_a$ (\cref{tab:priors}), together with the condition $w_0+w_a<0$ imposed to enforce a period of high-redshift matter domination. Since the parameter space we explore includes models whose equation of state crosses the $w=-1$ boundary, we use the parametrised post-Friedmann approach \cite{2008PhRvD..78h7303F} to compute the dark energy perturbations when calculating the CMB angular power spectrum. \Cref{fig:w0wa_1} shows the marginalised posteriors in the $w_0$--$w_a$ plane from DESI and combinations with other external datasets. DESI alone does not have sufficient power to break the degeneracy between $w_0$ and $w_a$ and thus the results are cut off by our priors (see \cref{tab:priors}), 
\twoonesig[1.5cm]
{w_0 &= -0.55^{+0.39}_{-0.21},}
{w_a &< -1.32,}
{DESI~BAO, \label{eq:DESI_w0waCDM}}
with the upper bound on $w_a$ referring to the 68\% limit.
This represents a mild pull away from the \lcdm\ value, with a $\Delta \chi_\mathrm{MAP}^2$ between the maximum a posteriori of the \wowacdm\ model and the MAP when fixing $(w_{0}, w_{a}) = (-1, 0)$ of just $-3.7$ for 2 additional degrees of freedom. The cause of this small preference for $w_0>-1$ (from DESI data alone) is primarily due to the $\FAP$ measurement from the $0.4<z<0.6$ LRG bin, which lies slightly higher (at the $\sim2\sigma$ level) than the best-fit \lcdm\ model can accommodate, as shown in \cref{fig:D_vs_z}. In order to better fit this data point, the equation of state $w(z\lesssim 0.5)$ of the best-fit \wowacdm\ model prefers to be high, thus preferentially pulling $w_0$ to less negative values than $-1$. On the other hand, to fit the other DESI points which are all fairly close to the \lcdm\ predictions, the parameter $w_a$ prefers to be strongly negative in order to compensate the integrated effect of $w(z)$ for those quantities at higher redshift. 

Given the small $\Delta \chi_\mathrm{MAP}^2$ it is clear there is no statistical preference for \wowacdm\ from DESI alone. CMB data alone also gives $\Delta \chi_\mathrm{MAP}^2=-3.7$ for the MAP \wowacdm\ model compared to fixing $(w_{0}, w_{a}) = (-1, 0)$, again showing no statistical preference. Nevertheless, given the overlap of the CMB and DESI contours in the $\left(w_0>-1, w_a<0\right)$ quadrant, the combined results give
\twoonesig[1.5cm]
{w_0 &= -0.45^{+0.34}_{-0.21}, }
{w_a &= -1.79^{+0.48}_{-1.0},}
{DESI~BAO+CMB, \label{eq:DESI_CMB_w0waCDM}}
and the $\Delta \chi_\mathrm{MAP}^2$ decreases to $-9.5$, indicating a preference for an evolving dark energy equation of state at the $\sim2.6\sigma$ level. The contours in this scenario are however still impacted by the priors we have assumed, thus care is required in interpreting these shifts. 

\begin{figure}
    \centering
    \includegraphics[width = 0.9\columnwidth]{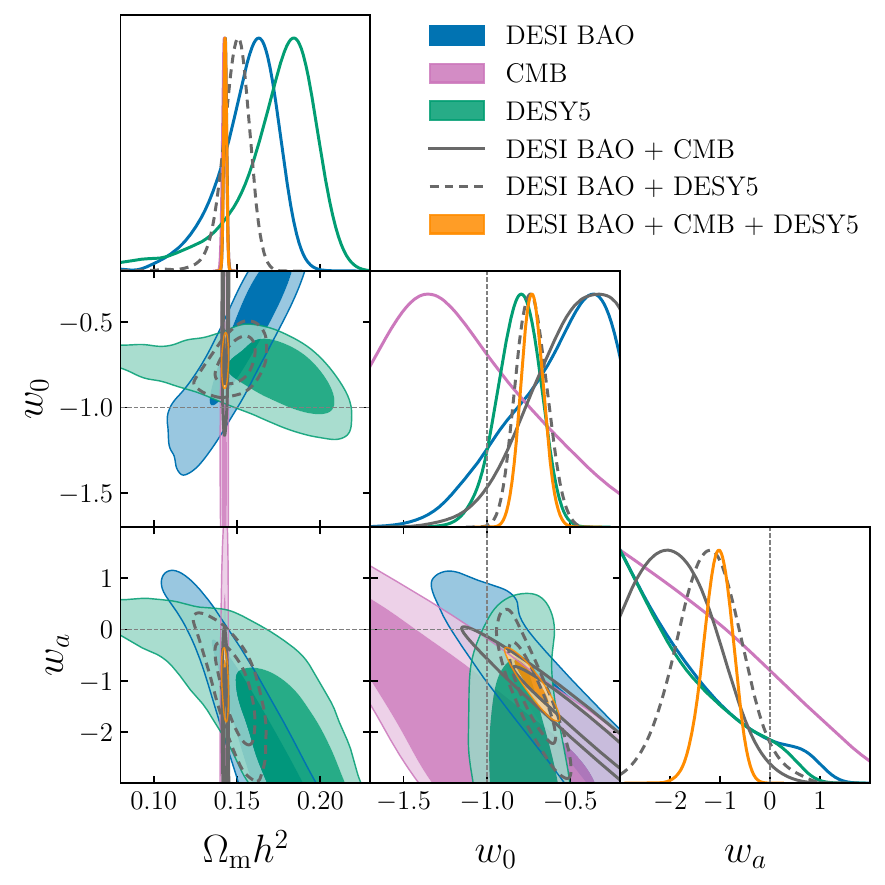} 
    \caption{Marginalised 68\% and 95\% posterior constraints on $\Om h^2$, $w_0$ and $w_a$ in the flat \wowacdm\ model from DESI BAO, CMB and DESY5 SN data individually and in different combinations, showing the roles of each observational probe in breaking degeneracies and contributing to the final joint constraints. Note that BAO and SN data are sensitive only to $\Om$, so the $\Om h^2$ constraints from these datasets shown here arise from this $\Om$ measurement combined with the $H_0\sim\mathcal{U}[20, 100]\kmsMpc$ prior from \cref{tab:priors}.}
    \label{fig:w0wa_triangle}
\end{figure}

SN~Ia data alone allow for $w_a<0$ and, as shown in the left panel of \cref{fig:w0wa_1}, in combination with DESI BAO they also marginally favor $w_0>-1$ although the statistical significance of this preference depends on the particular SN~Ia dataset and is not overwhelming in any case. In order to break the degeneracy in the $w_0$--$w_a$ plane it is necessary to look at the joint constraints from the combination of DESI, CMB and SN~Ia probes, shown in the right panel of \cref{fig:w0wa_1}. These constraints are now not prior-dominated in either parameter. We find that the results and the associated uncertainties again vary depending on the choice of supernova dataset, particularly when comparing PantheonPlus to the other two. However, in all cases, the results are compatible with each other. We find marginalised posterior means
\twoonesig[1.5cm]
{w_0 &= -0.827\pm 0.063,}
{w_a &= -0.75^{+0.29}_{-0.25},}
{DESI+CMB\dataplus PantheonPlus, \label{eq:DESI_CMB_Pantheon_w0waCDM}}
from combination with PantheonPlus,
\twoonesig[1.5cm]
{w_0 &= -0.64\pm 0.11, }
{w_a &= -1.27^{+0.40}_{-0.34},}
{DESI+CMB\dataplus Union3, \label{eq:DESI_CMB_Union3_w0waCDM}}
with Union3 SN~Ia, and 
\twoonesig[1.5cm]
{w_0 &= -0.727\pm 0.067, }
{w_a &= -1.05^{+0.31}_{-0.27},}
{DESI+CMB\dataplus DESY5, \label{eq:DESI_CMB_DESY5_w0waCDM}}
when using the DESY5 SN~Ia data. A better physical understanding of why the combined datasets result in these constraints can be obtained from examining the overlap of the individual posteriors in the expanded parameter space $(\Om h^2,w_0,w_a)$, as shown in \cref{fig:w0wa_triangle} for the example case of the DESY5 SN~Ia. In the \wowacdm\ model, CMB data is insufficient to measure either $\Om$ or $H_0$ individually, but still provides a tight constraint on $\Om h^2$. DESI BAO and SN both measure $\Om$ but not $H_0$: their constraints on $\Om h^2$ are therefore obtained from a combination of this measurement of $\Om$ and the prior on $H_0$. The region of overlap of all three probes is thus determined by the relative degeneracy directions between $\Om$, $w_0$ and $w_a$ obtained from BAO and SN, and the precise constraint on $\Om h^2$ from CMB data.

The $\Delta \chi_\mathrm{MAP}^2$ values between the maximum a posteriori of the \wowacdm\ model and the maximum of the posterior fixing $(w_{0}, w_{a}) = (-1, 0)$ are $-8.7$, $-15.2$ and $-18.1$ for the combinations of DESI and CMB with PantheonPlus, Union3 and DESY5 respectively. These correspond to preferences for a \wowacdm\ model over \lcdm\ model at the $2.5\sigma$, $3.5\sigma$ and $3.9\sigma$ significance levels, respectively.\footnote{Based on several independent MAP estimates, the uncertainty on these significance levels is of order $0.2 \sigma$.} A Bayesian model-selection analysis gives Bayes factors of $|\ln B_{21}|=0.65$, $2.4$, and $2.8$ in favor of the \wowacdm\ model over \lcdm\ for the combinations of DESI+CMB with PantheonPlus, Union3 and DESY5, respectively. On Jeffreys’ scale \cite{Jeffreys:1939xee,Trotta:2005ar}, these indicate a weak preference for \wowacdm\ over \lcdm\ by the first of these data combinations, and moderate preference by the latter two.

For this analysis, we ran additional nested-sampling chains using the \texttt{PolyChord} sampler \cite{Handley:2015fda} from which we derived the Bayesian evidence and Bayes factors using the \texttt{anesthetic} software \cite{Handley:2019mfs}. We have verified that the posteriors derived from these auxiliary chains match those derived from the main MCMC chains described in \cref{sec:inference}, and that the Bayes factors reported here provide consistent interpretations with those estimated from the main MCMC chains through the Savage-Dickey density ratio approximation \cite{Trotta:2005ar}. Not unexpectedly given the current constraints, we find the exact values of the Bayesian evidence (hence the Bayes factors) depend on the specific prior ranges for $w_0$ and $w_a$ (see \cref{tab:priors}). The Deviance Information Criterion (DIC) for model selection \cite{Liddle:2007fy} indicates similar preferences and in the same order as above: $\Delta(\mathrm{DIC}) = \mathrm{DIC}_{w_0w_a\mathrm{CDM}} -\mathrm{DIC}_{\Lambda\mathrm{CDM}}$ of $-5.5$, $-11.0$ and $-14.1$ for the same combinations, respectively (see Table~1 of \cite{Grandis:2016fwl} for a DIC reference scale and \cite{kass1995bayes} 
 for a more general discussion of its empirical thresholds).

Further useful information in the \wowacdm\ analysis is given by the pivot redshift, $\zpiv$, and the pivot equation of state, $\wpiv\equiv w(\zpiv)$, which inform us about the redshift and equation-of-state value at which $w(z)$ is best constrained by a given data set or combination of data sets, see e.g. \cite{Huterer:2000mj,Albrecht:2006um}. We use here the formalism and definitions of \cite{Albrecht:2006um} and find\footnote{Note that when going from the $(w_0, w_a)$ basis to $(\wpiv, w_a)$, the results in the parameter $w_a$ do not change.}
\begin{equation}
(\zpiv, \wpiv) = 
\left \{  
\begin{array}{cl}
(0.57, -1.094\pm 0.070)  & \quad\mbox{(DESI+CMB)} \\[0.2cm]
(0.26, -0.982\pm 0.028)  & \quad\mbox{(DESI+CMB+PantheonPlus)} \\[0.2cm]
(0.33, -0.960\pm 0.038)  & \quad\mbox{(DESI+CMB+Union3)} \\[0.2cm]
(0.26, -0.946\pm 0.026)  & \quad\mbox{(DESI+CMB+DESY5).} 
\end{array} 
\right .
\end{equation}

The results obtained here indicate a preference for an effective dark energy equation of state apparently in the phantom regime (i.e. $w(z)<-1$) at $z>1$, a conclusion that is also borne out by supporting analyses \cite{DESI:2024aqx,DESI:2024kob} which do not use the $(w_0, w_a)$ parametrisation.

 However, we caution against over-interpretation of this result, since at least some of the $\left(w_{0}, w_{a}\right)$ models preferred by our analysis could be mapped to quintessence models which reproduce the observable quantities $D_{M} / r_{d}$ and $D_{H} / r_{d}$ to high precision even though they do not violate the null energy condition at any redshift (e.g., see \cite{dePutter2008,Shlivko:2024}), and such an effective equation of state can also be obtained from models with modified gravity or multiple fields. It remains to be explored whether such a mapping towards models that avoid the phantom regime exists within the full range of $\left(w_{0}, w_{a}\right)$ values preferred by our analysis.
For this reason we opt not to impose any further priors on the $(w_0, w_a)$ parameter space to restrict $w(z)\geq-1$ and to keep our priors as model nonrestrictive as possible. 

\begin{figure}
\centering
    \includegraphics[width = 0.8\columnwidth]{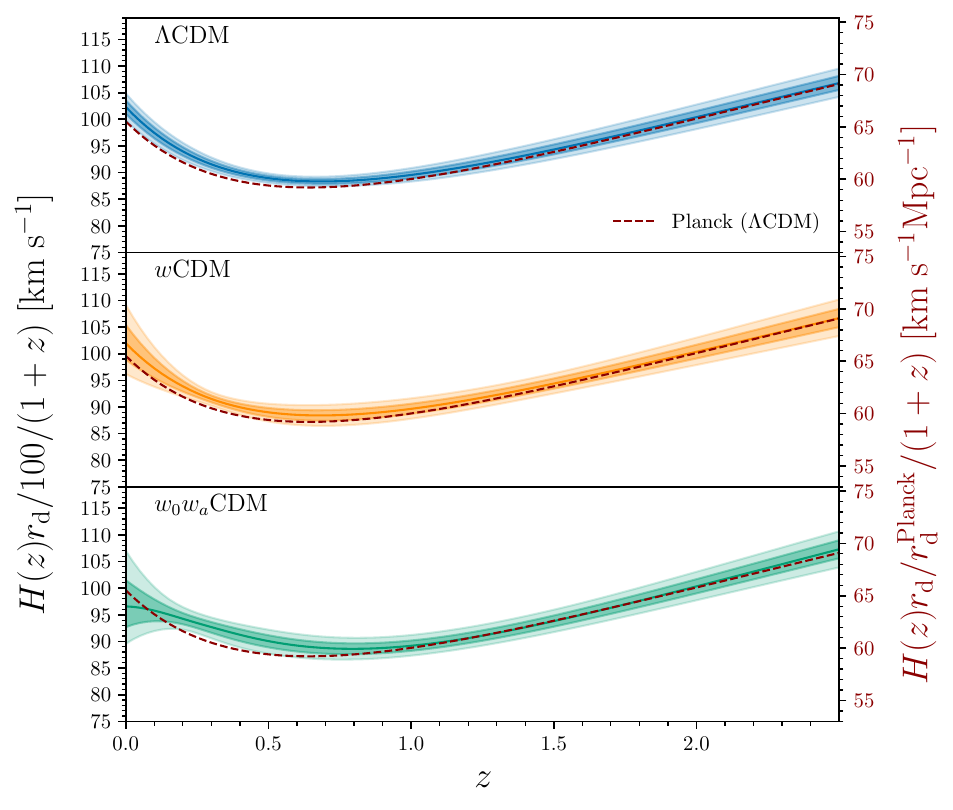} 
    \caption{Model-dependent constraints on the redshift dependence of the Hubble parameter times the sound horizon, $H(z)\rd\equiv r_\mathrm{d}/D_\mathrm{H}$ (scaled by $1/100/(1+z)$ for visual clarity) in three different classes of models fit to all DESI data from \cref{fig:D_vs_z}. From top to bottom, the panels show the 68\% and 95\% credible regions in the \lcdm, \wcdm\ and \wowacdm\ models respectively in the coloured bands. The dashed line in each panel shows the behaviour in the best fit \Planck\ \lcdm\ model. For convenience, the scale on the right-hand axis shows $(\rd/\rd^{\rm Planck}) H(z)/(1+z)$. 
    }
\label{fig:Hz_over_1plusz}
\end{figure}

\Cref{fig:Hz_over_1plusz} provides a summary of the constraints on the expansion history from the flat cosmological models that we have explored. Here, we show the quantity $r_\mathrm{d}/D_\mathrm{H}=H(z)\rd$ as a function of redshift $z$, rescaled for convenience by $1+z$. The top, middle, and bottom panels respectively show derived constraints on the temporal evolution of the Hubble parameter in \lcdm, \wcdm\ and \wowacdm\ models from fitting to DESI data, with the shaded regions indicating the 68\% and 95\% credible regions.
The same dashed line in all panels indicates the best fit \Planck\ \lcdm\ model.
Because $H(z)/(1+z)$ is directly proportional to $\dot a$, i.e.\ the expansion rate, its slope with redshift is directly proportional to minus the cosmic expansion acceleration, $-\ddot a$. The range of the variation of $H(z)$ shows that in each of the three cosmological models, a period of accelerated expansion (i.e. a negative slope) is required at $z\lesssim0.7$. \cref{fig:D_vs_z_with_w0wa} in \cref{sec:appendix_DESI+SDSS} also shows how individual redshift bins contribute to the model fits.

\subsection{$w_0w_a$CDM model with free spatial curvature}
\label{sec:kw0wa}

\begin{figure}
    \centering
    \includegraphics[width = \columnwidth]{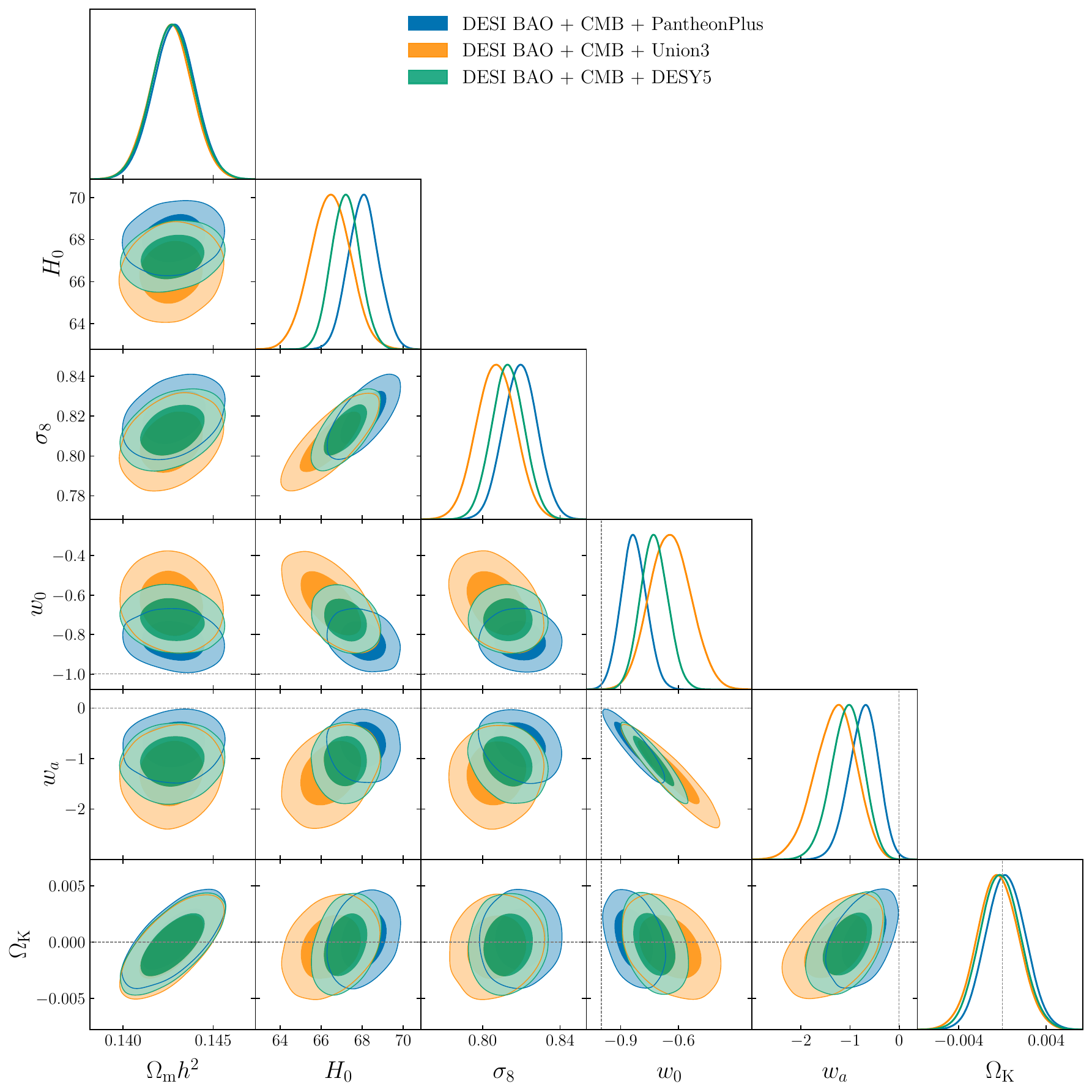}
    \caption{Marginalised posteriors on $w_0$, $w_a$ and $\Ok$ in a model with a time-varying dark energy equation of state and free spatial curvature, from DESI and CMB data combined with SN~Ia from PantheonPlus, Union3 and DESY5 in blue, orange and green respectively. We also show covariances with the physical matter density $\Om h^2$,  Hubble constant $H_0$ (in $\kmsMpc$) , and the amplitude of mass fluctuations $\sigma_8$. All combinations provide tight limits on $\Ok$. Constraints on $w_0$ and $w_a$ in each case broaden a little compared to those shown in the flat case (\cref{fig:w0wa_1}) but the overall trend remains the same.}
    \label{fig:kw0waCDM}
\end{figure}

Finally, we also explore the most general case of a time-varying equation of state $w(a)$ together with allowing free spatial curvature, in the so-called ``$\Ok w_0w_a$CDM model". Even in this broad parameter space, the combination of DESI, \Planck\ and SN~Ia is able to provide quite tight constraints. These are shown in \cref{fig:kw0waCDM} for each of the PantheonPlus, Union3 and DESY5 SN~Ia datasets. As seen before, the resultant central values and uncertainties vary depending on the choice of SN~Ia data, but all three are still broadly consistent with each other, and all combinations place very tight constraints on the allowed range of deviations from spatial flatness. We find
\threeonesig[1.5cm]
{w_0 &= -0.831\pm 0.066,}
{w_a &= -0.73^{+0.32}_{-0.28},}
{\Ok &= 0.0003\pm 0.0018,}{DESI+CMB\dataplus PantheonPlus, \label{eq:kw0wa_DESI_CMB_Pantheon}}
\threeonesig[1.5cm]
{w_0 &= -0.64\pm 0.11,}
{w_a &= -1.30^{+0.45}_{-0.39},}
{\Ok &= -0.0004\pm 0.0019,}{DESI+CMB\dataplus Union3, \label{eq:kw0wa_DESI_CMB_Union3}}
and
\threeonesig[1.5cm]
{w_0 &= -0.725\pm 0.071,}
{w_a &= -1.06^{+0.35}_{-0.31},}
{\Ok &= -0.0002\pm 0.0019,}{DESI+CMB\dataplus DESY5. \label{eq:kw0wa_DESI_CMB_DESY5}}

In comparison to the results for the flat \wowacdm\ model in \cref{sec:w0wa}, the addition of an extra degree of freedom in $\Ok$ leads to a slight broadening of the constraints in the $w_0$--$w_a$ plane, thus marginally reducing the significance of the tension with \lcdm. Nevertheless, the trend observed remains the same, with all combinations of DESI + CMB + SN~Ia preferring $w_0>-1$ and $w_a<0$.

\section{Hubble constant}
\label{sec:Hubble}

The determination of the Hubble constant has been contentious for many decades \cite{Hubble29,deVaucouleurs79,Sandage82}. By the turn of the century, a consensus value of around $70\,\kmsMpc$ had emerged \cite{Freedman01}.
However, since the first \Planck\ results \cite{Planck-2013-cosmology}, a growing tension has emerged between $H_0$ determinations based on physics of the early universe, which tend to cluster close to the \Planck\ preferred value of $67\,\kmsMpc$ (e.g., \cite{Planck-2018-cosmology,Alam-eBOSS:2021}), and local distance-ladder measurements based on Cepheids or other anchors, which mostly prefer larger values around $73\,\kmsMpc$ (e.g., \cite{Riess:2016,Riess:2018,Pesce:2020,Riess:2021jrx,Breuval:2024}). Although distance ladder measurements based on the Tip of the Red Giant Branch (TRGB) method \cite{CCHP} prefer a lower $H_0$ than Cepheid-based calibrations, they currently have larger uncertainties, hence do not yet provide a conclusive assessment.
This tension between the CMB and Cepheid-based determinations stands at the $\sim4$--$5\sigma$ level in \lcdm\ and, if not due to some unidentified residual systematics, may indicate deficiencies in the standard cosmological model. For reviews of the $H_0$ tension, see \cite{VTR19,DivalentinoHtension,RiessKamionH0}.

\begin{figure}
    {\includegraphics[width = 0.8\textwidth]{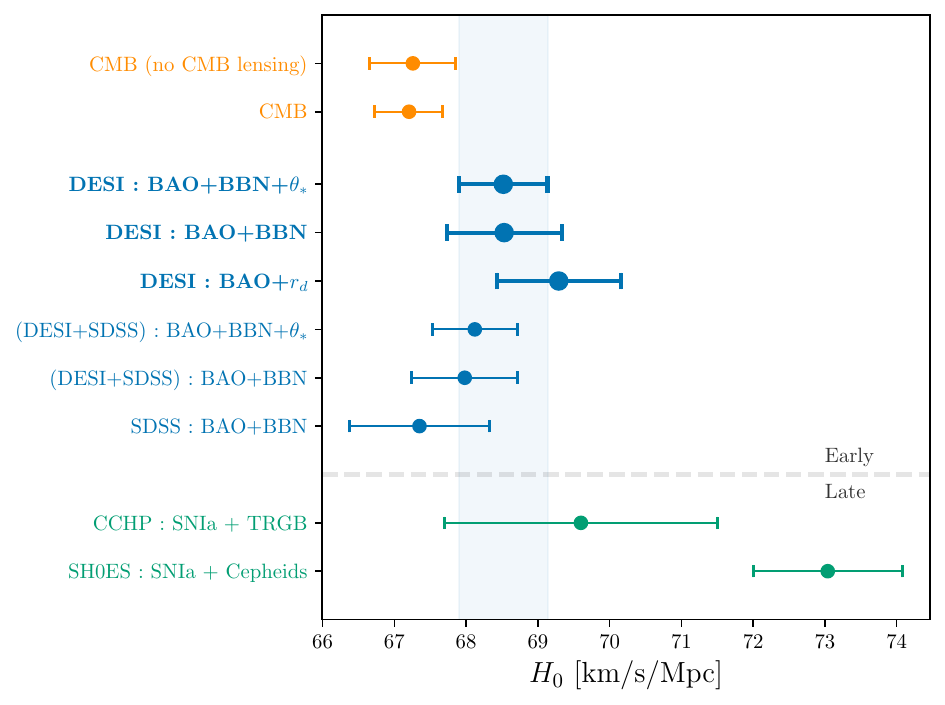}} 
    \caption{
        $68\%$ credible-interval constraints on the Hubble constant, assuming the flat \lcdm\ model. The blue, bold whiskers show DESI BAO measurements in combination with an external BBN prior on $\Ob h^2$ and measurement of the acoustic angular scale $\theta_\ast$, the BBN prior alone, or with the CMB measurement of the sound horizon, $r_{\rm d}^{Planck}$. The thin blue whiskers show the corresponding results from SDSS BAO and the combination of DESI+SDSS BAO results, as labelled.
        The orange whiskers show the results from CMB anisotropy measurements from Planck and ACT, while the green whiskers show measurements of $H_0$ from the distance ladder with either Cepheids or TRGB. }
\label{fig:H0}
\end{figure}

\begin{figure}
    \centering 
    \includegraphics[width=0.43\columnwidth]{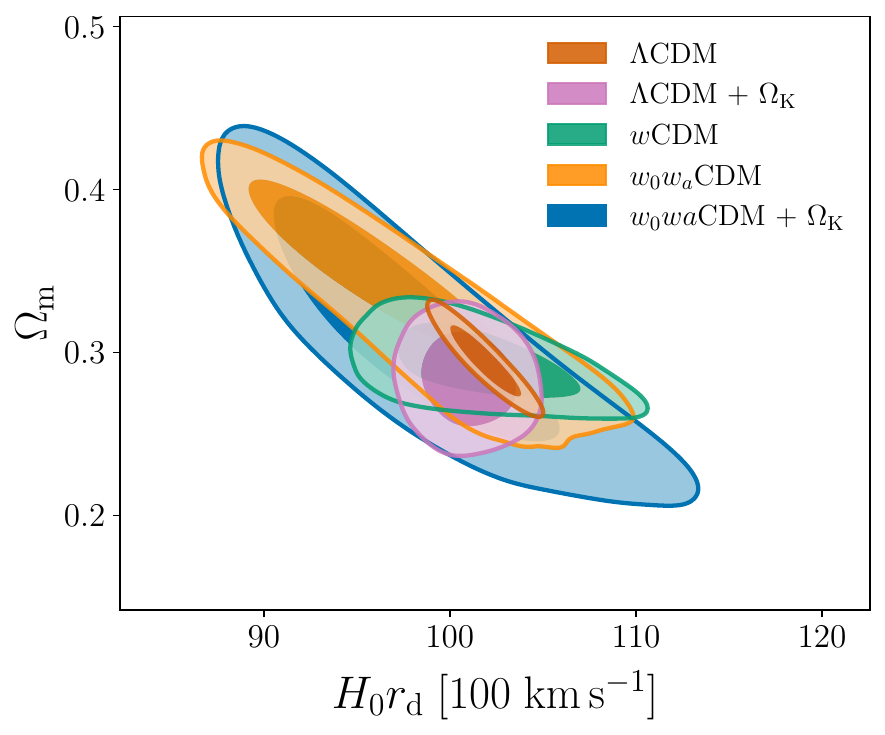}
    \includegraphics[width=0.49\columnwidth]{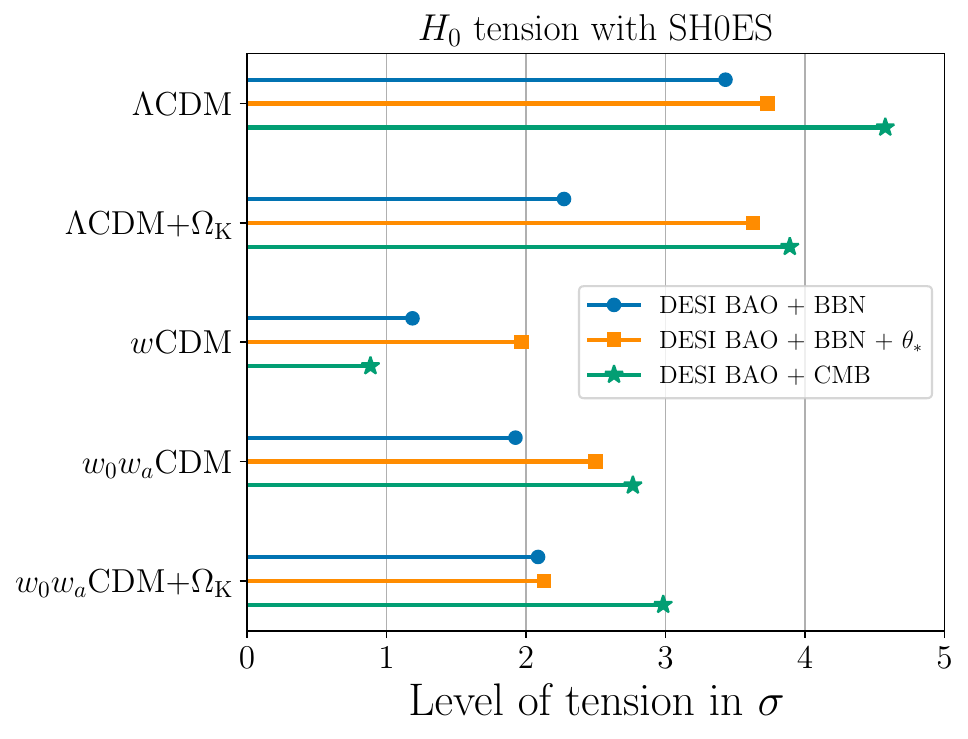}
    \caption{\emph{Left panel}: The 68\% and 95\% credible-interval contours for $\Om$ and $\Hrd$ obtained from fitting DESI DR1 BAO data in the base flat \lcdm\ model and in four extension models which modify the background geometry or late-time expansion history. \emph{Right panel}: A summary of the tension in the $H_0$ measurements obtained from the DESI BAO results combined with other data, and the SH0ES result of \cite{Riess:2021jrx}, assuming different cosmological models.
    }
    \label{fig:rdh_models}
\end{figure}

As discussed in \cref{sec:BBN,sec:flat-lcdm}, BAO alone cannot provide an absolute distance measurement, but can constrain $H_0$ when combined with some external information capable of calibrating the sound horizon scale and breaking the $\rd$--$h$ degeneracy. Within the flat \lcdm\ cosmological model, calibration of the DESI DR1 BAO with primordial deuterium abundances and BBN and the acoustic angular scale measurement leads to very precise $H_0$ determinations summarised in \cref{eq:H0_DESI-bbn,eq:H0_DESI-bbn-thetastar}, the tightest of which is $H_0=68.52\pm 0.62\;\kmsMpc$. As shown in \cref{fig:H0}, while this DESI+BBN+$\theta_\ast$ constraint is
slightly higher than the \Planck\ CMB value, it remains in tension (at the $3.7\sigma$ level) with the SH0ES Cepheid-based distance-ladder result $H_0=73.04\pm 1.04\;\kmsMpc$ (see \cref{fig:rdh_models}).

When the assumption of the flat \lcdm\ model is relaxed, the $H_0$ constraints obtained when adding DESI BAO measurements (in combination with external priors) are loosened too. There are two qualitatively different ways in which the more general cosmological models allow more freedom and lead to weaker $H_0$ constraints. 

First, the calibration of the sound horizon using BBN relies on assumptions about the physics at the time of BBN (as well as on the CMB temperature $T_0$, which however is measured very accurately \cite{2009ApJ...707..916F}). In particular, if the effective number of relativistic degrees of freedom $\Neff$ is allowed to vary from its value $\Neff=3.044$ in the base \lcdm\ model, the value of $\ob$ inferred from light element abundances is altered. When accounting for this in the $\ob$ prior as discussed in \cref{sec:BBN} using the results of \cite{Schoeneberg:2024}, but otherwise keeping the late-time geometry unchanged, the $H_0$ result from \cref{eq:H0_DESI-bbn-thetastar} relaxes slightly to
\begin{equation}
\label{eq:H0_DESI-bbn-thetastar_Neff}
H_0= (68.5\pm 1.4)\;\kmsMpc
\qquad
(\mbox{DESI\,+\,BBN\,+$\theta_\ast$, free~$\Neff$}),    
\end{equation}
a $2\%$ measurement. Keeping the BBN prior but dropping the $\theta_\ast$ information changes the direction of the contours in the 2D $(\Om,H_0)$ parameter space but the marginalised 1D constraint remains very similar, $H_0= (68.5\pm 1.5)\;\kmsMpc$. When adding the full CMB information to DESI BAO, this result slightly shifts to 
\begin{equation}
\label{eq:H0_DESI_CMB}
H_0=(68.3\pm1.1)\;\kmsMpc
\qquad
(\mbox{DESI~BAO+CMB,~free~$\Neff$}).    
\end{equation}

The second type of additional freedom that leads to weakened $H_0$ constraints is allowed by beyond-\lcdm{} models which change the geometry or the late-time expansion of the universe. This directly affects the BAO measurement of $\rd h$, thus degrading the precision with which $H_0$ can be determined, even when the early-universe physics remains the same. The left panel of \cref{fig:rdh_models} shows the joint constraint in the $\Omega_m$- $\rd h$ plane as the amount of freedom in the parametrisation of the expansion history is increased (for the models considered in the previous section). Clearly, the uncertainty on $\rd h$ increases dramatically as more model freedom is allowed. Nevertheless, the central values do not move much, and the results remain consistent with each other across models: we find $\rd h=(101.9\pm 1.3)\Mpc$ (\lcdm), $\rd h=(101.0\pm 1.6)\Mpc$ (\lcdm$+\Ok$), $\rd h=(101.7^{+2.9}_{-3.5})\Mpc$ (\wcdm), $\rd h=(96.4^{+3.3}_{-5.3})\Mpc$ (\wowacdm) and $\rd h=(98.2^{+3.9}_{-6.1})\Mpc$ (\wowacdm$+\Ok$). As a consequence, the $H_0$ values determined from combining DESI BAO with BBN or BBN+$\theta_\ast$ priors also remain stable or slightly decrease, as shown in \cref{tab:parameter_table1}. This conclusion remains true in most cases when combining DESI BAO with the CMB, or with the CMB and SN~Ia information (the one exception is in the \wcdm\ model, where the CMB pulls to higher $H_0$ in the absence of SN~Ia), i.e., independent of the late-time modifications studied, BAO data in combination with other probes always prefer low central values of $H_0$. However, as the $H_0$ uncertainties increase in these models, the level of tension with SH0ES still decreases, as summarised in the right panel of \cref{fig:rdh_models}. Further results on the value of $H_0$ in models which combine extensions to the background expansion history with changes to the neutrino sector are listed in \cref{tab:parameter_table2}.

\section{Neutrinos}
\label{sec:neutrinos}

A generic prediction of the hot Big Bang model is a relic neutrino background which leaves detectable imprints on cosmological observations. Neutrinos are the only known particles to behave as radiation in the early universe and as dark matter at late times, so they affect both the acoustic oscillations in the primordial plasma as well as the background evolution and structure formation.
Cosmological observations are sensitive to both the number of neutrino species and their total mass (e.g., \cite{DESI2016a.Science}), making cosmology constraints complementary to terrestrial neutrino experiments. 

\subsection{Sum of neutrino masses}
\label{sec:mnu}

\begin{figure}
    \centering
    \includegraphics[width=0.48\columnwidth]{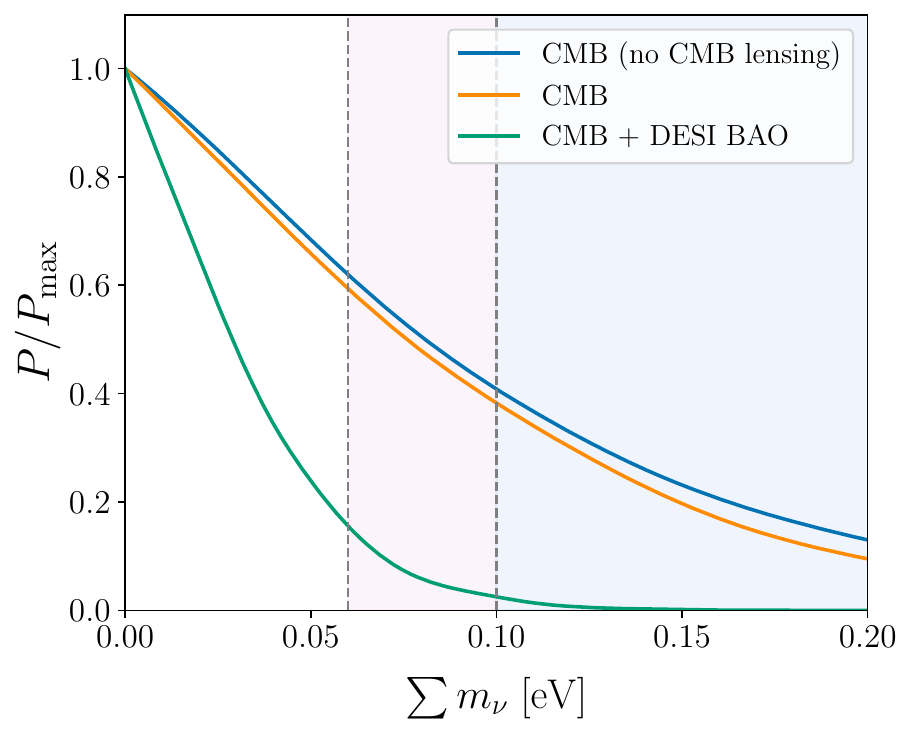}
    \includegraphics[width=0.465\columnwidth]{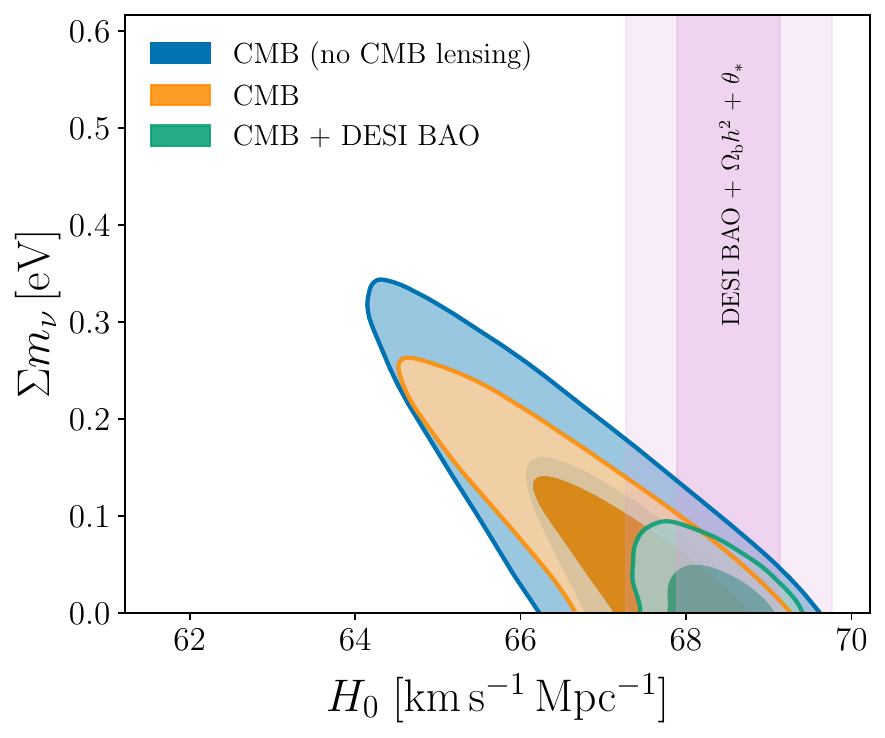}
     \caption{\emph{Left panel}: The marginalised 1D posterior constraints on $\sumnu$ from different combinations of datasets, in the single parameter extension flat \lcdm+$\sumnu$ model. As explained in the text, here we use a model with 3 degenerate mass eigenstates and with a minimal prior $\sumnu>0$ eV. The minimal masses for the normal or inverted mass ordering scenarios correspond to 
     $\sumnu>0.059$ 
     eV and $\sumnu>0.10$ eV respectively, shown by the vertical dashed lines and the shaded regions.
    \emph{Right panel}: Joint marginalised 68\% and 95\% credible intervals on $\sumnu$ and $H_0$ from \Planck, CMB and DESI+CMB data, illustrating the degeneracy between these parameters from the CMB, and how DESI BAO data contribute to breaking it. The vertical shaded regions indicate the 68\% and 95\% constraints on $H_0$ from DESI BAO data combined with knowledge of $\theta_\ast$ and $\Ob h^2$ in the \lcdm+$\sumnu$ model.
    This shows how DESI BAO breaks the primarily geometric degeneracy to place an upper limit on $\sumnu$.}
    \label{fig:mnu}
\end{figure}

\begin{table}
    \centering
    \small
    \begin{tabular}{lcccc}
    \toprule
    \midrule
    model / dataset & $\Omega_{\mathrm{m}}$ & $H_0$ [$\kmsMpc$] & $\Sigma m_\nu\,[\mathrm{eV}]$ & $N_{\mathrm{eff}}$\\
    \midrule
    $\mathbf{\Lambda}${\bf CDM+}$\boldsymbol{\sumnu}$ &&&&\\
    DESI+CMB & $0.3037\pm 0.0053$ & $68.27\pm 0.42$ & $< 0.072$ &---\\  
    \midrule
    $\mathbf{\Lambda}${\bf CDM+}$\boldsymbol{\Neff}$ &&&&\\
    DESI+CMB & $0.3058\pm 0.0060$ & $68.3\pm 1.1$ &---& $3.10\pm 0.17$  \\
    \midrule
    $\boldsymbol{w}${\bf CDM+}$\boldsymbol{\sumnu}$ &&&&\\
    DESI+CMB & $0.282\pm 0.013$ & $71.1^{+1.5}_{-1.8}$ & $< 0.123$ &---\\
    DESI+CMB+Panth. & $0.3081\pm 0.0067$ & $67.81\pm 0.69$ & $< 0.079$ &---\\
    DESI+CMB+Union3 & $0.3090\pm 0.0082$ & $67.72\pm 0.88$ & $< 0.078$ &---\\
    DESI+CMB+DESY5 & $0.3152\pm 0.0065$ & $67.01\pm 0.64$ & $< 0.073$ &---\\
    \midrule
    $\boldsymbol{w}${\bf CDM+}$\boldsymbol{\Neff}$ &&&&\\
    DESI+CMB & $0.281\pm 0.013$ & $71.0^{+1.6}_{-1.8}$ &---& $2.97\pm 0.18$ \\
    DESI+CMB+Panth. & $0.3090\pm 0.0068$ & $67.9\pm 1.1$ &---& $3.07\pm 0.18$ \\
    DESI+CMB+Union3 & $0.3097\pm 0.0084$ & $67.8\pm 1.2$ &---& $3.06\pm 0.18$ \\
    DESI+CMB+DESY5 & $0.3163\pm 0.0067$ & $67.2\pm 1.1$ &---& $3.09\pm 0.18$ \\
    \midrule
    $\boldsymbol{w_0w_a}${\bf CDM+}$\boldsymbol{\sumnu}$ &&&&\\
    DESI+CMB & $0.344^{+0.032}_{-0.026}$ & $64.7^{+2.1}_{-3.2}$ & $< 0.195$ \\
    DESI+CMB+Panth. & $0.3081\pm 0.0069$ & $68.07\pm 0.72$ & $< 0.155$ &--- \\
    DESI+CMB+Union3 & $0.3240\pm 0.0098$ & $66.48\pm 0.94$ & $< 0.185$ &---\\
    DESI+CMB+DESY5 & $0.3165\pm 0.0069$ & $67.22\pm 0.66$ & $< 0.177$ &---\\
    \midrule
    $\boldsymbol{w_0w_a}${\bf CDM+}$\boldsymbol{\Neff}$ &&&&\\
    DESI+CMB & $0.346^{+0.032}_{-0.026}$ & $63.9^{+2.2}_{-3.3}$ &---& $2.89\pm 0.17$ \\
    DESI+CMB+Panth. & $0.3093\pm 0.0069$ & $67.5\pm 1.1$ &---& $2.93\pm 0.18$ \\
    DESI+CMB+Union3 & $0.3245\pm 0.0098$ & $65.9\pm 1.3$ &---& $2.91\pm 0.18$ \\
    DESI+CMB+DESY5 & $0.3172\pm 0.0067$ & $66.6\pm 1.1$ &---& $2.92\pm 0.18$ \\
    \midrule
    \bottomrule
    \end{tabular}
    \caption{Cosmological parameter results from DESI DR1 BAO data in combination with external datasets when considering extensions to the baseline \lcdm\ model in the neutrino sector. Results with two-sided error bars refer to the marginalised means and 68\% credible intervals; upper bounds on $\sumnu$ refer to the 95\% limits. Note that the label ``CMB" includes CMB lensing from the combined \planckact\ likelihood. All constraints on $\sumnu$ assume a model with 3 degenerate mass eigenstates and a minimal prior $\sumnu>0$ eV.  (See \cref{eq:mnu_NH} and \cref{eq:mnu_IH} for the results using other priors.) The empty $\sumnu$ and $\Neff$ fields indicate that the fixed respective values of $\sumnu=0.06\eV$ and $\Neff=3.044$ were adopted.
    \label{tab:parameter_table2}}
\end{table}

The base model we have adopted so far assumes the sum of neutrino masses to be $\sumnu=0.06$ eV, with a single massive eigenstate and two massless ones. This is motivated by the lower bound for $\sumnu$ from neutrino oscillation experiments. In this section, we consider a single-parameter extension beyond this minimal model in which $\sumnu$ is allowed to freely vary, in order to explore the constraining power on $\sumnu$ of DESI data. Amongst terrestrial experiments directly measuring the neutrino mass, KATRIN \cite{KATRIN:2005} has produced the tightest constraints to date, from measuring the endpoint of the tritium $\beta$-decay spectrum. This gives an upper bound on the effective electron-neutrino mass that is independent of the cosmological model of $m_\beta < 0.8\eV$ (90\% CL) \cite{KATRIN:2021}, equivalent to $\sumnu \lesssim 2.4\eV$ (90\% CL).

Neutrino oscillation experiments have also shown that at least two of the three active neutrino masses are non-zero, but the ordering of these masses is not known.
In the normal ordering or normal hierarchy (NH), the two lowest mass neutrino eigenstates have the smallest mass splitting, implying that the total neutrino mass must be $\sumnu \geq 0.059\eV$, while in the inverted hierarchy (IH), however, the smallest mass splitting occurs between the two highest mass eigenstates, necessitating a total mass of $\sumnu \geq 0.10\eV$ \cite{NuFit}.
When allowing $\sumnu$ to be a free parameter, we adopt a model with three degenerate neutrino mass eigenstates. This degenerate mass model does not exactly correspond to either of the physically expected NH or IH scenarios; however, it produces a very good approximation of the observable effects of both \cite{Lesgourgues:2006}. In the event that a positive detection of non-zero neutrino mass is possible, an analysis using the degenerate mass model with a prior $\sumnu>0$ will recover the correct value of $\sumnu$ for both NH and IH scenarios with little reconstruction bias \cite{CORE:2018}. On the other hand, when a positive detection is not possible, using the degenerate model with appropriately modified priors on $\sumnu$ as above will also recover the correct upper bounds for both the NH and IH scenarios \cite{Choudhury:2020}.

The sum of the neutrino masses affects cosmology in two ways. First, the high-velocity dispersion of neutrinos implies that they free-stream over large distances, thus suppressing the late-time clustering and power spectrum of matter at small scales, below the free streaming length. However, if $\sumnu$ is varied with other cosmological parameters along the CMB degeneracy, the net effect of neutrino masses is an overall almost-scale-independent suppression of the amplitude of the matter power spectrum (together with a very small shift in the BAO scale). Second, at the low redshifts of relevance to DESI, massive neutrinos are non-relativistic ($\sumnu\gg T\simeq 10^{-3}\eV$) and therefore contribute to the total non-relativistic matter density $\omega_m=\omega_b+\omega_c+\omega_\nu$, where $\Omega_\nu=\sumnu/(93.14\eV h^2)$, see e.g. \cite{Lesgourgues:2006}. 
$\sumnu$ thus affects the background evolution and in particular the redshift of matter-$\Lambda$ equality. Forthcoming DESI analyses of the full shape of the galaxy power spectrum are sensitive to neutrino masses through the first effect, and constraints from them will be described in \cite{DESI2024.V.KP5,DESI2024.VII.KP7B}. The BAO data used in this paper only distinguish the background geometry. However, BAO constraints on the expansion history, when combined with the CMB and CMB lensing, are very helpful in improving neutrino mass constraints. This is because the CMB is sensitive to the neutrino mass via the angular diameter distance to recombination and via the effects of $\sumnu$ on the CMB lensing, and both of these effects can be mimicked in the CMB by varying other cosmological parameters such as $H_0$ and $\om$. 

Thus while BAO data are not directly sensitive to the suppression effects of neutrinos on the power spectrum, by determining the background geometry at low redshifts they help to break the CMB degeneracy and constrain $H_0$, thus greatly tightening the upper limit on $\sumnu$.
This effect is illustrated in the right panel of \cref{fig:mnu}.
In extensions of flat \lcdm{}, the power of BAO to pin down the late-time expansion history diminishes, resulting in notably weaker $\sumnu$ constraints.

For the Planck CMB alone, and assuming flat \lcdm, the constraints are
\onetwosig[3cm]{\sumnu < 0.21\eV}{CMB}{, \label{eq:mnu_CMB}}
when including CMB lensing from the combination of \Planck\ and ACT, while adding the DESI BAO data sharply reduces this to 
\onetwosig[4cm]{\sumnu < 0.072\eV}{DESI~BAO+CMB}{. \label{eq:mnu_DESI+ext}} 
The left panel of \cref{fig:mnu} shows the corresponding 1D marginalised posteriors. The posterior peaks at $\sumnu=0\eV$, which is excluded by terrestrial oscillation experiments, but the minimal mass $\sumnu=0.059\eV$ is not excluded by the cosmological fits. We find a $\Delta \chi_\mathrm{MAP}^{2}$ of $3.8$ for the scenario with $\sumnu=0.059\eV$ compared to $\sumnu=0\eV$ from DESI BAO and CMB. 

The result in \cref{eq:mnu_DESI+ext} represents a substantial tightening of the upper bound compared to previous state-of-the-art results \cite{2021PhRvD.103h3533A,PalanqueDelabrouille:2020,DiValentino:2021,Brieden:2022,2022PTEP.2022h3C01W,Madhavacheril:ACT-DR6}, of which the previous tightest reported upper bound was $\sumnu<0.082\eV$ \cite{Brieden:2022}. As can be seen from the right panel of \cref{fig:mnu},the improvement relative to CMB-only constraints is driven primarily by the tighter constraints on $H_0$ (and consequently $\Om$) obtained from BAO. The shift relative to the previous tightest upper bounds is instead a consequence of both the preference of DESI data for slightly higher $\rd h$ and thus higher $H_0$ (which is itself coupled to lower $\Om$), together with the improved lensing constraint from including the latest ACT (DR6) lensing data. Should the addition of more data in the future pull towards lower $H_0$, we may expect the neutrino mass limit to relax, even if the data achieve a higher precision. For the same reason, as discussed in \cref{sec:appendix_DESI+SDSS}, the upper bound obtained using the combined BAO data from DESI and SDSS (\cref{sec:DESI_SDSS_consistency}) instead of DESI alone is also slightly looser, $\sumnu<0.082\eV$, due to the slightly lower $\rd h$ and $H_0$ values seen in \cref{fig:Om_H0_rd,fig:H0}.

We also caution that the upper bound obtained is strongly dependent on the choice of prior for $\sumnu$. While we have deliberately chosen to use $\sumnu>0\eV$ for our primary analysis, one can also impose physically motivated priors corresponding specifically to the NH ($\sumnu>0.059\eV$) and IH ($\sumnu>0.1\eV$) scenarios. Applying these alternative priors we find the following upper limit for NH: 
\onetwosig[4cm]{\sumnu < 0.113\eV}{DESI~BAO+CMB; $\sumnu>0.059\eV$}{, \label{eq:mnu_NH}} 
and for IH: 
\onetwosig[4cm]{\sumnu < 0.145\eV}{DESI~BAO+CMB; $\sumnu>0.10$~eV}{. \label{eq:mnu_IH}}
 {Cosmological results alone do not yet strongly favour the NH over the IH: through comparing the best fits to the data obtained with $\sumnu=0.059$ eV and $\sumnu=0.10$ eV, we find $\Delta\chi^2_\mathrm{MAP}\simeq2$ in favour of the NH.}

It is possible to combine the cosmology result for $\sumnu$ with constraints from oscillation and $\beta$-decay experiments to compute the Bayesian evidence in favour of one hierarchy over the other, although the results depend strongly on the choice of prior for the individual masses. A full calculation is beyond the scope of this paper, but even with a minimally informative objective Bayesian prior the cosmological constraints we find---assuming the flat \lcdm\ background cosmology---when combined with terrestrial data give the Bayes factor for the normal hierarchy over the inverted one above 100 \cite{2018JCAP...04..047H,2022JCAP...09..006J,2022JCAP...10..010G}. 

However, we repeat the caveat that the limits are substantially relaxed in more extended dark energy models that affect the background geometry, such as those in \cref{sec:DE}. \cref{tab:parameter_table2} shows the corresponding upper bounds for $\sumnu$ when allowing for a \wcdm\ or \wowacdm\ background; as expected, these are significantly weaker: for instance, in a \wowacdm\ background, the upper limit from DESI BAO and CMB relaxes to $\sumnu<0.195$ eV (for the $\sumnu>0$ eV prior). Importantly though, while allowing a time-varying DE equation of state parameter significantly loosens the upper bound on $\sumnu$, allowing $\sumnu$ to vary freely only marginally affects the constraints on $w_0$ and $w_a$ reported in \cref{sec:w0wa} and therefore does not substantially change the conclusions of that section. For example, when allowing $\sumnu$ to vary freely, we find 
\twoonesig[1.5cm]
{w_0 &= -0.725^{+0.068}_{-0.076},}
{w_a &=  -1.07^{+0.39}_{-0.29},}
{DESI+CMB+DESY5, free~$\sumnu$}
which can be compared with \cref{eq:DESI_CMB_DESY5_w0waCDM}. Combinations with the other two SN~Ia datasets are similarly close to the results in \cref{eq:DESI_CMB_Pantheon_w0waCDM,eq:DESI_CMB_Union3_w0waCDM} obtained with fixed $\sumnu=0.06$ eV.

\subsection{Number of effective relativistic species}
\label{sec:Neff}

Under the standard assumption that three active neutrinos thermalise in the early universe, additional dark relativistic degrees of freedom can be parametrised in terms of a change to the effective number of neutrino species, $\Neff$. This is defined such that the total relativistic energy density, after the annihilation of electrons with positrons, is given by
\be
\label{eq:Neff_def}
\rho_\nu=\Neff\,\frac{7}{8}\left(\frac{4}{11}\right)^{4/3}\rho_\gamma,
\ee
where $\rho_\gamma$ is the photon energy density.
In the standard cosmological model with three massive species of neutrinos and no other particles that are relativistic at recombination (other than photons), $\Neff=3.044$ \cite{Froustey:2020,Bennett:2021}. Extended models with light sterile neutrinos or other dark relics (such as ``dark radiation") that are generated well before recombination produce effects very similar to that of active neutrinos and so can be usefully explored in terms of constraints on $\Neff$ in a \lcdm+$\Neff$ model.

Similar to the case of $\sumnu$, constraints on $\Neff$ from the CMB alone exhibit a geometrical degeneracy because changing the relativistic energy density before recombination shifts the sound horizon, and this effect can be absorbed through changes to $H_0$ (or $\Om$), such that increasing $\Neff$ corresponds to higher $H_0$ or lower $\Om$. Therefore the DESI BAO measurements at lower redshifts again contribute by breaking the geometric degeneracy through their ability to constrain $\Om$.

In particular, while the combination of CMB anisotropies and lensing power spectra from \Planck\ and ACT give \oneonesig[2.cm]{\Neff=2.98\pm0.20}{CMB}{, \label{eq:nnu_CMB}} the addition of DESI BAO changes this to \oneonesig[3.5cm]{\Neff=3.10\pm0.17}{CMB+DESI~BAO}{, \label{eq:nnu_CMB_DESI}} i.e., a small shift of the central value due to the DESI preference for lower $\Om$, and a $\sim15\%$ reduction in the uncertainty. As shown in \cref{tab:parameter_table2}, despite the additional freedom allowed in the neutrino sector, the recovered value of $H_0$ remains compatible with that from \Planck. This remains true even when allowing the temporal variation of the dark energy equation of state.

Note that changes in $\Neff$ can also produce damping of the BAO amplitude and shifts to the scale and phase of the BAO oscillations due to the neutrino dragging effect \cite{2017JCAP...11..007B,2019NatPh..15..465B}. In principle these effects could introduce biases in the BAO measurements, if the template used for the fitting is drawn from a cosmological model that does not contain the same physical effects. The implications of this for BAO measurements are studied in \cite{KP4s9-Perez-Fernandez}, where it is found that even for large values $\Neff\simeq3.70$ that are excluded at high significance by the joint constraints above, the systematic offset in the recovered BAO scale $\DVrd$ is, at most, $\sim0.1\%$.

\section{Conclusions}
\label{sec:conclusions}

In this paper we have presented the first cosmological results from DESI, the first Stage-IV galaxy survey in operation, marking the start of a new era of dark energy experiments. These results are based on samples of bright galaxies, LRGs, ELGs, quasars and \lya\ forest tracers in the redshift range $0.1<z<4.2$. These data include a total of over 6 million unique extragalactic spectroscopic redshifts from the first year of DESI observations alone (out of the full five-year survey program), already representing an increase of more than a factor of two over the number of tracers used in the previous largest such dataset assembled \cite{Alam-eBOSS:2021}, which was the culmination of two decades of observations with SDSS. This remarkable achievement is made possible by the speed and quality of the DESI instrument \cite{DESI2022.KP1.Instr}, whose increased spectral resolution also delivers significantly improved redshift accuracy.

The results in this paper are based on the measurement of the BAO scale using different DESI tracers of the matter density (galaxies, quasars, and the Lyman-$\alpha$ forest) in seven redshift bins, presented in detail in \cite{DESI2024.III.KP4,DESI2024.IV.KP6}. The observed data were analysed using a state-of-the-art blind analysis pipeline to protect against confirmation bias and included end-to-end validation. This methodology enabled us to unleash the potential of the BAO method to, in conjunction with other probes, powerfully constrain cosmological parameters.

Having demonstrated the internal consistency of the DESI BAO measurements over the full redshift range of observations and shown that the results are in agreement with previous measurements from SDSS, we proceeded to examine the constraints and implications of these data for a range of cosmological models. 

In the standard flat \lcdm\ cosmology, we determine the matter density parameter $\Om =  0.295\pm 0.015$, and the product of the drag-epoch sound horizon and the scaled Hubble constant of $\rd h = (101.8\pm 1.3)\Mpc$. These values are slightly different from those measured from the combination of CMB anisotropies from \Planck\ plus CMB lensing from \Planck\ and ACT, giving a somewhat lower value of $\Om$ and a higher $\rd h$, although the discrepancy is not statistically significant. In combination with a conservative prior on the baryon abundance from BBN, DESI BAO data determine the Hubble constant value to be $H_0=(68.53\pm0.80)\;\kmsMpc$, the most precise measurement to date that does not rely on information from CMB anisotropies. Combining the conservative BBN prior with an equally conservative prior on the extremely precise and model-independent measurement of the acoustic angular scale $\theta_\ast$ gives $H_0=(68.52\pm0.62)\;\kmsMpc$, approaching the precision from \Planck.   
Each of these two $H_0$ constraints is in a $>3\sigma$ tension with the SH0ES Cepheid-based distance-ladder result.
In combination with the full CMB information, DESI results give $H_0=(67.97\pm 0.38)\;\kmsMpc$---a $0.6\%$ precision measurement.

BAO distance measurements are particularly important in constraining model extensions to \lcdm, where they help break geometric degeneracies that limit the power of the CMB. We have examined several models of dark energy, allowing spatial curvature to vary. Together with CMB information, DESI BAO data provide extremely tight limits on the spatial curvature in all scenarios, with $\Ok=-0.0024\pm 0.0016$ in the simplest single-parameter extension of \lcdm\ and almost equally tight constraints in more extended model variations.

The measurement of the dark energy equation of state $w$ is a key science goal of DESI. Assuming the \wcdm\ model where the equation-of-state parameter is constant in time, 
we find $w=-0.99^{+0.15}_{-0.13}$ from DESI alone, and $w=-0.997\pm0.025$ from the combination of DESI BAO, CMB, and SN~Ia results from the Pantheon+ compilation, in good consistency with \lcdm. This result does not appreciably change when the Pantheon+ data are replaced by those from other recent SN~Ia releases from Union3 and the Dark Energy Survey (DES-SN5YR). 

However, when the equation of state is allowed to vary with time, $w(a)=w_0 + (1-a)w_a$, DESI data favour solutions with $w_0>-1$ and $w_a<0$. The combination of DESI and CMB gives $w_0=-0.45^{+0.34}_{-0.21}$ and $w_a=-1.79^{+0.48}_{-1.00}$, and indicates a $\sim2.2\sigma$ difference to \lcdm. When adding information from SN~Ia, all combinations prefer $w_0>-1$ and $w_a<0$, with the level of the tension with \lcdm\ remaining at the $\sim2.5\sigma$ level for combining DESI and CMB information with Pantheon+, but increasing to $3.5\sigma$ and $3.9\sigma$ levels for the Union3 and DESY5SN  SN~Ia datasets respectively. All three of the primary BAO, CMB and SN~Ia probes contribute partially to this tension and the results including the 3 SN~Ia datasets are mutually consistent with each other. Moreover, combining any two of the DESI BAO, CMB or SN data sets shows some level of departure from the $\Lambda$CDM model. Relaxing the assumption of a spatially flat geometry through varying $\Ok$ marginally increases the uncertainties but does not change the overall picture. It remains important to thoroughly examine unaccounted-for sources of systematic uncertainties or inconsistencies between the different datasets that might be contributing to these results. Nevertheless, these findings provide a tantalizing suggestion of deviations from the standard cosmological model that motivate continued study and highlight the potential of DESI and other Stage-IV surveys to pin down the nature of dark energy.

Neutrinos are the only particles of the Standard Model of particle physics whose mass parameters are unknown. DESI, and Stage-IV surveys more generally, will improve cosmological constraints on neutrino mass parameters and provide key insights into their mass hierarchy.
Allowing the sum of the neutrino masses to vary, the combination of DESI BAO and CMB information breaks a geometric degeneracy (in CMB constraints) between the Hubble constant $H_0$ and the amplitude of matter fluctuations and thus places an extremely tight upper bound on the total mass, $\sumnu<0.072\eV$ (95\% CL). We however caution that this substantial tightening of the upper bound compared to the previous state-of-the-art measurements is partly driven by the preference of DESI data for a higher value of $H_0$. Moreover, the upper bound obtained depends on the priors chosen for $\sumnu$ and, because DESI contributes through breaking a geometrical degeneracy, the upper limit is relaxed in extended models that alter the background geometry. We examined this point explicitly in the paper by considering changes to the upper bound on $\sumnu$ in models with a varying dark energy equation of state. Finally, we have also reported an updated constraint $\Neff=3.10\pm0.17$ on the effective number of extra relativistic degrees of freedom from the combination of DESI BAO and CMB data, and showed that this constraint only marginally shifts to $\Neff=2.89\pm0.17$ even when varying the dark energy equation of state.

As the first set of cosmological results from DESI, this paper, together with the accompanying results in \cite{DESI2024.III.KP4, DESI2024.IV.KP6}, demonstrates the enormous power of the DESI instrument and survey. Subsequent papers will examine the implications of measurement of the full clustering broadband shape of DESI \cite{DESI2024.V.KP5, DESI2024.VII.KP7B} and constraints on primordial non-Gaussianity \cite{DESI2024.VIII.KP7C}. With data quality and the speed of survey completion continuing to match or exceed expectations, future data releases will soon be able to provide even better insights into the hints of the exciting cosmological findings presented here.

\section{Data Availability}
The data used in this analysis will be made public along the Data Release 1 (details in \url{https://data.desi.lbl.gov/doc/releases/}).

\acknowledgments

This material is based upon work supported by the U.S.\ Department of Energy (DOE), Office of Science, Office of High-Energy Physics, under Contract No.\ DE–AC02–05CH11231, and by the National Energy Research Scientific Computing Center, a DOE Office of Science User Facility under the same contract. Additional support for DESI was provided by the U.S. National Science Foundation (NSF), Division of Astronomical Sciences under Contract No.\ AST-0950945 to the NSF National Optical-Infrared Astronomy Research Laboratory; the Science and Technology Facilities Council of the United Kingdom; the Gordon and Betty Moore Foundation; the Heising-Simons Foundation; the French Alternative Energies and Atomic Energy Commission (CEA); the National Council of Humanities, Science and Technology of Mexico (CONAHCYT); the Ministry of Science and Innovation of Spain (MICINN), and by the DESI Member Institutions: \url{https://www.desi. lbl.gov/collaborating-institutions}. 

The DESI Legacy Imaging Surveys consist of three individual and complementary projects: the Dark Energy Camera Legacy Survey (DECaLS), the Beijing-Arizona Sky Survey (BASS), and the Mayall z-band Legacy Survey (MzLS). DECaLS, BASS and MzLS together include data obtained, respectively, at the Blanco telescope, Cerro Tololo Inter-American Observatory, NSF NOIRLab; the Bok telescope, Steward Observatory, University of Arizona; and the Mayall telescope, Kitt Peak National Observatory, NOIRLab. NOIRLab is operated by the Association of Universities for Research in Astronomy (AURA) under a cooperative agreement with the National Science Foundation. Pipeline processing and analyses of the data were supported by NOIRLab and the Lawrence Berkeley National Laboratory. Legacy Surveys also uses data products from the Near-Earth Object Wide-field Infrared Survey Explorer (NEOWISE), a project of the Jet Propulsion Laboratory/California Institute of Technology, funded by the National Aeronautics and Space Administration. Legacy Surveys was supported by: the Director, Office of Science, Office of High Energy Physics of the U.S. Department of Energy; the National Energy Research Scientific Computing Center, a DOE Office of Science User Facility; the U.S. National Science Foundation, Division of Astronomical Sciences; the National Astronomical Observatories of China, the Chinese Academy of Sciences and the Chinese National Natural Science Foundation. LBNL is managed by the Regents of the University of California under contract to the U.S. Department of Energy. The complete acknowledgments can be found at \url{https://www.legacysurvey.org/}.

Any opinions, findings, and conclusions or recommendations expressed in this material are those of the author(s) and do not necessarily reflect the views of the U.S.\ National Science Foundation, the U.S.\ Department of Energy, or any of the listed funding agencies.

The authors are honored to be permitted to conduct scientific research on Iolkam Du’ag (Kitt Peak), a mountain with particular significance to the Tohono O’odham Nation.



\bibliographystyle{JHEP}
\bibliography{refs_key_paper,DESI2024_update17Jun,extra_references}


\appendix

\section{Comparison of results from DESI and combined DESI+SDSS BAO data}
\label{sec:appendix_DESI+SDSS}

As discussed in \cref{sec:DESI_SDSS_consistency}, it is possible to define a combined ``DESI+SDSS" BAO dataset across the redshift range covered by both surveys which uses: 
1) SDSS results at $z_{\rm eff}=0.15, 0.38$ and $0.51$ in place of the DESI BGS and lowest-redshift LRG points;
2) DESI results from LRGs in $0.6<z<0.8$, the combination of LRGs and ELGs in $0.8<z<1.1$, and ELGs and QSOs at higher redshifts; and
3) the combined DESI+SDSS result from \cref{eq:DESI+SDSSLya1,eq:DESI+SDSSLya2} for the \lya\ BAO. The combined DESI+SDSS dataset therefore 
leverages observations from SDSS and DESI in the redshift regimes where each survey respectively covers a larger effective volume, and combines the \lya\ BAO measurements from both. We emphasise again that unlike the DESI DR1 BAO, this combination of measurements does not come from uniform analysis methods or data processing pipelines; nevertheless, they are of interest as, by maximising the effective volume at each redshift, they provide the most precise combination of BAO data currently available. 

In this section we compare the results using this combination to the DESI ones in the main body of the paper, in the cases of most interest.
We note again that DESI BAO measurements are visually compared to those from SDSS in Figure~15 of Ref.~\cite{DESI2024.III.KP4}.

We start with the flat \lcdm\ model. DESI+SDSS constraints alone give
\twoonesig[1.5cm]
{\Om &= 0.297\pm 0.012,}
{\rd h &= (101.3\pm 1.1) \Mpc,}
{(DESI+SDSS).
\label{eq:bestBAO_LCDM_results}} 
This is in excellent agreement with the DESI-only results in \cref{eq:DESI_LCDM_results}, with about 20\% smaller errors. Combining BAO information with BBN, we obtain
\begin{equation} 
\label{eq:H0_bestBAO-bbn}
H_0= (67.98\pm 0.75)\kmsMpc
\qquad
(\mbox{(DESI+SDSS)~BAO+BBN}), 
\end{equation}
and further adding the acoustic angular scale constraint,
\begin{equation} 
\label{eq:H0_bestBAO-bbn-thetastar}
H_0= (68.13\pm 0.59)\kmsMpc
\qquad
(\mbox{(DESI+SDSS)+BBN+$\theta_\ast$}), 
\end{equation}
both of which are in excellent agreement with the corresponding DESI results (\cref{eq:H0_DESI-bbn,eq:H0_DESI-bbn-thetastar}), and with those from the CMB. These results are also shown in \cref{fig:H0}. The combination with the full CMB data gives
\twoonesig[3.5cm]
{\Om &= 0.309\pm 0.0048,}
{H_0 &= (67.80\pm 0.37) \kmsMpc}
{(DESI+SDSS)~BAO+CMB. \label{eq:DESI-SDSS+CMB_LCDM_results}}
Finally, the 95\% upper limit on the neutrino mass in flat \lcdm\ for the combination of this BAO data with the CMB, and with the $\sumnu>0$ eV prior is
\onetwosig[6cm]{\sumnu < 0.082\eV}{(DESI+SDSS)~BAO+CMB}{. \label{eq:mnu_bestBAO+ext}} 
This is slightly weaker than the upper bound in \cref{eq:mnu_DESI+ext}: as discussed in \cref{sec:mnu}, it is because the BAO precision in both cases is close enough that the primary determinant of the limit obtained on $\sumnu$ is actually the \emph{central value} of $H_0$, a lower $H_0$ allowing a larger $\sumnu$. As can be seen from the results quoted above, the DESI+SDSS combination favours very slightly lower values of $H_0$.

\begin{figure}
    \centering
    \includegraphics[width = 0.6\columnwidth]{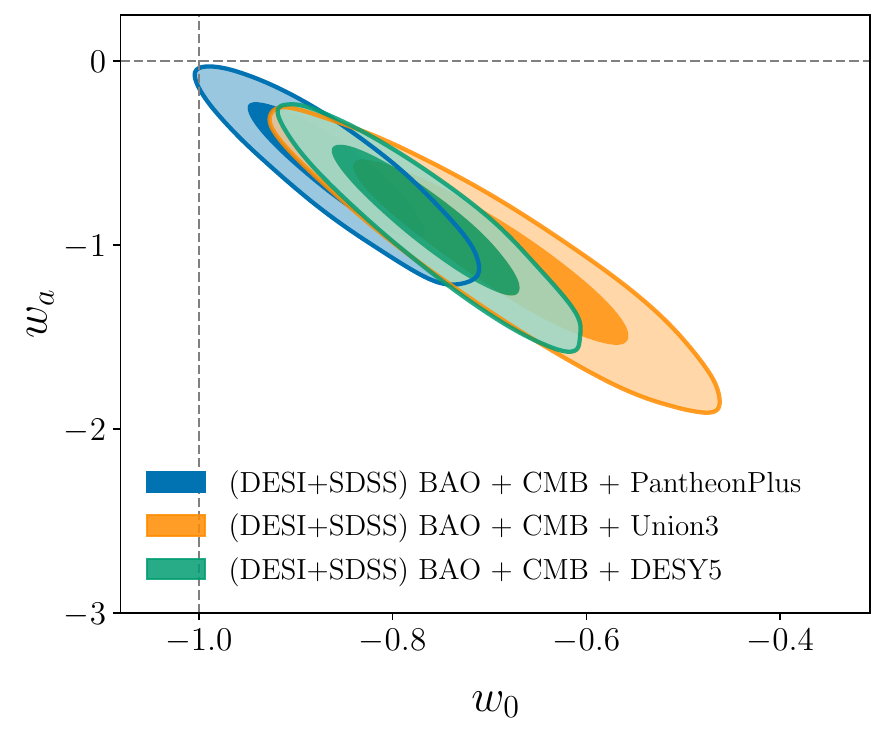}
    \caption{The 68\% and 95\% marginalised posterior constraints from DESI+SDSS BAO combined with CMB and each of the PantheonPlus, Union3 and DESY5 SN~Ia datasets. Compared to cases with DESI DR1 BAO shown in the right panel of \cref{fig:w0wa_1} all contours are very consistent but the significances of the tensions with \lcdm\ marginally decrease.}
    \label{fig:w0wa_appendix}
\end{figure}

In terms of extensions to the base \lcdm\ model, the case of most interest is naturally the $(w_0,w_a)$ parametrisation of the dark energy equation of state parameter. \cref{fig:w0wa_appendix} shows the marginalised posterior constraints in $w_0$ and $w_a$ for the same combinations with CMB and SN~Ia datasets as in the right panel of \cref{fig:w0wa_1}, but replacing the DESI DR1 BAO data with the DESI+SDSS equivalents. We find
\twoonesig[1.5cm]
{w_0 &= -0.855\pm 0.060,}
{w_a &= -0.60^{+0.26}_{-0.23},}
{(DESI+SDSS)+CMB\dataplus PantheonPlus, \label{eq:bestBAO_CMB_Pantheon_w0waCDM}}
from combination with PantheonPlus,
\twoonesig[1.5cm]
{w_0 &= -0.692\pm 0.095, }
{w_a &= -1.06^{+0.36}_{-0.31},}
{(DESI+SDSS)+CMB\dataplus Union3, \label{eq:bestBAO_CMB_Union3_w0waCDM}}
with Union3 SN~Ia, and 
\twoonesig[1.5cm]
{w_0 &= -0.761\pm 0.064, }
{w_a &= -0.88^{+0.29}_{-0.25},}
{(DESI+SDSS)+CMB\dataplus DESY5, \label{eq:DESIbestBAO_CMB_DESY5_w0waCDM}}
when using the DESY5 SN~Ia data, all in excellent agreement with the DESI results provided in the main text. All these results shift marginally closer to the \lcdm\ expectation: the $\Delta \chi_\mathrm{MAP}^{2}$ values between the maximum a posteriori of the \wowacdm\ model and the maximum of the posterior fixing $(w_{0}, w_{a}) = (-1, 0)$ are $-7.0$, $-12.6$ and $-15.5$ for the combinations with PantheonPlus, Union3 and DESY5 respectively, corresponding to discrepancies at the $2.2\sigma$, $3.1\sigma$ and $3.5\sigma$ significance levels. These are all comparable to, though slightly lower than, the equivalent $2.5\sigma$, $3.5\sigma$ and $3.9\sigma$ discrepancies reported earlier using DESI DR1 BAO.

\begin{figure}[t]
    \centering
    \includegraphics[width = \columnwidth]{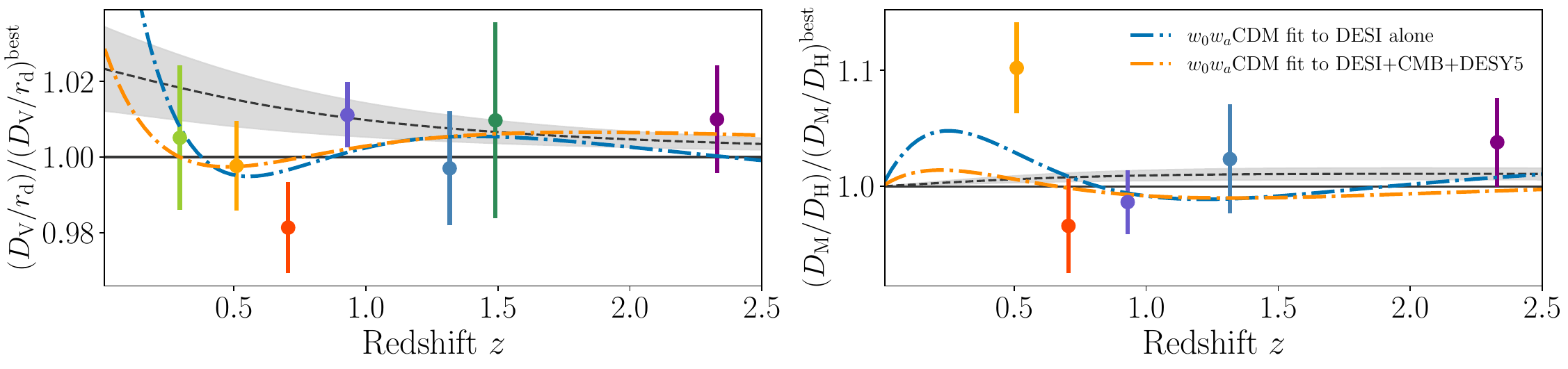}
    \caption{Same as the lower panels of Fig.~\ref{fig:D_vs_z}, except with model lines for two \wowacdm\ models shown. The blue dot-dashed line shows the best-fit \wowacdm\ model from DESI data alone, while the orange line shows the best fit from DESI+CMB+DESY5. As before, the solid line (at value of unity) is the best-fit \lcdm\ model to DESI data alone, while the dashed line and the grey shaded region showing the \Planck\ best-fit \lcdm\ model and the corresponding 68\% uncertainties.
    }
    \label{fig:D_vs_z_with_w0wa}
\end{figure}

Since replacing all the DESI points at $z<0.6$ with SDSS does not change the qualitative conclusions of tension with \lcdm, this indicates that---despite being a $\sim2\sigma$ offset from the \lcdm\ model---the single DESI BAO measurement of $\DM/D_\mathrm{H}(z=0.51)$ from LRGs is \emph{not} significantly driving the final dark energy results in \cref{sec:w0wa}. 

A better understanding of why this is the case can be obtained from \cref{fig:D_vs_z_with_w0wa}, which shows the DESI BAO measurements of $\DVrd$ and $\DM/D_\mathrm{H}$ relative to the best-fit \lcdm\ model, and overlays the model predictions for two representative \wowacdm\ models. The $z_\mathrm{eff}=0.51$ DESI BAO measurement is anomalous in the less well-measured quantity $\DM/D_\mathrm{H}$ (at the $\lesssim2\sigma$ level), although perfectly consistent with \lcdm\ in $\DVrd$. The blue dash-dot lines show the model which has the minimum $\chi^2_\mathrm{MAP}$ for a fit to the DESI BAO alone: this is a rather extreme case, with $\Om=0.385$, $w_0=-0.159$ and $w_a=-3.0$, the $w_a$ value hitting our prior $w_a\in[-3, 2]$ (\cref{tab:priors}). (However, as noted in \cref{sec:w0wa}, although these best-fit parameters are extreme, the level of ``tension" with \lcdm\ actually does not even reach the $2\sigma$ level, with $\Delta\chi^2_\mathrm{MAP}=-3.7$.) The orange dash-dot lines instead show the model predictions for the model with $\Om=0.316$, $w_0=-0.733$ and $w_a=-1.010$, which minimises $\chi^2_\mathrm{MAP}$ for the fit to the combination of DESI BAO+CMB+DESY5. This is the data combination which also gives the highest significance tension with \lcdm, of $3.9\sigma$.

What is clear from \cref{fig:D_vs_z_with_w0wa} is that the slightly anomalous DESI measurement of $\DM/D_\mathrm{H}(z=0.51)$ does not strongly affect the \wowacdm\ fits: even the $(w_0,w_a)$ values at the edge of our prior range fail to fit this data point well, so it contributes very little to the $\Delta\chi^2$ between \lcdm\ and \wowacdm\ even for the fit to DESI BAO alone (in which case the statistical evidence for variation of the equation of state is very low anyway). In the more interesting case of combining BAO, CMB and SN~Ia data the difference between the \lcdm\ and \wowacdm\ model predictions in $\DM/D_\mathrm{H}$ is even smaller, due to the tight constraints from SN~Ia at $z<0.5$.

In conclusion, the DESI result for $\DM/D_\mathrm{H}(z=0.51)$ appears to be simply a mild ($\sim2\sigma$) statistical fluctuation. It does not appear to have a simple explanation in terms of any model, and it consequently also does not strongly drive the results in any model that we studied, especially when DESI data are combined with CMB and SN~Ia. We have explicitly shown that our conclusions about dark energy still hold when all DESI measurements at $z<0.6$ are completely replaced by the corresponding SDSS values.

\section{Comparison to Planck PR4 likelihoods}
\label{sec:appendix_PR4}

\begin{figure}[t]
    \centering
    \includegraphics[width = 0.8\columnwidth]{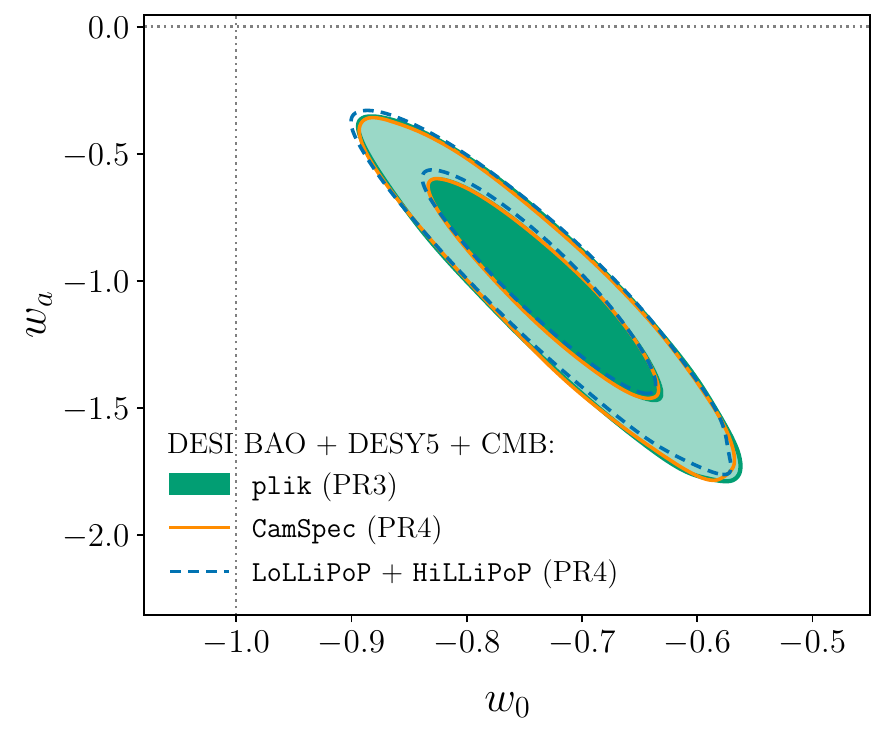}
    \caption{The 68\% and 95\% marginalised posterior constraints on the dark energy equation of state parameters $w_0$ and $w_a$ from the combination of DESI BAO, DESY5 SN Ia and CMB data, comparing three different CMB likelihoods as described in the text. The differences arising from the choice of CMB likelihood are negligible.
    }
    \label{fig:w0wa_PR3_vs_PR4}
\end{figure}

Throughout the main text of this paper, we have used the \texttt{plik} likelihood from the official \Planck\ PR3 data release as the default for the high-$\ell$ CMB temperature and polarisation (TTTEEE) power spectrum likelihood analysis. As mentioned in \cref{sec:CMB}, more recently other likelihoods have been released by teams that make use of updated data from the \Planck\ PR4 release: \texttt{CamSpec} \cite{Efstathiou:2021,Rosenberg:2022} and \texttt{HiLLiPoP} \cite{Tristram:2023}. The \texttt{LoLLiPoP} likelihood \cite{Tristram:2021} provides another alternative likelihood for the $\ell<30$ EE power spectrum, replacing our default choice of \texttt{simall}. The use of these likelihoods leads to small shifts in parameter estimates compared to \texttt{plik}, but in most cases of interest in this paper, after combination with CMB lensing likelihoods from \Planck\ and ACT \cite{Carron:2022,Madhavacheril:ACT-DR6}, these shifts become very small, and negligibly so when our DESI BAO likelihoods are also included. Nevertheless, for completeness we report the shifts resulting from using these alternative likelihoods in this section, focusing on only those cases where the parameter shifts are non-negligible. For conciseness, in this appendix we will refer to shifts in the mean value of a parameter that amount to $x\%$ of the original statistical uncertainty as a shift of ``$x\,\sigma$", but this is not a statement about the level of discrepancy between the results (if any), and performing such a calculation accounting for the overlap in the data used by the different CMB likelihoods is far beyond the scope of this paper.

In the base \lcdm\ model, we find
\oneonesig[4cm]
{H_0 = 67.73\pm 0.36 \, \kmsMpc}
{DESI~BAO+\Planck [\texttt{CamSpec}]\dataplus CMB~lensing,}{\label{eq:LCDM_DESI+CamSpec_results}}
corresponding to a $0.6\sigma$ shift in the mean, and marginal improvement in the uncertainties compared to the result in \cref{tab:parameter_table1} using \texttt{plik}. Although not quantities we have focused on in this paper, the baryon density $\ob$ and matter power spectrum amplitude $\sigma_8$ change slightly as well, from 
\twoonesig[3cm]
{\ob &= 0.02248\pm 0.00013,}
{\sigma_8 &= 0.8135\pm 0.0053,}
{DESI~BAO+\Planck [\texttt{plik}]\dataplus CMB~lensing, \label{eq:Obh2_plik}}
to
\twoonesig[3cm]
{\ob &= 0.02226\pm 0.00012,}
{\sigma_8 &= 0.8103\pm 0.0051,}
{DESI+\Planck [\texttt{CamSpec}]\dataplus CMB~lensing, \label{eq:Obh2_CamSpec}}
a shift of $\sim1.8\sigma$ in $\ob$.
This causes a smaller shift in the sound horizon value, from $\rd=147.34\pm 0.22$ to $\rd=147.59\pm 0.21$ when swapping from \texttt{plik} to \texttt{CamSpec}, although with negligible change to $\rd h$. Changes in $\Om$ are also negligible ($<0.5\sigma$).

Although it has been noted \cite{Rosenberg:2022} that the \texttt{CamSpec} likelihood somewhat alleviates the apparent preference in \texttt{plik} for a non-zero spatial curvature, related to the $A_L$ lensing amplitude parameter anomaly, this preference is in any case also removed by combination with CMB lensing and BAO, and we find the result in \cref{eq:DESI_CMB_OmegaK} is completely unaffected by the change in \Planck\ likelihood.
There is a small shift in the result for $\Neff$ in flat \lcdm,
\oneonesig[3cm]
{\Neff = 3.20\pm0.19,}
{DESI~BAO+\Planck [\texttt{CamSpec}]+ CMB~lensing},{\label{eq:Neff_CamSpec}}
a shift of slightly more than $0.5\sigma$ and a marginal change in the uncertainty compared to \cref{eq:nnu_CMB_DESI}. 

The upper bounds on $\sumnu$ shift slightly relative to our baseline result $\sumnu<0.072\;\eV$ (\cref{eq:mnu_DESI+ext}) depending on the choice of CMB likelihood. In flat \lcdm\ and assuming a prior $\sumnu>0  \; \eV$, replacing the \texttt{plik} likelhood with \texttt{CamSpec} gives the marginally tighter constraint $\sumnu<0.071 \; \eV$ while using \texttt{HiLLiPoP} gives a marginally looser one, $\sumnu<0.074\;\eV$. Replacing both \texttt{simall} and \texttt{plik} with \texttt{LoLLiPoP} and \texttt{HiLLiPoP} respectively gives $\sumnu<0.079\;\eV$.

In models with changes to the dark energy equation of state, we find that changing between \texttt{plik} and the newer PR4 likelihoods has a negligible difference on $w_0$ and $w_a$. A demonstration of this is provided in \cref{fig:w0wa_PR3_vs_PR4}, which shows the posterior constraints in the $w_0$-$w_a$ plane obtained from the combination of DESI BAO, CMB and DESY5 SN Ia data, comparing cases using the default \texttt{plik} likelihood to \texttt{CamSpec} (both cases also using \texttt{simall}) and \texttt{HiLLiPoP} (additionally replacing \texttt{simall} with \texttt{LoLLiPoP}).


\section{Author Affiliations}
\label{sec:affiliations}

\noindent \hangindent=.5cm $^{1}${Instituto de F\'{\i}sica Te\'{o}rica (IFT) UAM/CSIC, Universidad Aut\'{o}noma de Madrid, Cantoblanco, E-28049, Madrid, Spain}

\noindent \hangindent=.5cm $^{2}${Lawrence Berkeley National Laboratory, 1 Cyclotron Road, Berkeley, CA 94720, USA}

\noindent \hangindent=.5cm $^{3}${Physics Dept., Boston University, 590 Commonwealth Avenue, Boston, MA 02215, USA}

\noindent \hangindent=.5cm $^{4}${Tata Institute of Fundamental Research, Homi Bhabha Road, Mumbai 400005, India}

\noindent \hangindent=.5cm $^{5}${Centre for Extragalactic Astronomy, Department of Physics, Durham University, South Road, Durham, DH1 3LE, UK}

\noindent \hangindent=.5cm $^{6}${Institute for Computational Cosmology, Department of Physics, Durham University, South Road, Durham DH1 3LE, UK}

\noindent \hangindent=.5cm $^{7}${Department of Physics, University of Michigan, Ann Arbor, MI 48109, USA}

\noindent \hangindent=.5cm $^{8}${Leinweber Center for Theoretical Physics, University of Michigan, 450 Church Street, Ann Arbor, Michigan 48109-1040, USA}

\noindent \hangindent=.5cm $^{9}${IRFU, CEA, Universit\'{e} Paris-Saclay, F-91191 Gif-sur-Yvette, France}

\noindent \hangindent=.5cm $^{10}${Institut de F\'{i}sica d’Altes Energies (IFAE), The Barcelona Institute of Science and Technology, Campus UAB, 08193 Bellaterra Barcelona, Spain}

\noindent \hangindent=.5cm $^{11}${Instituto Avanzado de Cosmolog\'{\i}a A.~C., San Marcos 11 - Atenas 202. Magdalena Contreras, 10720. Ciudad de M\'{e}xico, M\'{e}xico}

\noindent \hangindent=.5cm $^{12}${Instituto de Ciencias F\'{\i}sicas, Universidad Aut\'onoma de M\'exico, Cuernavaca, Morelos, 62210, (M\'exico)}

\noindent \hangindent=.5cm $^{13}${Universit\'{e} Claude Bernard Lyon 1, CNRS/IN2P3, IP2I, Lyon, France}

\noindent \hangindent=.5cm $^{14}${Physics Department, Yale University, P.O. Box 208120, New Haven, CT 06511, USA}

\noindent \hangindent=.5cm $^{15}${Department of Physics and Astronomy, University of California, Irvine, 92697, USA}

\noindent \hangindent=.5cm $^{16}${Department of Physics, Kansas State University, 116 Cardwell Hall, Manhattan, KS 66506, USA}

\noindent \hangindent=.5cm $^{17}${Department of Physics \& Astronomy, University of Rochester, 206 Bausch and Lomb Hall, P.O. Box 270171, Rochester, NY 14627-0171, USA}

\noindent \hangindent=.5cm $^{18}${Department of Physics, The University of Texas at Dallas, Richardson, TX 75080, USA}

\noindent \hangindent=.5cm $^{19}${Institute for Astronomy, University of Edinburgh, Royal Observatory, Blackford Hill, Edinburgh EH9 3HJ, UK}

\noindent \hangindent=.5cm $^{20}${Dipartimento di Fisica ``Aldo Pontremoli'', Universit\`a degli Studi di Milano, Via Celoria 16, I-20133 Milano, Italy}

\noindent \hangindent=.5cm $^{21}${Centre for Astrophysics \& Supercomputing, Swinburne University of Technology, P.O. Box 218, Hawthorn, VIC 3122, Australia}

\noindent \hangindent=.5cm $^{22}${NSF NOIRLab, 950 N. Cherry Ave., Tucson, AZ 85719, USA}

\noindent \hangindent=.5cm $^{23}${Department of Physics and Astronomy, The University of Utah, 115 South 1400 East, Salt Lake City, UT 84112, USA}

\noindent \hangindent=.5cm $^{24}${Department of Physics \& Astronomy, University College London, Gower Street, London, WC1E 6BT, UK}

\noindent \hangindent=.5cm $^{25}${Department of Astronomy and Astrophysics, University of Chicago, 5640 South Ellis Avenue, Chicago, IL 60637, USA}

\noindent \hangindent=.5cm $^{26}${Fermi National Accelerator Laboratory, PO Box 500, Batavia, IL 60510, USA}

\noindent \hangindent=.5cm $^{27}${Korea Astronomy and Space Science Institute, 776, Daedeokdae-ro, Yuseong-gu, Daejeon 34055, Republic of Korea}

\noindent \hangindent=.5cm $^{28}${Institute of Cosmology and Gravitation, University of Portsmouth, Dennis Sciama Building, Portsmouth, PO1 3FX, UK}

\noindent \hangindent=.5cm $^{29}${Departamento de Astrof\'{\i}sica, Universidad de La Laguna (ULL), E-38206, La Laguna, Tenerife, Spain}

\noindent \hangindent=.5cm $^{30}${Instituto de Astrof\'{\i}sica de Canarias, C/ V\'{\i}a L\'{a}ctea, s/n, E-38205 La Laguna, Tenerife, Spain}

\noindent \hangindent=.5cm $^{31}${Department of Physics and Astronomy, University of Sussex, Brighton BN1 9QH, U.K}

\noindent \hangindent=.5cm $^{32}${Departamento de F\'{i}sica, Instituto Nacional de Investigaciones Nucleares, Carreterra M\'{e}xico-Toluca S/N, La Marquesa,  Ocoyoacac, Edo. de M\'{e}xico C.P. 52750,  M\'{e}xico}

\noindent \hangindent=.5cm $^{33}${Institute for Advanced Study, 1 Einstein Drive, Princeton, NJ 08540, USA}

\noindent \hangindent=.5cm $^{34}${Center for Cosmology and AstroParticle Physics, The Ohio State University, 191 West Woodruff Avenue, Columbus, OH 43210, USA}

\noindent \hangindent=.5cm $^{35}${NASA Einstein Fellow}

\noindent \hangindent=.5cm $^{36}${School of Mathematics and Physics, University of Queensland, 4072, Australia}

\noindent \hangindent=.5cm $^{37}${Instituto de F\'{\i}sica, Universidad Nacional Aut\'{o}noma de M\'{e}xico,  Cd. de M\'{e}xico  C.P. 04510,  M\'{e}xico}

\noindent \hangindent=.5cm $^{38}${CIEMAT, Avenida Complutense 40, E-28040 Madrid, Spain}

\noindent \hangindent=.5cm $^{39}${Department of Physics \& Astronomy and Pittsburgh Particle Physics, Astrophysics, and Cosmology Center (PITT PACC), University of Pittsburgh, 3941 O'Hara Street, Pittsburgh, PA 15260, USA}

\noindent \hangindent=.5cm $^{40}${Department of Astronomy, School of Physics and Astronomy, Shanghai Jiao Tong University, Shanghai 200240, China}

\noindent \hangindent=.5cm $^{41}${Space Sciences Laboratory, University of California, Berkeley, 7 Gauss Way, Berkeley, CA  94720, USA}

\noindent \hangindent=.5cm $^{42}${University of California, Berkeley, 110 Sproul Hall \#5800 Berkeley, CA 94720, USA}

\noindent \hangindent=.5cm $^{43}${Universities Space Research Association, NASA Ames Research Centre}

\noindent \hangindent=.5cm $^{44}${Center for Astrophysics $|$ Harvard \& Smithsonian, 60 Garden Street, Cambridge, MA 02138, USA}

\noindent \hangindent=.5cm $^{45}${Department of Physics, The Ohio State University, 191 West Woodruff Avenue, Columbus, OH 43210, USA}

\noindent \hangindent=.5cm $^{46}${The Ohio State University, Columbus, 43210 OH, USA}

\noindent \hangindent=.5cm $^{47}${Kavli Institute for Particle Astrophysics and Cosmology, Stanford University, Menlo Park, CA 94305, USA}

\noindent \hangindent=.5cm $^{48}${SLAC National Accelerator Laboratory, Menlo Park, CA 94305, USA}

\noindent \hangindent=.5cm $^{49}${Instituto de Astrof\'{i}sica de Andaluc\'{i}a (CSIC), Glorieta de la Astronom\'{i}a, s/n, E-18008 Granada, Spain}

\noindent \hangindent=.5cm $^{50}${Ecole Polytechnique F\'{e}d\'{e}rale de Lausanne, CH-1015 Lausanne, Switzerland}

\noindent \hangindent=.5cm $^{51}${Departamento de F\'isica, Universidad de los Andes, Cra. 1 No. 18A-10, Edificio Ip, CP 111711, Bogot\'a, Colombia}

\noindent \hangindent=.5cm $^{52}${Observatorio Astron\'omico, Universidad de los Andes, Cra. 1 No. 18A-10, Edificio H, CP 111711 Bogot\'a, Colombia}

\noindent \hangindent=.5cm $^{53}${Institut d'Estudis Espacials de Catalunya (IEEC), 08034 Barcelona, Spain}

\noindent \hangindent=.5cm $^{54}${Institute of Space Sciences, ICE-CSIC, Campus UAB, Carrer de Can Magrans s/n, 08913 Bellaterra, Barcelona, Spain}

\noindent \hangindent=.5cm $^{55}${Departament de F\'{\i}sica Qu\`{a}ntica i Astrof\'{\i}sica, Universitat de Barcelona, Mart\'{\i} i Franqu\`{e}s 1, E08028 Barcelona, Spain}

\noindent \hangindent=.5cm $^{56}${Institut de Ci\`encies del Cosmos (ICCUB), Universitat de Barcelona (UB), c. Mart\'i i Franqu\`es, 1, 08028 Barcelona, Spain.}

\noindent \hangindent=.5cm $^{57}${Consejo Nacional de Ciencia y Tecnolog\'{\i}a, Av. Insurgentes Sur 1582. Colonia Cr\'{e}dito Constructor, Del. Benito Ju\'{a}rez C.P. 03940, M\'{e}xico D.F. M\'{e}xico}

\noindent \hangindent=.5cm $^{58}${Departamento de F\'{i}sica, Universidad de Guanajuato - DCI, C.P. 37150, Leon, Guanajuato, M\'{e}xico}

\noindent \hangindent=.5cm $^{59}${Centro de Investigaci\'{o}n Avanzada en F\'{\i}sica Fundamental (CIAFF), Facultad de Ciencias, Universidad Aut\'{o}noma de Madrid, ES-28049 Madrid, Spain}

\noindent \hangindent=.5cm $^{60}${Excellence Cluster ORIGINS, Boltzmannstrasse 2, D-85748 Garching, Germany}

\noindent \hangindent=.5cm $^{61}${University Observatory, Faculty of Physics, Ludwig-Maximilians-Universit\"{a}t, Scheinerstr. 1, 81677 M\"{u}nchen, Germany}

\noindent \hangindent=.5cm $^{62}${Department of Astrophysical Sciences, Princeton University, Princeton NJ 08544, USA}

\noindent \hangindent=.5cm $^{63}${Kavli Institute for Cosmology, University of Cambridge, Madingley Road, Cambridge CB3 0HA, UK}

\noindent \hangindent=.5cm $^{64}${Department of Astronomy, The Ohio State University, 4055 McPherson Laboratory, 140 W 18th Avenue, Columbus, OH 43210, USA}

\noindent \hangindent=.5cm $^{65}${Department of Physics, Southern Methodist University, 3215 Daniel Avenue, Dallas, TX 75275, USA}

\noindent \hangindent=.5cm $^{66}${Department of Physics and Astronomy, University of Waterloo, 200 University Ave W, Waterloo, ON N2L 3G1, Canada}

\noindent \hangindent=.5cm $^{67}${Perimeter Institute for Theoretical Physics, 31 Caroline St. North, Waterloo, ON N2L 2Y5, Canada}

\noindent \hangindent=.5cm $^{68}${Waterloo Centre for Astrophysics, University of Waterloo, 200 University Ave W, Waterloo, ON N2L 3G1, Canada}

\noindent \hangindent=.5cm $^{69}${Graduate Institute of Astrophysics and Department of Physics, National Taiwan University, No. 1, Sec. 4, Roosevelt Rd., Taipei 10617, Taiwan}

\noindent \hangindent=.5cm $^{70}${Sorbonne Universit\'{e}, CNRS/IN2P3, Laboratoire de Physique Nucl\'{e}aire et de Hautes Energies (LPNHE), FR-75005 Paris, France}

\noindent \hangindent=.5cm $^{71}${Department of Astronomy and Astrophysics, UCO/Lick Observatory, University of California, 1156 High Street, Santa Cruz, CA 95064, USA}

\noindent \hangindent=.5cm $^{72}${Department of Astronomy and Astrophysics, University of California, Santa Cruz, 1156 High Street, Santa Cruz, CA 95065, USA}

\noindent \hangindent=.5cm $^{73}${Department of Astronomy \& Astrophysics, University of Toronto, Toronto, ON M5S 3H4, Canada}

\noindent \hangindent=.5cm $^{74}${University of Science and Technology, 217 Gajeong-ro, Yuseong-gu, Daejeon 34113, Republic of Korea}

\noindent \hangindent=.5cm $^{75}${Departament de F\'{i}sica, Serra H\'{u}nter, Universitat Aut\`{o}noma de Barcelona, 08193 Bellaterra (Barcelona), Spain}

\noindent \hangindent=.5cm $^{76}${Laboratoire de Physique Subatomique et de Cosmologie, 53 Avenue des Martyrs, 38000 Grenoble, France}

\noindent \hangindent=.5cm $^{77}${Instituci\'{o} Catalana de Recerca i Estudis Avan\c{c}ats, Passeig de Llu\'{\i}s Companys, 23, 08010 Barcelona, Spain}

\noindent \hangindent=.5cm $^{78}${Max Planck Institute for Extraterrestrial Physics, Gie\ss enbachstra\ss e 1, 85748 Garching, Germany}

\noindent \hangindent=.5cm $^{79}${Department of Physics and Astronomy, Siena College, 515 Loudon Road, Loudonville, NY 12211, USA}

\noindent \hangindent=.5cm $^{80}${Department of Physics \& Astronomy, University  of Wyoming, 1000 E. University, Dept.~3905, Laramie, WY 82071, USA}

\noindent \hangindent=.5cm $^{81}${National Astronomical Observatories, Chinese Academy of Sciences, A20 Datun Rd., Chaoyang District, Beijing, 100012, P.R. China}

\noindent \hangindent=.5cm $^{82}${Aix Marseille Univ, CNRS, CNES, LAM, Marseille, France}

\noindent \hangindent=.5cm $^{83}${Ruhr University Bochum, Faculty of Physics and Astronomy, Astronomical Institute (AIRUB), German Centre for Cosmological Lensing, 44780 Bochum, Germany}

\noindent \hangindent=.5cm $^{84}${Departament de F\'isica, EEBE, Universitat Polit\`ecnica de Catalunya, c/Eduard Maristany 10, 08930 Barcelona, Spain}

\noindent \hangindent=.5cm $^{85}${Aix Marseille Univ, CNRS/IN2P3, CPPM, Marseille, France}

\noindent \hangindent=.5cm $^{86}${University of California Observatories, 1156 High Street, Sana Cruz, CA 95065, USA}

\noindent \hangindent=.5cm $^{87}${Department of Physics \& Astronomy, Ohio University, Athens, OH 45701, USA}

\noindent \hangindent=.5cm $^{88}${Department of Physics and Astronomy, Sejong University, Seoul, 143-747, Korea}

\noindent \hangindent=.5cm $^{89}${Abastumani Astrophysical Observatory, Tbilisi, GE-0179, Georgia}

\noindent \hangindent=.5cm $^{90}${Faculty of Natural Sciences and Medicine, Ilia State University, 0194 Tbilisi, Georgia}

\noindent \hangindent=.5cm $^{91}${Space Telescope Science Institute, 3700 San Martin Drive, Baltimore, MD 21218, USA}

\noindent \hangindent=.5cm $^{92}${Centre for Advanced Instrumentation, Department of Physics, Durham University, South Road, Durham DH1 3LE, UK}

\noindent \hangindent=.5cm $^{93}${Physics Department, Brookhaven National Laboratory, Upton, NY 11973, USA}

\noindent \hangindent=.5cm $^{94}${Beihang University, Beijing 100191, China}

\noindent \hangindent=.5cm $^{95}${Department of Astronomy, Tsinghua University, 30 Shuangqing Road, Haidian District, Beijing, China, 100190}

\noindent \hangindent=.5cm $^{96}${Physics Department, Stanford University, Stanford, CA 93405, USA}

\noindent \hangindent=.5cm $^{97}${Department of Physics, University of California, Berkeley, 366 LeConte Hall MC 7300, Berkeley, CA 94720-7300, USA}

\end{document}